\documentclass[journal]{new-aiaa}

\usepackage[utf8]{inputenc}
\usepackage{textcomp}

\usepackage{graphicx}
\usepackage{amsmath, mathtools}
\usepackage{siunitx}
\usepackage{longtable,tabularx}
\usepackage{enumitem}
\usepackage{longtable,tabularx}
\usepackage{algorithm}
\usepackage{algpseudocode}
\usepackage{subfigure}

\usepackage{epstopdf}
\usepackage{booktabs}

\DeclarePairedDelimiter\floor{\lfloor}{\rfloor}

\title{Orbit-attitude predictive control in the vicinity of asteroids with in-situ gravity estimation}

\author{Julio C. Sanchez\footnote{Ph.D. Candidate, Aerospace Engineering Department; jsanchezm@us.es.} and Rafael Vazquez\footnote{Associate Professor, Aerospace Engineering Department; rvazquez1@us.es.}}
\affil{University of Seville, 41092, Seville, Spain}
\author{James D. Biggs\footnote{Associate Professor, Department of Aerospace Science and Technology; biggs\_james@yahoo.co.uk.} and Franco Bernelli-Zazzera\footnote{Full Professor, Department of Aerospace Science and Technology; franco.bernelli@polimi.it. Senior Member AIAA.}}
\affil{Polytechnic University of Milan, 20156, Milan, Italy}
	
\begin{document}
	
\maketitle

\begin{abstract}
This paper presents an integrated model-learning predictive control scheme for spacecraft orbit-attitude station-keeping in the vicinity of asteroids. The orbiting probe relies on optical and laser navigation while attitude measurements are provided by star trackers and gyroscopes. The asteroid gravity field inhomogeneities are assumed to be unkown a priori. The state and gravity model parameters are estimated simultaneously using an unscented Kalman filter. The proposed gravity model identification enables the application of a learning-based predictive control methodology. The predictive control allow for a high degree of accuracy since the predicted model is progressively identified in-situ. Consequently, the tracking errors decrease over time as the model accuracy increases. Finally, a constellation mission concept is analyzed in order to speed up the model identification process. Numerical results are shown and discussed.
\end{abstract}

\section*{Nomenclature}

{\renewcommand\arraystretch{1.0}
	\noindent\begin{longtable*}{@{}l @{\quad=\quad} l@{}}
		$\mathbf{A}$ & state matrix\\
		$\mathbf{a}$ & acceleration, m/s$^2$\\
		$a$ & semi-major axis, m\\
		$\mathbf{B}$ & control matrix\\
		$\mathbf{C}$ & attitude kinematics matrix\\
		$C_{ij}$ & spherical harmonics coefficients\\
		$e$ & eccentricity\\
		$i$ & inclination, rad\\
		$\mathbf{J}$ & inertia matrix, kg$\cdot$m$^2$\\
		$m$ & mass, kg\\
		$\mathbf{p}$ & pixel\\
		$\mathbf{R}$ & rotation matrix\\
		$R_e$ & normalization radius, m\\
		$r$ & orbital radius, m\\
		$S_{ij}$ & spherical harmonics coefficients\\
		$\mathbf{x}$ & state\\
		$\mathbf{T}$ & external torque, N$\cdot$m\\
		$t$ & time, s\\
		$\Delta\mathbf{x}$ & tracking error\\
		$\eta$ & satellite number within the constellation\\
		$\lambda$ & longitude, rad\\
		$\mu$ & standard gravitational parameter, kg$^3$/m$^2$\\
		$\nu$ & true anomaly, rad\\
		$\rho$ & ranging distance, m\\
		$\pmb{\sigma}$ & modified Rodrigues parameters\\
		$\pmb{\Phi}(t,t_0)$ & state transition matrix from $t_0$ to $t$\\
		$\phi$ & latitude, rad\\ 
		$\Omega$ & right ascension of the ascending node, rad\\
		$\omega$ & argument of periapsis, rad\\
		$\pmb{\omega}$ & angular velocity, rad/s\\
		$\omega_T$ & asteroid rotation rate, rad/s\\
\end{longtable*}}

\section{Introduction}

Future small bodies exploration can enable a deeper understanding of the early solar system and planetary processes \cite{Castillo2012}. Currently, OSIRIS-REX \cite{Berry2013} and Hayabusa 2 \cite{Watanabe2017} sample return missions, to the asteroids 101955 Bennu and 162173 Ryugu respectively, are underway. Future small body exploration missions include Lucy which will explore five Jupiter trojan asteroids at L4 \cite{Stanbridge2017}, the Psyche orbiter \cite{Williams2018} which plans to visit the rare metallic asteroid 16 Psyche and the DART mission which will see the spacecraft impact the minor body (Didymoon) of the binary system 65803 Didymos \cite{Cheng2018}. 

The spacecraft dynamics in the vicinity of an asteroid are complex and provide challenges for fuel efficient station-keeping. This is mainly due to the asteroid's inhomogeneous gravity field which yields large deviations from the Keplerian dynamics of spherical bodies \cite{Ceccaroni2013}. Moreover, the inhomogeneous gravity field could lead the spacecraft to escape trajectories or collision with the asteroid. To prevent this, natural trajectories maintaining some orbit parameters constant in average (also known as frozen orbits) have been identified \cite{Scheeres2012, Yu2016, Jean2019, Hanlun2019}. In addition, controlled spacecraft which hover above a point on the small body have also been proposed  \cite{Broschart2005}. Such asteroid orbit station-keeping requires a closed-loop control strategy. Furthermore, the frozen orbits are usually determined using simplified gravity and solar radiation pressure models. Consequently, this reference orbit will never be in accordance with real dynamics, thus requiring active closed-loop tracking control. To this end, several tracking methods have been proposed. Reference \cite{Kenshiro2019} designed a Lyapunov stable feedback controller, \cite{Biggs2019} proposed an active disturbance rejection control, \cite{Yu2020} developed an impulsive targeting method through state transition matrix exploitation (around a binary asteroid) and \cite{Taniguchi2020} considered a local proportional derivative control which recursively updates its gains via a learning method. However, these control approaches are based on disturbance attenuation or state targeting without optimizing the required control effort. A suitable methodology balancing control effort with respect to the tracking error is model predictive control (MPC), see \cite{Camacho2004}. This technique recursively updates the control sequence by solving an optimization problem (based on the state prediction). Model predictive control has been previously employed for low Earth orbits station-keeping \cite{Tavakoli2014}, orbit-attitude rendezvous control \cite{Malladi2019} and asteroid soft landing \cite{Carson2008, AlandiHallaj2017} amongst other applications. Additionally, its generic formulation allows to consider different thrusters models such as continuous \cite{Vazquez2017} or impulsive ones \cite{Sanchez2020_bis}. However, MPC relies on an accurate model of the system. Therefore, in the case of operating under model uncertainty, a direct MPC-based approach will not be optimal neither accurate.

Mission design to non-visited small bodies, such as asteroids and comets, is challenging since limited data of the target object (orbit, spin-rate and pole orientation) is usually known prior to the orbiting phase. Typically, a long in-situ characterization campaign \cite{Miller2002}, largely relying on Earth ground segment data processing, is carried out to accurately determine the body shape and its gravity field. In order to eliminate the ground segment dependency, recent research aimed to demonstrate autonomous gravity field estimation. Reference \cite{Hesar2015} combined optical navigation with satellite-to-satellite radiometric measurements for gravity field determination. A solar sail navigation technique was developed in \cite{Biggs2019} by system disturbance quantification, with an extended state observer, and a subsequent regression process with gravity and solar sail degradation. Unscented Kalman filtering (UKF), developed by \cite{Wan2000}, has been employed for orbit determination in \cite{Vetrisano2016, Dietrich2017, GilFernandez2018} while \cite{Stacey2018} demonstrated the feasibility of joint orbit and model parameters estimation. Following that trend, this work employs the UKF for a joint state and gravity inhomogeneities estimation, thus providing a navigation solution while reinforcing the control model prediction at the same time. For the sake of mission autonomy, the spacecraft solely relies on its own navigation devices. These are a camera and a laser imaging detection and ranging (LIDAR) for orbit measurements. The attitude measurements are provided by star-trackers and gyroscopes.

The optical and laser devices have to point to the asteroid surface for close-proximity navigation \cite{Wibben2012, Gaudet2020}. As a consequence, the spacecraft attitude must be accounted for in order to ensure navigation. Moreover, the orbit-attitude dynamics is coupled due to the gravity-gradient torque. This torque depends on the spacecraft position, with respect to the asteroid, and the spacecraft orientation with respect to the orbit frame \cite{Wie2008}. Attitude station-keeping around small bodies may possibly require active control since classical (assuming a circular orbit in an homogeneous gravity field) passive gravity-gradient stabilization regions may be altered under gravity field inhomogeneities \cite{Wang2013}. In that line, \cite{Kumar2007} counteracted the linearized higher-order gravity disturbance using feedback control, \cite{Zhu2019} proposed a non-linear attitude control law based on pulse-width modulation and \cite{Lee2020} designed a quaternion-based adaptive controller.   

This paper presents an integrated guidance, navigation and control (GNC) scheme for orbit-attitude station-keeping while simultaneously estimating the asteroid gravity. The main control objective is to maintain a closed orbit (though other options like constraining the orbital radius are also possible), for safety and to enable a good operation of the sensors (camera and laser). In that sense, the camera line-of-sight with the asteroid surface has to be maintained in order to recognize the landmarks. The guidance logic generates, by integration over the control horizon, a reference where only controlled states are prescribed by design \cite{Gazzino2017}. Then, a continuous optimization problem to track this reference is posed. This control program is reduced to a tractable quadratic programming (QP) form by means of linearization and discretization. The previous algorithms are embedded within a MPC scheme. Since the inhomogeneous gravity field is assumed unknown, the model serves only as an approximation leading to an initially inefficient control. However, by in-situ estimation of the gravity field inhomogenities, through UKF, the model is updated and the controller performance improves. The previous strategy follows the emerging paradigm of learning-based predictive control \cite{Hewing2020}. This novel approach tackles uncertain dynamics through model learning, thus improving the control accuracy.

The main contribution of this work is a learning-based GNC scheme with the distinctive feature of removing uncertainty from the system, thus recursively improving the predicted model accuracy. The numerical results and computational times assure this is a first step towards demonstrating the feasibility and autonomy of the proposed mission concept. Finally, the previous methodology is extended to a constellation of spacecraft which significantly improves in-situ gravity estimation accuracy and convergence (this novel mission concept could be possible due to recent advances in CubeSats for space exploration \cite{Benedetti2017}).

The structure of the paper is as follows. Section II introduces the orbit and attitude dynamics around an inhomogeneous gravity field. Section III presents the UKF based navigation strategy. Section IV develops the predictive guidance and control algorithm. Section V presents the integrated GNC scheme. Section VI shows numerical results of interest. Finally, Section VII concludes the paper with some remarks.

\section{Spacecraft dynamics in the vicinity of an asteroid}

\begin{figure}[] 
	\begin{center}
		\includegraphics[width=7.5cm,height=7.5cm,keepaspectratio]{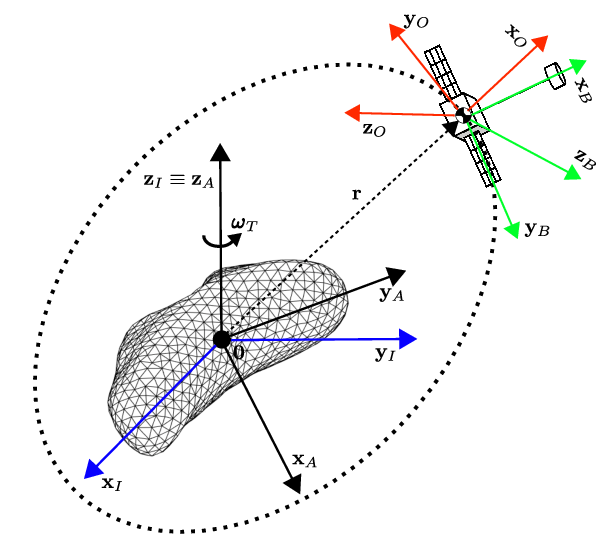}
		\caption{Inertial, asteroid, orbit, body and camera frames of reference.}	
		\label{fig:reference_frames}
	\end{center}
\end{figure}

In this section, the spacecraft translational and rotational motions, under the influence of an inhomogeneous gravity field, are presented. The main body is assumed to be uniformly rotating around its major inertia axis as this is the usual case for the majority of asteroids. Let denote the asteroid-centered inertial frame as $I\equiv\{\mathbf{0},\mathbf{i}_I,\mathbf{j}_I,\mathbf{k}_I\}$ being the origin, $\mathbf{0}$, the asteroid center of mass. Let denote the rotating asteroid frame as $A\equiv\{\mathbf{0},\mathbf{i}_A,\mathbf{j}_A,\mathbf{k}_A\}$ where $\mathbf{k}_A$ is aligned with the asteroid major inertia axis while $\mathbf{i}_A$ and $\mathbf{j}_A$ define its equatorial plane. Assuming the typical case of an asteroid uniformly rotating around its major inertia axis, the frame $A$ rotates with angular velocity $\pmb{\omega}_{A/I}=\omega_T \mathbf{k}_A$ ($\omega_T\equiv\text{constant}$) with respect to the inertial frame. The orbit frame is denoted by $O\equiv\{\mathbf{r},\mathbf{i}_O,\mathbf{j}_O,\mathbf{k}_O\}$ being $\mathbf{r}$ the satellite center of mass position, $\mathbf{i}_O$ the radial component (positive outwards the main body), $\mathbf{k}_O$ the out-of-plane component (parallel to the spacecraft angular momentum) and $\mathbf{j}_O$, the cross-track component, completes the right-hand system. Finally, let denote the spacecraft body frame as $B\equiv\{\mathbf{r},\mathbf{i}_B,\mathbf{j}_B,\mathbf{k}_B\}$ and the camera frame as $C\equiv\{\mathbf{r},\mathbf{i}_C,\mathbf{j}_C,\mathbf{k}_C\}$. The camera boresight is assumed to be aligned with the $\mathbf{k}_C$ direction. These frames of reference are depicted in Fig. \ref{fig:reference_frames}.

\subsection{Translational motion}

This paper employs the modified equinoctial elements (\text{MEE}), see \cite{Walker1985} for MEE details, for the translational state representation $\mathbf{x}_{\text{orb}}=[p,f,g,h,k,L]^T$. This parameterization provides smoother time variations than the cartesian representation while also avoiding the classic orbital elements, $\{a,e,\omega,i,\Omega,\nu\}$, singularities for circular ($e=0$) and equatorial orbits ($i=0^{\circ},~180^{\circ}$). However, the retrograde equatorial orbits, which are not considered in this paper, are still singular for the MEE. The Gauss variational equations (GVE) for the modified equinoctial elements are written as as $\dot{\mathbf{ x}}_{\text{orb}}=f(\mathbf{x}_{\text{orb}},\mathbf{a})$\cite{Walker1985}, where
%  follows  
%\begin{equation}
%	\left\lbrace{
%		\begin{aligned}
%			\dot{p}&=\frac{2p}{w}\sqrt{\frac{p}{\mu}}F^O_r,\\
%			\dot{f}&=\sqrt{\frac{p}{\mu}}\left[F^O_r\sin{L}+[(w+1)\cos{L}+f]\frac{F^O_t}{w}-(h\sin{L}-k\cos{L})\frac{gF^O_n}{w}\right],\\
%			\dot{g}&=\sqrt{\frac{p}{\mu}}\left[-F^O_r\cos{L}+[(w+1)\sin{L}+g]\frac{F^O_t}{w}+(h\sin{L}-k\cos{L})\frac{fF^O_n}{w}\right],\\
%			\dot{h}&=\sqrt{\frac{p}{\mu}}\frac{s^2F^O_n}{2w}\cos{L},\\
%			\dot{k}&=\sqrt{\frac{p}{\mu}}\frac{s^2F^O_n}{2w}\sin{L},\\
%			\dot{L}&=\sqrt{\mu p}\left(\frac{w}{p}\right)^2+\frac{1}{w}\sqrt{\frac{p}{\mu}}(h\sin{L}-k\cos{L})F^O_n,
%	\end{aligned}}\right.\label{eq:MEE_dynamics}
%\end{equation}
%where $s=\sqrt{1+h^2+k^2}$ and $w=1+f\cos L+g\sin L$. 
%T
the variable $\mathbf{a}=[a_r,~a_t,~a_n]^T$ is the non-Keplerian acceleration expressed in the orbit frame $O$. This term is composed of natural perturbations and the thrusters control
\begin{equation}
	\mathbf{a}=\underbrace{\mathbf{a}_{\text{grav}}}_{\text{inhomogeneous gravity}}+\underbrace{\mathbf{a}_{\text{sun}}}_{\text{Sun gravity}}+\underbrace{\mathbf{a}_{\text{SRP}}}_{\text{solar radiation pressure}}+\underbrace{\mathbf{a}_u.}_{\text{control acceleration}}
\end{equation}

\subsubsection{Inhomogeneous gravity field}

Assuming the spacecraft is placed in a low-asteroid orbit, the main natural non-Keplerian acceleration is the one related to the asteroid inhomogeneous gravity field. Several gravity models, such as spherical harmonics expansion series \cite{Balmino1994}, polyhedron shape \cite{Scheeres1996} or mass concentrations \cite{Chanut2015}, the so called mascons, can be employed to describe an inhomogeneous gravity field with adequate accuracy. Each model presents advantages and disadvantages. The spherical harmonics expansion series is able to account for both asteroid density inhomogeneities and shape irregularities in an indirect way. It is also relatively efficient in terms of computational burden since the series could be truncated at a certain degree. However, the spherical harmonics expansion diverges from the real gravity if the vehicle is very close to the asteroid surface (Brillouin zone). This is a concern if a descent and landing operation is considered. The polyhedron shape model could potentially account for shape irregularities with almost exact accuracy by refining the asteroid shape increasing the number of facets and vertexes. On the other hand, its computational burden is high since all the polyhedron facets and vertexes have to be evaluated. Moreover, it entails strong assumptions on the density distribution (constant in \cite{Scheeres1996} and linear in \cite{Urso2014}), thus failing to model contact binary asteroids such as 25143 Itokawa. The mass concentrations model describe the gravity field in terms of several point masses smartly distributed within the asteroid volume. It provides enough flexibility, being capable of accounting for both density and shape irregularities, with a low computational burden. However, its main drawback is the determination of an accurate mascons distribution as it typically requires to minimize the mismatch of the mascons predicted gravity (or potential) with respect to another gravity model.      

Since this paper aims to control a bounded orbit, the spherical harmonics model provides a convenient representation for estimation. The spherical harmonics non-Keplerian gravity is given by, see \cite{Howell2007}, the following expansion series truncated at $i_{\text{max}}$ degree
\begin{equation}
	\mathbf{a}^{S}_{\text{grav}}=\sum^{i_{\text{max}}}_{i=2}\sum^{i}_{j=0}\frac{\mu}{r^2}\left(\frac{R_e}{r}\right)^i\begin{bmatrix}
		-(i+1)P^{(j)}_i(C_{ij}\cos(j\lambda)+S_{ij}\sin(j\lambda))\\
		\dfrac{j}{\cos\phi}P^{(j)}_i(-C_{ij}\sin(j\lambda)+S_{ij}\cos(j\lambda))\\
		\cos{\phi}P^{(j)'}_i(C_{ij}\cos(j\lambda)+S_{ij}\sin(j\lambda))
	\end{bmatrix},\label{eq:aspherical_force_spherical_frame}
\end{equation}      
where $\mu$ is the asteroid main gravitational parameter, $r=\lVert\mathbf{r}\rVert_2$ is the orbital radius, $\lambda=\arctan{(y_A/x_A)}$ is the longitude (measured counter clockwise in the equatorial plane $x_Ay_A$) and $\phi=\arcsin(z_A/r)$ is the latitude. Note that Eq.\eqref{eq:aspherical_force_spherical_frame} is expressed in the $S$ frame. The $S$ frame denotes the spherical frame as $S\equiv\{\mathbf{r},\mathbf{i}_S,\mathbf{j}_S,\mathbf{k}_S\}$ with $\mathbf{i}_S$ being the radial direction, $\mathbf{j}_S$ pointing to the east and $\mathbf{k}_S$ to the north pole. The spherical harmonics coefficients $C_{ij}$ and $S_{ij}$ are normalized with respect to the normalization radius $R_e$ (which is usually taken as the asteroid maximum elongation). The term $P^{(j)}_i$ is the $i\text{th}$ degree normalized Legendre polynomial of the first kind in $\sin\phi$ and $P^{(j)'}_i$ is its first derivative with respect to $\sin\phi$, see \cite{Howell2007}.
%
%which can be computed recursively as 
%\begin{equation}
%	P^{(j)}_i=\left\lbrace{
%		\begin{aligned}
%			&\sqrt{\frac{2i+1}{(i+j)(i-j)}}\left(\sqrt{2i-1}\sin\phi P^{(j)}_{i-1}-\sqrt{\frac{(i+j-1)(i-j-1)}{2i-3}}P^{(j)}_{i-2}\right),\quad &i<j,\\
%			&\sqrt{\frac{2i+1}{2i}}\cos\phi P^{(j-1)}_{i-1},\quad &i=j,\\
%			&0,\quad &i>j,
%	\end{aligned}}\right.\label{eq:normalized_legendre_polynomials}
%\end{equation}
%initiating the sequence with $P^{(0)}_0=1$, $P^{(0)}_1=\sqrt{3}\sin\phi$ and $P^{(1)}_1=\sqrt{3}\cos\phi$. The term 
%$P^{(j)'}_i$ is the $i\text{th}$ degree normalized Legendre polynomial with respect to its argument, $\sin\phi$,
%\begin{equation}
%	P^{(j)'}_i=\frac{1}{\cos^2\phi}\left(-i\sin\phi P^{(j)}_i+\sqrt{\frac{(2i+1)(i+j)(i-j)}{2i-1}}P^{(j)}_{i-1}\right).
%\end{equation}

To insert the non-Keplerian gravity into the GVE, the expression given by Eq.\eqref{eq:aspherical_force_spherical_frame} has to be projected in the orbit frame
\begin{equation}
	\mathbf{a}_{\text{grav}}(\mathbf{x}_{\text{orb}})=\mathbf{R}^O_I(\mathbf{x}_{\text{orb}})\mathbf{R}^I_A\mathbf{R}^A_S(\lambda,\phi)\mathbf{a}^S_{\text{grav}}(r,\lambda,\phi),
\end{equation}
where $\mathbf{R}^{A}_{S}$ denotes the rotation matrix from the spherical to the asteroid frame, $\mathbf{R}^{I}_{A}$ is the rotation matrix from the asteroid to the inertial frame (it only depends on the asteroid rotation rate $\omega_T$ which is constant) and $\mathbf{R}^{O}_{I}$ denotes the rotation matrix from the inertial to the orbit frame. Note that $[r,\lambda,\phi]^T\equiv\mathbf{f}_{\text{orb}}(\mathbf{x}_{\text{orb}})$ where $\mathbf{f}_{\text{orb}}(\mathbf{x}_{\text{orb}}):\mathbb{R}^6\rightarrow\mathbb{R}^3$ is the mapping function between MEE and spherical coordinates in the asteroid frame.

\subsubsection{Solar perturbations}
	The solar perturbations are the Sun third-body effect and its radiation pressure. Note that these perturbations will be expressed in the orbit frame $O$. The Sun third-body perturbation is
	\begin{equation}
		\mathbf{a}_{\text{sun}}(\mathbf{x}_{\text{orb}})=\mathbf{R}^{O}_{I}(\mathbf{x}_{\text{orb}})\mu_{\odot}\left(\frac{\mathbf{r}_{\odot}-\mathbf{r}}{\lVert\mathbf{r}_{\odot}-\mathbf{r}\rVert^3_2}-\frac{\mathbf{r}_{\odot}}{\lVert\mathbf{r}_{\odot}\rVert^3_2}\right),
	\end{equation}
	where $\mathbf{r}_{\odot}$ is the Sun position in the asteroid-centered inertial frame. The term $\mu_{\odot}=1.3271244\cdot 10^{11}~\text{km}^3/\text{s}^2$ is the Sun standard gravity parameter. The solar radiation pressure is considered in a simplified way as
	\begin{equation}
		\mathbf{a}_{\text{SRP}}(\mathbf{x}_{\text{orb}})=-\mathbf{R}^O_I(\mathbf{x}_{\text{orb}})\frac{C_Rp_{\text{1AU}}A}{m}\left(\frac{r_{\text{1AU}}}{r_{\odot}}\right)^2\frac{\mathbf{r}_{\odot}-\mathbf{r}}{\lVert\mathbf{r}_{\odot}-\mathbf{r}\rVert_2},
	\end{equation}
	where $C_R$, $A$ and $m$ are the spacecraft reflectivity coefficient, exposed surface and mass respectively. The term $p_{\text{1AU}}=4.5~\mu \text{Pa}$ is the solar radiation pressure at the distance of one astronomical unit $r_{\text{1AU}}=1~\text{AU}$ \cite{Modest2003}.

\subsection{Rotational motion}

In this paper, the modified Rodrigues parameters (MRP), see \cite{Schaub1996}, $\pmb{\sigma}=[\sigma_1,\sigma_2,\sigma_3]^T$  are chosen to represent the spacecraft attitude. They are preferred over the classical quaternions since they do not need to account for the unit-norm constraint, thus easing optimization constraints. The MRP relation with the rotation axis $\mathbf{e}_{\text{rot}}\in\mathbb{R}^3$ and angle $\theta_{\text{rot}}\in\mathbb{R}$ is $\pmb{\sigma}=\mathbf{e}_{\text{rot}}\tan(\theta_{\text{rot}}/4)$. Note that singularities arise when $\theta_{\text{rot}}=\pm2\pi$. However, these singularities could be avoided by constraining $\theta_{\text{rot}}\in[-\pi,\pi]$ since $\{\mathbf{e}_{\text{rot}},~\theta_{\text{rot}}\}\equiv\{-\mathbf{e}_{\text{rot}},~2\pi-\theta_{\text{rot}}\}$ represent the same attitude. The rotation matrix $\mathbf{R}$, as a function of the MRP, is given by
\begin{equation}
	\mathbf{R}(\pmb{\sigma})=\mathbf{I}+\frac{8\pmb{\sigma}^{\times}\pmb{\sigma}^{\times}-4(1-||\pmb{\sigma}||^2_2)\pmb{\sigma}^{\times}}{(1+||\pmb{\sigma}||^2_2)^2},
\end{equation}
being $\pmb{\sigma}^{\times}\in\mathbb{R}^{3\times3}$ the cross-product matrix associated to a MRP, see \cite{Wie2008}. The MRP atttitude composition rule is given by
\begin{equation}
	\pmb{\sigma}_0\xrightarrow{\pmb{\sigma}_{\text{rot}}}\pmb{\sigma}_f,\quad\pmb{\sigma}_f =\frac{\left(1-\lVert \pmb{\sigma}_{\text{rot}} \rVert^2_2\right)\pmb{\sigma}_0+\left(1-\lVert \pmb{\sigma}_0 \rVert^2_2\right)\pmb{\sigma}_{\text{rot}}+2\pmb{\sigma}_{0}\times\pmb{\sigma}_{\text{rot}}}{1+\left(\lVert \pmb{\sigma}_{\text{rot}} \rVert_2 \lVert \pmb{\sigma}_0 \rVert_2\right)^2 - 2\pmb{\sigma}^T_{\text{rot}}\pmb{\sigma}_0}. \label{eq:att_rot}
\end{equation}
The MRP attitude kinematics is as follows
\begin{equation}
	\dot{\pmb{\sigma}}=\frac{1}{4}\mathbf{C}(\pmb{\sigma})\pmb{\omega},\label{eq:attitude_kinematics_inertial}
\end{equation}
where $\pmb{\omega}=[\omega_1,\omega_2,\omega_3]^T$ is the body angular velocity with respect to the inertial frame $I$ expressed in the body frame $B$. The matrix $\mathbf{C}$ is given by
\begin{equation}
\mathbf{C}(\pmb{\sigma})=\begin{bmatrix}
1+\sigma^2_1-\sigma^2_2-\sigma^2_3 & 2(\sigma_1\sigma_2-\sigma_3) & 2(\sigma_1\sigma_3+\sigma_2)\\
2(\sigma_1\sigma_2+\sigma_3) & 1-\sigma^2_1+\sigma^2_2-\sigma^2_3 & 2(\sigma_2\sigma_3-\sigma_1)\\
2(\sigma_1\sigma_3-\sigma_2) & 2(\sigma_2\sigma_3+\sigma_1) & 1-\sigma^2_1-\sigma^2_2+\sigma^2_3\\
\end{bmatrix}.
\end{equation}
As one of the control objectives is to maintain the body orientation with respect to the orbit frame, $\pmb{\sigma}_{B/O}$, the following kinematics equation is also employed \cite{Wie2008}
\begin{equation}
	\dot{\pmb{\sigma}}_{B/O}=\frac{1}{4}\mathbf{C}(\pmb{\sigma}_{B/O})\left[\pmb{\omega}-\mathbf{R}(\pmb{\sigma}_{B/O})\pmb{\omega}^O_{O/I}(\mathbf{x}_{\text{orb}})\right],\label{eq:attitude_kinematics_orbit}
\end{equation}
where $\pmb{\omega}^O_{O/I}$ is the angular velocity of the orbit frame with respect to the inertial frame. In that line, let define the attitude state as $\mathbf{x}_{\text{att}}=[\pmb{\sigma}_{B/O}^T,~\pmb{\omega}^T]^T$.

The attitude dynamics is as follows
\begin{equation}
	\mathbf{J}\dot{\pmb{\omega}}=\mathbf{T}-\pmb{\omega}\times\mathbf{J}\pmb{\omega},\quad\quad\mathbf{T}=\underbrace{\mathbf{T}_{\text{grav}}}_{\text{gravity-gradient}}+\underbrace{\mathbf{T}_u}_{\text{control torque}},\label{eq:attitude_dynamics}
\end{equation}
where $\mathbf{J}\in\mathbb{R}^{3\times3}$ is the probe inertia matrix and the external torque $\mathbf{T}\in\mathbb{R}^3$ is composed of the gravity-gradient torque, $\mathbf{T}_{\text{grav}}$ and the control torque $\mathbf{T}_u$. In this manuscript, the gravity-gradient torque model is based on a discrete mass distribution. Alternatively, there are also analytic models considering higher order inertia moments along with second order gravity terms \cite{Wang2013_bis}. However, the discrete mass distribution is more accurate (since both higher order gravity and inertia terms are considered), at the expense of higher computational cost. Let define the spacecraft mass distribution through $l_{\text{max}}$ discrete masses $m_l$ ($l=1\hdots l_{\text{max}}$) placed at $\Delta\mathbf{r}_l$ relative positions with respect to the spacecraft center of mass $\mathbf{r}$. Then, the gravity gradient torque is computed as
\begin{equation}
	\mathbf{T}_{\text{grav}}=\sum^{l_{\text{max}}}_{l=1}m_l\Delta\mathbf{r}^B_l\times\mathbf{a}^B(\mathbf{r}+\Delta\mathbf{r}_l,~\pmb{\sigma}),\label{eq:gravity_gradient_torque}
\end{equation}
where the term $\mathbf{a}^B$ collects the Keplerian and non-Keplerian gravity as
\begin{equation}
	\mathbf{a}^B(\mathbf{r},~\pmb{\sigma})=\mathbf{R}^B_O(\pmb{\sigma})\left[\mathbf{a}_{\text{main}}(\mathbf{r})+\mathbf{a}_{\text{grav}}(\mathbf{r})\right],
\end{equation}
being $\mathbf{a}_{\text{main}}=[-\mu/r^2,0,0]^T$ the Keplerian gravity. Note that $\mathbf{r}\equiv\mathbf{f}(\mathbf{x}_{\text{orb}})$.

\section{Asteroid navigation with in-situ gravity estimation}

This section describes the asteroid navigation strategy which relies on the unscented Kalman filter with process noise estimation. The orbit and attitude estimation are treated separately since their state variations and measurements frequencies are different. The orbit navigation is assumed to rely on optical devices (camera and LIDAR) whereas a star-tracker and gyroscopes are considered for the attitude estimation. In both cases, the state and gravity parameters are jointly estimated through the extended state.  

\subsection{UKF with process noise estimation}

The UKF is a sub-optimal state-of-the-art non-linear estimator. Its main advantage is that it does not rely on linearization so the probabilistic distributions are not affected by linearization error when propagated (as this is the case for the extended Kalman filter). However, the UKF is more computationally intensive with respect to the extended Kalman filter as it requires propagation of several statistical realizations. 

The UKF assumes that both the estimation variable and measurements are statistically distributed as multivariate gaussians $N_n$ with dimension $n$. Define the extended state (including both state and model parameters) as $\mathbf{y}\sim N_n(\pmb{\mu},\pmb{\Sigma})$ where $\pmb{\mu}\in\mathbb{R}^n$ is the mean and $\pmb{\Sigma}\in\mathbb{R}^{n\times n}$ the covariance matrix. Define the measurements as $\mathbf{z}\sim N_m(\pmb{\epsilon},\mathbf{Q}'_z)$ where $\pmb{\epsilon}\in\mathbb{R}^m$ is the mean and $\mathbf{Q}'_z\in\mathbb{R}^{m\times m}$ the covariance matrix. Define the extended state propagation between UKF calls, named as process function, as $\mathbf{g}(\mathbf{y}):\mathbb{R}^n\rightarrow\mathbb{R}^n$. Define the transformation function, from extended state space to measurement space, as $\mathbf{h}(\mathbf{y}):\mathbb{R}^n\rightarrow\mathbb{R}^m$. Define the UKF process and measurement noises covariance matrices as $\mathbf{Q}_y\in\mathbb{R}^{n\times n}$ and $\mathbf{Q}_z\in\mathbb{R}^{m\times m}$ respectively. The process noise is unknown as it quantifies the process error with respect to the reality, thus it is a tuning parameter. Alternatively, measurements noise covariance is typically provided by sensor datasheets as $\mathbf{Q}_z$ though some discrepancies with the real noise may arise, $\mathbf{Q}_z\neq\mathbf{Q}'_z$. On the other hand, the process noise covariance matrix is a hard to tune parameter, even more in the case of model parameters estimation, since it highly depends on the process propagation mismatch with respect to the real dynamics. A workaround to solve this issue consists in adding an extra step to the nominal UKF algorithm in order to obtain an estimation of the process noise covariance $\mathbf{Q}_y$, see \cite{Akhlaghi2017}, for the next UKF call.  

The UKF with process noise estimation is presented in Algorithm \ref{alg1}. Each time a new measurement $\mathbf{z}$ is available, the UKF is called. The first step is to generate $2n+1$ initial propagation conditions (sigma points $\pmb{\chi}$) from the last extended state mean, $\pmb{\mu}_0$, and covariance, $\pmb{\Sigma}_0$, estimation, see Eq.\eqref{eq:UKF_gen_sigmapoints}. Then, each one of these sigma points is propagated through the process function, $\mathbf{g}$, and the final propagated extended state mean $\pmb{\mu}'$ and covariance $\pmb{\Sigma}'$ are reconstructed by means of averaging, see Eq.\eqref{eq:UKF_reconstruct_mean_covariance}. The previous steps are known as the extended state unscented transform. The next step consists in transforming the sigma points from state space to measurement space through the transformation function $\mathbf{h}$. Then, the predicted measurement mean $\hat{\mathbf{z}}$ and covariance $\mathbf{S}$ are computed by averaging, see Eq.\eqref{eq:UKF_space_to_measurement}. Joining the propagated extended state and its associated measurements prediction, the cross-correlation covariance $\mathbf{H}\in\mathbb{R}^{n\times m}$ and Kalman gain $\mathbf{K}\in\mathbb{R}^{n\times m}$ matrices are computed via Eq.\eqref{eq:UKF_crosscorrelation_kalmangain}. Applying the Kalman gain to the measurements, $\mathbf{z}$, a new extended state prediction, as a gaussian mean ($\pmb{\mu}$) and covariance ($\pmb{\Sigma}$) is obtained through Eq.\eqref{eq:UKF_state_prediction}. Finally, the state innovation $\hat{\mathbf{w}}$, which is related to the mismatch between the process and real dynamics, is computed, through Eq.\eqref{eq:UKF_process_noise_covariance}, and employed to obtain a process noise covariance prediction $\hat{\mathbf{Q}}_y$ for the next UKF call. To ease the UKF computational load, the process noise covariance is computed by applying a fading factor $\alpha\in[0,1]$ to the last $\mathbf{Q}_y$ estimation.
\begin{algorithm}[t]
	\caption{(UKF with process noise estimation)}
	\textbf{Input:} $\pmb{\mu}_0$, $\pmb{\Sigma}_0$, $\mathbf{z}$, $\mathbf{Q}_y$, $\mathbf{Q}_z$\\
	\textbf{Output:} $\pmb{\mu}$, $\pmb{\Sigma}$, $\hat{\mathbf{Q}}_y$
	\begin{algorithmic}[1]
		\State Generate sigma points:
		\begin{equation}
			\pmb{\chi}^{[k]}=\pmb{\mu}_0+\left(\sqrt{(n+\lambda)\pmb{\Sigma}_0}\right)_{k+n}, \>\> k=-n \hdots n,\label{eq:UKF_gen_sigmapoints}
		\end{equation}
		where the subindex $k+n$ denotes the column.
		\State Propagate sigma points (function $\mathbf{g}$) and compute process mean and covariance:
		\begin{equation}
			\pmb{\mu}'=\sum^{n}_{k=-n}w_m^{[k]}\mathbf{g}(\pmb{\chi}^{[k]}),~
			\pmb{\Sigma}'=\sum^{n}_{k=-n}w_c^{[k]}\left(\mathbf{g}(\pmb{\chi}^{[k]})-\pmb{\mu}'\right)\left(\mathbf{g}(\pmb{\chi}^{[k]})-\pmb{\mu}'\right)^T+\mathbf{Q}_y.\label{eq:UKF_reconstruct_mean_covariance}
		\end{equation}
		\State Transform the propagated sigma points to measurement space (function $\mathbf{h}$):
		\begin{equation}
			\begin{aligned}
				\mathbf{Z}^{[k]}=\mathbf{h}(\mathbf{g}(\pmb{\chi}^{[k]})),\quad\hat{\mathbf{z}}=\sum^{n}_{k=-n}w_m^{[k]}\mathbf{Z}^{[k]},\\
				\mathbf{S}=\sum^{n}_{k=-n}w_c^{[k]}\left(\mathbf{Z}^{[k]}-\hat{\mathbf{z}}\right)\left(\mathbf{Z}^{[k]}-\hat{\mathbf{z}}\right)^T+\mathbf{Q}_z.\label{eq:UKF_space_to_measurement}
			\end{aligned}
		\end{equation}
		\State Compute the cross-correlation covariance matrix and Kalman gain:
		\begin{equation}
			\mathbf{H}=\sum^{n}_{k=-n}w_c^{[k]}\left(\mathbf{g}(\pmb{\chi}^{[k]})-\pmb{\mu}'\right)\left(\mathbf{Z}^{[k]}-\hat{\mathbf{z}}\right)^T,\quad\mathbf{K}=\mathbf{H}\mathbf{S}^{-1}.\label{eq:UKF_crosscorrelation_kalmangain}
		\end{equation}
		\State Predict the state and its covariance:
		\begin{equation}
			\pmb{\mu}=\pmb{\mu}'+\mathbf{K}(\mathbf{z}-\hat{\mathbf{z}}),\quad\pmb{\Sigma}=\pmb{\Sigma}'-\mathbf{K}\mathbf{H}\pmb{\Sigma}'.\label{eq:UKF_state_prediction}
		\end{equation}
		\State Estimate the process noise and its covariance:
		\begin{equation}
			\hat{\mathbf{w}}=\mathbf{K}(\mathbf{z}-\hat{\mathbf{z}}), \quad \hat{\mathbf{Q}}_y=(1-\alpha)\hat{\mathbf{w}}\hat{\mathbf{w}}^T+\alpha\mathbf{Q}_y.\label{eq:UKF_process_noise_covariance}
		\end{equation}
	\end{algorithmic}\label{alg1}
\end{algorithm}

Apart from the process and measurement noises covariances, the UKF has several tuning parameters $\{\alpha,\theta,\beta,\lambda\}$. The variable $\alpha$, as explained before, controls the pace at which the process noise is updated. The variables $\theta$, $\beta$ and $\lambda$ arise through the mean $w^{[k]}_m$, and covariance, $w^{[k]}_c$ weights which are usually chosen following \cite{Wan2000} guidelines
\begin{equation}
	\begin{aligned}
		w^{[0]}_{m}&=\frac{\lambda}{n+\lambda},\quad w^{[0]}_{c}=\frac{\lambda}{n+\lambda}+(1-\theta^2+\beta),
		%\\
		\quad
		w^{[k]}_{c}=w^{[k]}_{m}
		%&
		=\frac{1}{2(n+\lambda)}~~\text{for}~~k\neq0.
	\end{aligned}
\end{equation}
The parameter $\beta$ encodes information about the underlying statistical distribution. Under a Gaussian assumption, its optimal value is $\beta=2$. The parameters $\lambda$ and $\theta$ control the spread of sigma points and weights. These two parameters affect the UKF transient response.

\subsection{Orbit estimation}

The orbit extended state is taken as $\mathbf{y}_{\text{orb}}\in\mathbb{R}^{6+\sum^{n_{\text{orb}}}_{i=2}2i+1}$ where $n_{\text{orb}}$ represents the higher order gravity degree to be estimated
\begin{equation}\label{eq:orb_extended_state}
	\mathbf{y}_{\text{orb}}=[p,~f,~g,~h,~k,~L,~C_{ij},~S_{ij}]^T,\quad i=2\hdots n_{\text{orb}},\quad j=0\hdots i.
\end{equation}
The predicted orbit extended state is its expected value $\pmb{\mu}_{\text{orb}}=[\hat{\mathbf{x}}^T_{\text{orb}},~\hat{C}_{ij},~\hat{S}_{ij}]^T$, where the hat denotes each component mean. The orbit process function $\mathbf{g}_{\text{orb}}:\mathbb{R}^{6+\sum^{n_{\text{orb}}}_{i=2}2i+1}\rightarrow\mathbb{R}^{6+\sum^{n_{\text{orb}}}_{i=2}2i+1}$ is as follows
\begin{equation}
	\mathbf{g}_{\text{orb}}=[(\varphi^{t,t_0}_{\text{orb}}(\mathbf{x}_{\text{orb},0}))^T, ~C_{ij},~S_{ij}]^T,\label{eq:orbit_process}
\end{equation}
where $\varphi^{t,t_0}_{\text{orb}}(\mathbf{x}_{\text{orb},0}):\mathbb{R}^6\rightarrow\mathbb{R}^6$ is the MEE dynamical flow for the GVE with inhomogeneous gravity field perturbation up to $n_{\text{orb}}\times n_{\text{orb}}$ degree and order. Let recall that the input to the the process function $\mathbf{g}_{\text{orb}}$ are the orbit extended state sigma points $\pmb{\chi}_{\text{orb}}$, see steps 1-2 of algorithm \ref{alg1}. The commanded control acceleration is known but solar perturbations are not included in the process. The gravitational parameters are held constant along the process.

In this work, the orbit measurements are provided by means of a camera and LIDAR. The camera is able to track previously identified landmarks on the asteroid surface \cite{Miller1995}. This previous phase could also enable a preliminary estimation of the inhomogeneous gravity field which has not been considered available for the sake of generality. A landmark is a surface feature which is easily distinguishable (e.g. craters). The landmark positions dataset is assumed available as they are surveyed during asteroid approach and high orbit operations phases. In particular, the camera provides the pixel row and column of a number of $q_{\text{max}}$ tracked landmarks \cite{Pellacani2018}, on the image plane, as $\mathbf{p}_q=[p_{x_q},p_{y_q}]^T$ . The sub-index $q$ refers to each tracked landmark identifed by the camera landmark recognizition algorithn (not considered in this work). Guided by the camera line-of-sight, the LIDAR provides ranging pseudodistance, $\rho_{q}$, to the tracked landmarks. In view of the previous facts, the orbit navigation measurement is 
	\begin{equation}
		\mathbf{z}_{\text{orb}}=\left[\mathbf{p}^T_1,\rho_1,\hdots,\mathbf{p}^T_{q_{\text{max}}},\rho_{q_{\text{max}}}\right]^T, \quad q=1\hdots q_{\text{max}}.
	\end{equation}
	Following \cite{Vetrisano2016}, a pinhole camera model is employed to describe the orbit state transformation to landmarks pixels. Let express the landmark-spacecraft relative distance $\pmb{\rho}_q$ in the camera frame as
	\begin{equation}
		\pmb{\rho}^C_{q}=\mathbf{R}^{C}_{B}\mathbf{R}^{B}_{I}(\pmb{\sigma}_{B/I})\left[\mathbf{R}^{I}_{A}\mathbf{r}^{A}_q-\mathbf{f}_{\text{orb}}(\mathbf{x}_{\text{orb}})\right],\label{eq:rho_lmk}
	\end{equation}
	where the function $\mathbf{f}_{\text{orb}}(\mathbf{x}_{\text{orb}}):\mathbb{R}^6\rightarrow\mathbb{R}^3$ transforms the MEE to cartesian position in the inertial frame $\mathbf{r}^I$. The term $\mathbf{r}^{A}_q$ denotes the landmark position in the asteroid frame. Note that the body orientation arises through $\pmb{\sigma}_{B/I}$ in the rotation matrix from inertial to body frame. The previous orientation is obtained through the attitude filter. The rotation matrix $\mathbf{R}^{C}_{B}$ is the camera orientation with respect to the body. Assuming the camera boresight aligned with the $z_C$ direction, a landmark is projected on the image plane as
	\begin{equation}
		u_q=\frac{f_{\text{foc}}}{\rho_{z_q}}\rho_{x_q},\quad v_q=\frac{f_{\text{foc}}}{\rho_{z_q}}\rho_{y_q},
	\end{equation}
	where $f_{\text{foc}}$ is the focal length of the camera. The pixel row and column are obtained as
	\begin{equation}
		\mathbf{p}=[p_{x_q},p_{y_q}]^T=\left[\floor*{u_q/p_{\text{width}}},\floor*{v_q/p_{\text{width}}}\right]^T,\label{eq:pixel_lmk}
	\end{equation}
	where $p_{\text{width}}$ is the pixel width and $\floor*{}$ denotes the floor operator (since $p_{x_q},p_{y_q}\in\mathbb{Z}$). Under the previous formulation, the camera boresight is the pixels origin. Let define $\mathbf{h}_{\text{cam},q}(\mathbf{x}_{\text{orb}},\pmb{\sigma}_{B/I}):\mathbb{R}^6\rightarrow\mathbb{R}^2$ as the function transforming MEE to camera pixel $\mathbf{p}_q$, through application of Eq.\eqref{eq:rho_lmk}-\eqref{eq:pixel_lmk}. The fact that the camera orientation, with respect to inertial frame, arises in Eq.\eqref{eq:rho_lmk} leads to the control requirement of guaranteeing camera line-of-sight directed to the asteroid surface (e.g. by aligning the camera boresight with the orbit radial direction towards the asteroid). The orbit extended state transformation function, to camera and LIDAR measurements, is defined as $\mathbf{h}_{\text{orb}}(\mathbf{y}_{\text{orb}},\pmb{\sigma}_{B/I}):\mathbb{R}^{6+\sum^{n_{\text{orb}}}_{i=2}2i+1}\rightarrow\mathbb{R}^{3q_{\text{max}}}$
	\begin{equation}
		\mathbf{h}_{\text{orb}}(\mathbf{y}_{\text{orb}})=\left[\mathbf{h}^T_{\text{cam},1},\rho_1,\hdots,\mathbf{h}^T_{\text{cam},q_{\text{max}}},
		\rho_{q_{\text{max}}}\right]^T,
	\end{equation}
	where the right-hand side dependencies in terms of the orbit state are ommited for the sake of clarity. Similarly, the body orientation dependency has not been explicitly stated as it is not updated within the orbit filter. The LIDAR ranging distance is obtained as the landmark-spacecraft distance $\rho_q=\lVert\pmb{\rho}^{C}_q\rVert_2$.

\subsection{Attitude estimation}

The attitude extended state is defined as $\mathbf{y}_{\text{att}}\in\mathbb{R}^{9+\sum^{n_{\text{att}}}_{i=2}2i+1}$ where $n_{\text{att}}$ represents the higher spherical harmonics order and degree to be estimated. Usually $n_{\text{att}}\leq n_{\text{orb}}$ since the asteroid gravity field inhomogeneities influence the gravity-gradient torque less than the gravity acceleration. Moreover, the attitude filter works at a higher sampling rate than the orbit one, hence the computational load should be reduced. 
\begin{equation}
	\begin{aligned}
	\mathbf{y}_{\text{att}}=[\pmb{\sigma}^T,~\pmb{\omega}^T,~C_{ij},~S_{ij},~\Delta\pmb{\omega}^T_{\text{gyro}}]^T,\quad i=2\hdots n_{\text{att}},\quad j=0\hdots i,
	\end{aligned}\label{eq:att_extended_state}
\end{equation}
where $\Delta\pmb{\omega}_{\text{gyro}}=[\Delta\omega_{\text{gyro},1},~\Delta\omega_{\text{gyro},2},~\Delta\omega_{\text{gyro},3}]^T$ represents the gyroscope bias. The gyroscope bias is assumed as constant for the sake of simplicity. Note that the body orientation with respect to the inertial frame, $\pmb{\sigma}$, is the one considered for attitude navigation since, as detailed below, attitude sensors measurements are taken with respect to the inertial frame. The orientation of the body with respect to the orbit frame $\pmb{\sigma}_{B/O}$ is subsequently reconstructed with the orbit filter output using Eq.(10) composition rule as $\pmb{\sigma}_{B/I}\xrightarrow{-\pmb{\sigma}_{O/I}}\pmb{\sigma}_{B/O}$. Note that $\pmb{\sigma}_{O/I}\equiv\pmb{\sigma}_{O/I}(\mathbf{x}_{\text{orb}})$. Then, the predicted attitude extended state is given by its expectation $\pmb{\mu}_{\text{att}}=[\hat{\pmb{\sigma}}^T,~\hat{\pmb{\omega}}^T,~\hat{C}_{ij},~\hat{S}_{ij},~\Delta\hat{\pmb{\omega}}^T_{\text{gyro}}]^T$. The attitude process function, $\mathbf{g}_{\text{att}}:\mathbb{R}^{9+\sum^{n_{\text{att}}}_{i=2}2i+1}\rightarrow\mathbb{R}^{9+\sum^{n_{\text{att}}}_{i=2}2i+1}$, is as follows
\begin{equation}
	\mathbf{g}_{\text{att}}=\left[\left(\varphi^{t,t_0}_{\text{att}}(\pmb{\sigma}_{0},~\pmb{\omega}_0)\right)^T, ~C_{ij},~S_{ij},~\Delta\pmb{\omega}^T_{\text{gyro}}\right]^T,\label{eq:attitude_process}
\end{equation}
where $\varphi^{t,t_0}_{\text{att}}(\pmb{\sigma}_{0},~\pmb{\omega}_0):\mathbb{R}^6\rightarrow\mathbb{R}^6$ is the flow for the attitude dynamics given by Eq.\eqref{eq:attitude_kinematics_inertial} and Eq.\eqref{eq:attitude_dynamics} with the gravity-gradient torque assuming $n_{\text{att}}\times n_{\text{att}}$ degree and order of spherical harmonics. Let recall that the input to the the process function $\mathbf{g}_{\text{att}}$ are the attitude extended state sigma points $\pmb{\chi}_{\text{att}}$, see steps 1-2 of algorithm \ref{alg1}. The gravity-gradient torque depends on the orbit state, see Eq.\eqref{eq:gravity_gradient_torque}. Consequently, an orbit propagation integrating the GVE is required within the attitude process. Both the gravity parameters and gyroscopes bias are held constant along the process.

In this paper the attitude sensors are assumed to be a star-tracker and gyroscopes. The star-tracker directly measures the body attitude with respect to the inertial frame, $\pmb{\sigma}_{\text{star}}$. On the other hand, gyroscopes are able to provide the body angular velocity with respect to the inertial frame, $\pmb{\omega}_{\text{gyro}}$. Consequently the following measurements are available when the attitude filter is called
\begin{equation}
	\mathbf{z}_{\text{att}}=\left[\left(\pmb{\sigma}_{\text{star}}\right)^T,~\left(\pmb{\omega}_{\text{gyro}}\right)^T\right]^T.
\end{equation}
The attitude transformation function $\mathbf{h}_{\text{att}}(\pmb{\mu}_{\text{att}}):\mathbb{R}^{9+\sum^{n_{\text{att}}}_{i=2}2i+1}\rightarrow\mathbb{R}^6$ transforms the attitude extended state to the measurement space
\begin{equation}
	\mathbf{h}_{\text{att}}=\left[\pmb{\sigma}^T,~(\pmb{\omega}+\Delta\pmb{\omega}_{\text{gyro}})^T\right]^T.
\end{equation}

\section{Model-learning predictive control}\label{MPC}

The guidance and control module objective is to station-keep the probe in a bounded orbit while maintaining a stationary attitude with respect to the orbit frame. To this end, MPC is employed for both attitude and orbit guidance and control. The guidance algorithm generates a reference by integrating the dynamical model (up to the filters knowledge) with a predefined control policy. Note that the reference is inherently affected by model errors which will diminish progressively. Then, in order to track the guidance reference, a control program minimizing a weighted sum of the tracking error and control effort is posed.

\subsection{Guidance} 

The guidance algorithm computes a reference to be subsequently tracked by a control program. The orbit guidance cancels the perturbing acceleration in order to maintain a closed orbit. On the other hand, the attitude guidance assumes a null reference torque, thus its reference is fictitious but close to the truth one.

\subsubsection{Orbit guidance}

The orbit MPC guides the probe to a closed orbit. In terms of orbital elements, the controlled variables are the semi-major axis $a$ and eccentricity $e$ (since they are associated to the orbit size and shape). The orbit inclination ($i$), right ascension of the ascending node ($\Omega$), argument of periapsis ($\omega$) and true anomaly ($\nu$) are let to evolve freely. The guidance output is a reference orbit which has to be tracked by a subsequent control program. Let consider the simple case of maintaining a circular orbit such that $\{\bar{a},~\bar{e}\}\equiv\{\text{constant},0\}$. Note that the bar is employed to denote a reference variable. The semi-major axis and eccentricity are related to MEE as $a=p/(1-e^2)$ and $e=\sqrt{f^2+g^2}$, thus this translates to
\begin{equation}\label{eq:p_f_g_orbit_reference}
	\bar{p}=\bar{a},\quad\bar{f}=0,\quad\bar{g}=0,
\end{equation}
in terms of modified equinoctial elements. To prescribe Eq.\eqref{eq:p_f_g_orbit_reference} reference, the control acceleration follows a cancellation policy of the inhomogeneous gravity field effects on the controlled variables as $\bar{\mathbf{a}}_u(t)=-\left[ a_{\text{grav},r}(\bar{\mathbf{x}}_{\text{orb}}),~
			a_{\text{grav},t}(\bar{\mathbf{x}}_{\text{orb}}),~
			0\right]^T$, so that
	\begin{equation}
			\begin{aligned}
				\dot{\bar{p}}=&\dot{\bar{f}}=\dot{\bar{g}}=0,\\
				\dot{\bar{h}}=&\sqrt{\frac{\bar{p}}{\mu}}\frac{\bar{s}^2\bar{a}_{\text{grav},n}}{2\bar{w}}\cos{\bar{L}},\\
				\dot{\bar{k}}=&\sqrt{\frac{\bar{p}}{\mu}}\frac{\bar{s}^2\bar{a}_{\text{grav},n}}{2\bar{w}}\sin{\bar{L}},\\
				\dot{\bar{L}}=&\sqrt{\mu \bar{p}}\left(\frac{\bar{w}}{\bar{p}}\right)^2+\frac{1}{\bar{w}}\sqrt{\frac{\bar{p}}{\mu}}(\bar{h}\sin{\bar{L}}-\bar{k}\cos{\bar{L}})\bar{a}_{\text{grav},n},
		\end{aligned}
		\label{eq:ode_orbit_guidance}
	\end{equation}
	where the fact that the in-plane motion is decoupled from the out-of-plane perturbation and variables in the GVE, has been exploited. Note that $\bar{a}_{\text{grav},n}\equiv a_{\text{grav},n}(\bar{\mathbf{x}}_{\text{orb}})$. This allows to only cancel radial and tangential perturbations thus avoiding unnecessary out-of-plane control. Under such control the state subset $h$, $k$ and $L$ is let to evolve freely. The reference is generated through numerical integration, over the control horizon, of the ordinary differential equation (ODE) system given by Eq.\eqref{eq:ode_orbit_guidance}. As a consequence, a time-varying orbit reference is obtained
	\begin{equation}
		\bar{\mathbf{x}}_{\text{orb}}(t)=[\bar{p},~\bar{f},~\bar{g},~\bar{h}(t),~\bar{k}(t),~\bar{L}(t)]^T.\label{eq:orb_reference}
	\end{equation}
A target elliptic orbit would also be possible by designing a reference fulfilling $\bar{e}=\sqrt{\bar{p}^2+\bar{g}^2}\equiv\text{constant}$. It can be demonstrated that the previous relation holds by just cancelling the radial and tangential non-Keplerian perturbations but some complexity is added since $\dot{\bar{f}}, \dot{\bar{g}}\neq0$ for an elliptic orbit.

\subsubsection{Attitude guidance}

In order to keep the camera pointing to the asteroid surface, the attitude MPC has to ensure a stationary body orientation with respect to the orbit frame. As a consequence the attitude MPC state, $\mathbf{x}_{\text{att}}(t)=[\pmb{\sigma}_{B/O}^T,~\pmb{\omega}^T(t)]^T$, is composed of the MRP representing the body orientation with respect to the orbit frame, evolving as per Eq.\eqref{eq:attitude_kinematics_orbit}, and the angular velocity of the body with respect to the inertial frame, evolving as per Eq.\eqref{eq:attitude_dynamics}. Let recall that this orientation is obtained by composing the body (attitude filter) and orbit (orbit filter) orientations with respect to the inertial frame. Neglecting the required torque to maintain a stationary orientation with respect to the orbit frame, $\bar{\pmb{\sigma}}_{B/O}\equiv\text{constant}$, leads to the following reference
\begin{equation}
	\bar{\mathbf{x}}_{\text{att}}(t)=\left[\bar{\pmb{\sigma}}_{B/O}^T,~\left(\mathbf{R}(\bar{\pmb{\sigma}}_{B/O})\pmb{\omega}^{O}_{O/I}(\bar{\mathbf{x}}_{\text{orb}}(t))\right)^T\right]^T,\quad \bar{\mathbf{T}}_u(t)\approx\mathbf{0},\label{eq:att_reference}
\end{equation}
where the angular velocity reference has to nullify the attitude kinematics in Eq.\eqref{eq:attitude_kinematics_orbit} as $\dot{\bar{\pmb{\sigma}}}_{B/O}=\mathbf{0}$. By design, this is a ficticious reference which shall be close to the real one. However, let recall that the orbit reference is neither true as the model is only known to a certain degree of accuracy. In particular, the attitude fictitious reference will cause a drift that has to be compensated in the attitude control program. This approach eases the reference computation load as no integration other than the orbit reference is required in light of Eq.\eqref{eq:att_reference}.

\subsection{Control}   

The orbit-attitude controllers follow the same formulation: firstly, the continuous reference tracking problem is posed; secondly, the dynamics are linearized around the reference; finally, the continuous problem is transformed to finite tractable form (QP) by means of discretization.

For the sake of compactness, the MPC formulation is presented in a generic way.  Nonetheless, the orbit-attitude controllers formulation, with their proper notation, is also described in appendix \ref{appendix}.  Let define the tracking error as $\Delta\mathbf{x}_{(\cdot)}(t)=\mathbf{x}_{(\cdot)}(t)-\bar{\mathbf{x}}_{(\cdot)}(t)$ where the subscript $(\cdot)\equiv\{\text{att},~\text{orb}\}$ refers to either the attitude or orbit case. Let denote the control as $\mathbf{u}_{(\cdot)}\equiv\{\mathbf{T}_u,~\mathbf{a}_u\}$ which refers to the torque (attitude) or control acceleration (orbit) respectively. The subscript $(\cdot)$ is omitted in the sequel as it is understood that the formulation applies for both cases. 

\subsubsection{Continuous tracking problem}

The MPC controller aims to obtain a solution of
% this reference tracking optimization problem
\begin{equation}
	\begin{array}{rrclcl}
		\displaystyle \min_{\Delta\mathbf{x}(t),\Delta\mathbf{u}(t)} && \multicolumn{3}{l} {J=\frac{1}{t_f-t_0}\int^{t_f}_{t_0}\left(\gamma\Delta\mathbf{x}^T(t)\mathbf{P}_{x}\Delta\mathbf{x}(t)+\Delta\mathbf{u}^T(t)\Delta\mathbf{u}(t)\right)dt,} \\
		\textrm{s.t.} &&&\Delta\dot{\mathbf{x}}(t)=\dot{\mathbf{x}}(\mathbf{x}(t),\mathbf{u}(t))-\dot{\bar{\mathbf{x}}}(t)+\Delta\dot{\bar{\mathbf{x}}}(t),\\
		&&& \mathbf{u}(t)=\bar{\mathbf{u}}(t)+\Delta\mathbf{u}(t),\\
		&&&-\mathbf{u}_{\text{max}} \leq \mathbf{u}(t) \leq \mathbf{u}_{\text{max}},\\
		&&&\Delta u_n(t)=0,~~~~~\text{if}~~~~(\cdot)\equiv\text{orb},\\
	\end{array}\label{eq:control_problem_continuous}
\end{equation}
where, following \cite{Tavakoli2014}, the out-of-plane control acceleration, $a_{u_n}$, is nullified as it does not induce direct changes on the orbit size and shape but can lead to long-term inefficient (though short-term efficient) indirect changes on them through the GVEs orbital elements coupling. Note that $\Delta\dot{\bar{\mathbf{x}}}(t)=\dot{\mathbf{x}}(\mathbf{x}(t),\bar{\mathbf{u}}(t))-\dot{\bar{\mathbf{x}}}(t)$ is the reference drift which accounts for the possibility of a fictitious guidance reference. The variable  $\gamma>0$ is a weight parameter balancing the tracking accuracy with respect to control effort. The matrix $\mathbf{P}_{x}$ is given as
\begin{equation}
	\mathbf{P}_{x}=\begin{bmatrix}
		\mathbf{I}_{3\times3} & \mathbf{0}_{3\times3}\\
		\mathbf{0}_{3\times3} & \mathbf{0}_{3\times3}\\
	\end{bmatrix},~~~~
\end{equation}
for both cases since $\{p,~f,~g\}$ are the orbit controlled variables whereas the reference attitude is stationary, thus if tracked perfectly its variation $\dot{\pmb{\sigma}}_{B/O}$ will be implicitly null. Note that $\mathbf{I}$ is the identity matrix and $\mathbf{0}_{3\times3}$ is a matrix full of zeros. The term $\mathbf{u}_{\text{max}}$ denotes the maximum available control. The tracking control problem \eqref{eq:control_problem_continuous} is a non-linear continuous optimization problem with infinite degrees of freedom. This problem is transformed to a finite tractable static program (QP) by means of dynamics linearization and discretization.

\subsubsection{Dynamics linearization}

The orbit-attitude tracking error dynamics is highly non-linear due to the GVE and Eq.\eqref{eq:attitude_kinematics_orbit}-\eqref{eq:attitude_dynamics} respectively. Since the control objective is to station-keep the guidance reference, the position tracking errors are expected to be low compared to the reference semi-major axis $\lVert\Delta\mathbf{r}\rVert_2/\bar{a}<<1$. Consequently, the gravity acceleration and gravity-gradient torque effects will be very similar to the reference. Under the previous facts, it is reasonable to linearize orbit-attitude dynamics as
\begin{equation}\label{eq:dynamics_linearization}
	\dot{\mathbf{x}}(\mathbf{x}(t),\mathbf{u}(t))\approx\dot{\bar{\mathbf{x}}}(t)+\mathbf{A}(\bar{\mathbf{x}}(t),\bar{\mathbf{u}}(t))\Delta\mathbf{x}(t)+\mathbf{B}(\bar{\mathbf{x}}(t),\bar{\mathbf{u}}(t))\Delta\mathbf{u}(t),
\end{equation}
where $\mathbf{A}\in\mathbb{R}^{6\times6}$ is the linearized tracking error matrix and $\mathbf{B}\in\mathbb{R}^{6\times3}$ the control matrix as
\begin{equation}
	\mathbf{A}(\bar{\mathbf{x}}(t),\bar{\mathbf{u}}(t))=\frac{\partial \dot{\mathbf{x}}}{\partial \mathbf{x}}\bigg\rvert_{\bar{\mathbf{x}}(t),\bar{\mathbf{u}}(t)}+\frac{\partial \dot{\mathbf{x}}}{\partial\mathbf{M}_{\text{grav}}}\frac{\partial\mathbf{M}_{\text{grav}}}{\partial{\mathbf{x}}}\bigg\rvert_{\bar{\mathbf{x}}(t),\bar{\mathbf{u}}(t)},\quad\mathbf{B}(\bar{\mathbf{x}}(t),\bar{\mathbf{u}}(t))=\frac{\partial \dot{\mathbf{x}}}{\partial \mathbf{u}}\bigg\rvert_{\bar{\mathbf{x}}(t),\bar{\mathbf{u}}(t)},
\end{equation}
where $\mathbf{M}_{\text{grav}}\equiv\{\mathbf{T}_{\text{grav}},~\mathbf{a}_{\text{grav}}\}$ denotes the natural disturbance (gravity-gradient torque and non-Keplerian gravity as known by the filters) for each case. Introducing Eq.\eqref{eq:dynamics_linearization} linearized dynamics into the control problem \eqref{eq:control_problem_continuous}, the tracking error dynamics yields
\begin{equation}\label{eq:err_state_dynamics_linear}
	\Delta\dot{\mathbf{x}}(t)=\mathbf{A}(\bar{\mathbf{x}}(t),\bar{\mathbf{u}}(t))\Delta\mathbf{x}(t)+\mathbf{B}(\bar{\mathbf{x}}(t),\bar{\mathbf{u}}(t))\Delta\mathbf{u}(t)+\Delta\dot{\bar{\mathbf{x}}}(t),
\end{equation}
which is a linear time-varying system (LTV) due to the varying reference. The general solution to Eq.\eqref{eq:err_state_dynamics_linear} LTV system can be expressed by means of the state transition matrix $\pmb{\Phi}\in\mathbb{R}^{6\times6}$
\begin{equation}\label{eq:err_state_LTV_solution}
	\Delta\mathbf{x}(t)=\pmb{\Phi}(t,t_0)\Delta\mathbf{x}_{0}+\int^{t}_{t_0}\pmb{\Phi}(t,\tau)\mathbf{B}(\tau)\Delta\mathbf{u}(\tau)d\tau+\Delta\bar{\mathbf{x}}(t),
\end{equation}  
where the transition matrix is obtained by integrating its own dynamics
\begin{equation}
	\dot{\pmb{\Phi}}(t,t_0)=\mathbf{A}(\bar{\mathbf{x}}(t))\pmb{\Phi}(t,t_0),\quad\pmb{\Phi}(t_0,t_0)=\mathbf{I}.\label{eq:STM_dynamics}
\end{equation}
Note that for both attitude and orbit cases, Eq.\eqref{eq:STM_dynamics} is an ODE system with 36 differential equations.

\subsubsection{Discrete tracking problem} 

To convert the continuous control problem \eqref{eq:control_problem_continuous} into a discrete form, the control horizon is divided in $N$ sampling intervals of duration $\Delta t=(t_f-t_0)/N$. The state is evaluated at discrete instants $t_{k}=t_0+k\Delta t~(k=1\hdots N)$ whereas the control is assumed constant within each sampling interval $k$. The previous fact implies the conversion of the reference continuous control to a discrete form as
\begin{equation}
	\bar{\mathbf{u}}_k=\frac{1}{t_k-t_{k-1}}\int^{t_k}_{t_{k-1}}\bar{\mathbf{u}}(t)dt.
\end{equation} 

Under the previous assumptions, the tracking error propagation of Eq.\eqref{eq:err_state_LTV_solution} is transformed to a discrete form as follows
\begin{equation}
	\Delta\mathbf{x}_{k}=\pmb{\Phi}(t_k,t_0)\Delta\mathbf{x}_{0}+\sum^{k}_{i=1}\pmb{\Phi}(t_k,t_i)\left(\int^{t_i}_{t_{i-1}}\pmb{\Phi}(t_i,\tau)\mathbf{B}(\tau)d\tau\right)\Delta\mathbf{u}_i+\Delta\bar{\mathbf{x}}_{k}.
\end{equation} 
The objective function states in a discrete form as
\begin{equation}
	J=\sum^{N}_{k=1}\left(\gamma\Delta\mathbf{x}^T_{k}\mathbf{P}_{x}\Delta\mathbf{x}_{k}+\Delta\mathbf{u}^T_k\Delta\mathbf{u}_k\right).
\end{equation}
To ease the notation, following \cite{Gavilan2012}, a compact formulation is developed. Let define the following stack vectors associated to the tracking error and reference drift, $\Delta\mathbf{x}_{\mathbf{S}},\Delta\bar{\mathbf{x}}_{\mathbf{S}}\in\mathbb{R}^{6N}$, as well as the control increment and reference, $\Delta\mathbf{u_S},\bar{\mathbf{u}}_{\mathbf{S}}\in\mathbb{R}^{3N}$, as
\begin{equation}
	\begin{aligned}
		\Delta\mathbf{x}_{\mathbf{S}}=[\Delta\mathbf{x}^T_{1},\hdots,\Delta\mathbf{x}^T_{N}]^T,&\quad\Delta\bar{\mathbf{x}}_{\mathbf{S}}=[\Delta\bar{\mathbf{x}}^T_1,\hdots,\Delta\bar{\mathbf{x}}^T_N]^T,\\
		\Delta{\mathbf{u}}_{\mathbf{S}}=[\Delta\mathbf{u}^T_1,\hdots,\Delta\mathbf{u}^T_{N}]^T,&\quad\bar{\mathbf{u}}_{\mathbf{S}}=[\mathbf{u}^T_1,\hdots,\mathbf{u}^T_{N}]^T,
	\end{aligned}\label{eq:stack_vectors}
\end{equation} 
and the following stack matrices $\mathbf{D}\in\mathbb{R}^{6N\times6}$ and $\mathbf{G}\in\mathbb{R}^{6N\times3N}$
\begin{equation}
	\begin{aligned}
		\mathbf{D}&=[\pmb{\Phi}^T(t_1,t_0),\hdots,\pmb{\Phi}^T(t_N,t_0)]^T,\\
		\mathbf{G}&=\begin{bmatrix}
			\int^{t_1}_{t_0}\pmb{\Phi}(t_1,\tau)\mathbf{B}(\tau)d\tau & \hdots & \mathbf{0}_{6\times3}\\
			\vdots & \ddots & \vdots\\
			\pmb{\Phi}(t_N,t_1)\int^{t_1}_{t_{0}}\pmb{\Phi}(t_1,\tau)\mathbf{B}(\tau)d\tau & \hdots & \int^{t_N}_{t_{N-1}}\pmb{\Phi}(t_N,\tau)\mathbf{B}(\tau)d\tau\\
		\end{bmatrix}.
	\end{aligned}\label{eq:stack_matrices}
\end{equation}
Using the previously defined stack vectors and matrices, see Eq.\eqref{eq:stack_vectors}-\eqref{eq:stack_matrices}, the tracking error propagation can be posed in compact form as
\begin{equation}\label{eq:err_state_compact_propagation}
	\Delta\mathbf{x}_{\mathbf{S}}=\mathbf{D}\Delta\mathbf{x}_{0}+\mathbf{G}\Delta\mathbf{u_S}+\Delta\bar{\mathbf{x}}_{\mathbf{S}},
\end{equation}
and the objective function as
\begin{equation}\label{eq:obj_function_compact}
	J=\gamma\Delta\mathbf{x}^T_{\mathbf{S}}\mathbf{P}_{\mathbf{S}x}\Delta\mathbf{x}_{\mathbf{S}}+\Delta\mathbf{u}^T_{\mathbf{S}}\Delta\mathbf{u}_{\mathbf{S}}.
\end{equation}
Introducing the compact discrete propagation, see Eq.\eqref{eq:err_state_compact_propagation}, into the objective function given by Eq.\eqref{eq:obj_function_compact}, the discrete optimization problem is only dependant on the control increment stack vector, $\Delta\mathbf{u_S}$, as
\begin{equation}
	\begin{array}{rrclcl}
		\displaystyle \min_{\Delta\mathbf{u_S}} && \multicolumn{3}{l} {J=2\gamma(\mathbf{D}\Delta\mathbf{x}_{0}+\Delta\bar{\mathbf{x}}_{\mathbf{S}})^T\mathbf{P}_{\mathbf{S}x}\mathbf{G}\Delta\mathbf{u_S}+\Delta\mathbf{u}^T_{\mathbf{S}}(\gamma\mathbf{G}^T\mathbf{P}_{\mathbf{S}x}\mathbf{G}+\mathbf{I})\Delta\mathbf{u}_{\mathbf{S}},} \\
		\textrm{s.t.}
		&&&-\mathbf{u}_{\mathbf{S}\text{max}} \leq \bar{\mathbf{u}}_{\mathbf{S}}+\Delta\mathbf{u_S} \leq \mathbf{u}_{\mathbf{S}\text{max}},\\
		&&& \mathbf{W}_{\mathbf{S}u_n}\Delta\mathbf{u_S}=\mathbf{0}_{N\times1}~~~~~\text{if}~~~~(\cdot)\equiv\text{orb},\\
	\end{array}\label{eq:control_problem_compact_form}
\end{equation}
where the objective function constant terms have been disregarded. The stack matrix $\mathbf{W}_{\mathbf{S}u_n}\in\mathbb{R}^{N\times3N}$ has the following structure 
\begin{equation}
	\mathbf{W}_{\mathbf{S}u_n}=\begin{bmatrix}
		0 & 0 & 1 & \hdots & 0 & 0 & 0\\
		\vdots & \vdots & \vdots & \ddots & \vdots & \vdots & \vdots\\
		0 & 0 & 0 & \hdots & 0 & 0 & 1\\
	\end{bmatrix}.
\end{equation}
The stack vector $\mathbf{u}_{\mathbf{S}\text{max}}=[\mathbf{u}^T_{\text{max}},\hdots,\mathbf{u}^T_{\text{max}}]^T$ stacks the maximum available control. Note that the static discrete program \eqref{eq:control_problem_compact_form} is a QP problem with $3N$ decision variables.  

\section{Integrated model-learning predictive control}\label{sec:constellation}

\begin{figure*}[t] 
	\begin{center}
		\includegraphics[width=12cm,height=12cm,keepaspectratio]{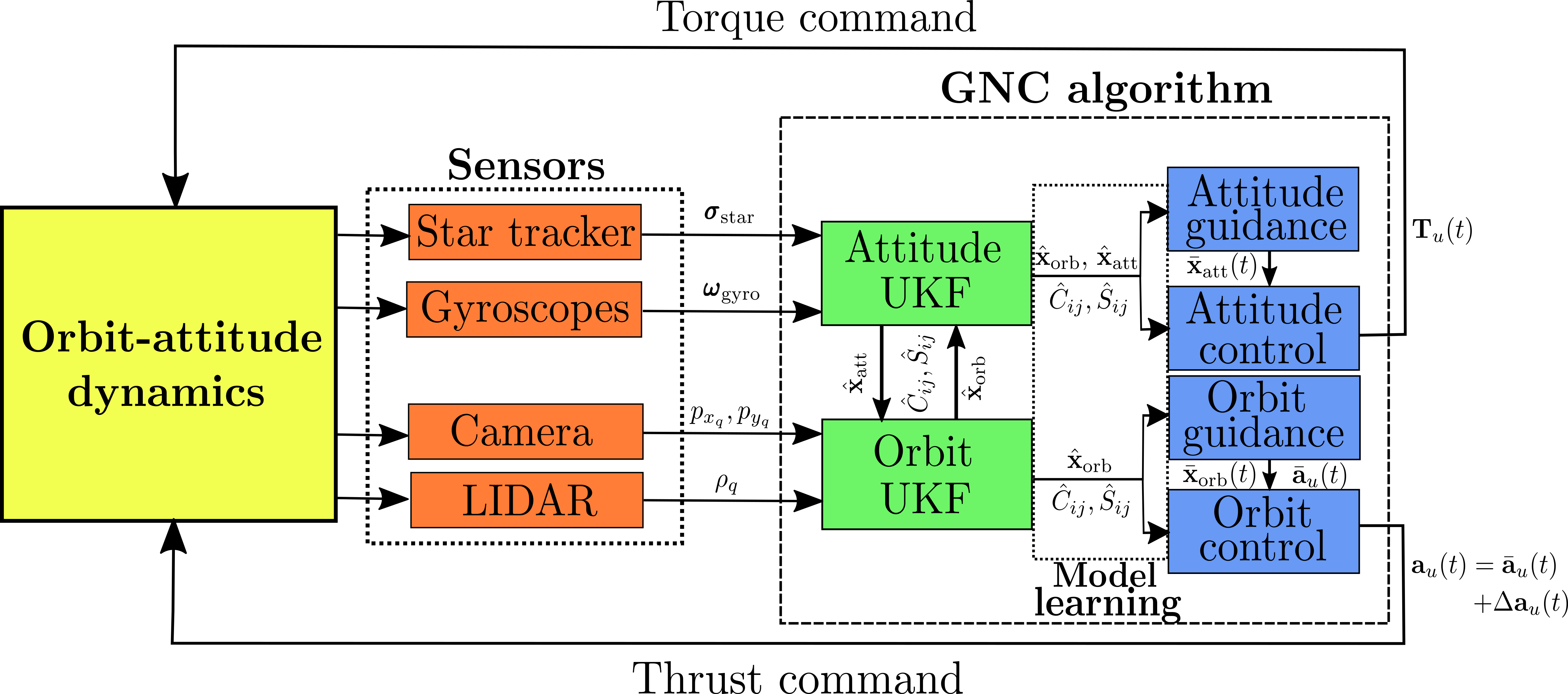}
		\caption{Integrated model-learning predictive control scheme for a standalone spacecraft around the asteroid.}	
		\label{fig:GNC_scheme}
	\end{center}
\end{figure*}

The integrated model-learning predictive control scheme is summarized in Fig.\ref{fig:GNC_scheme}. Four modules can be distinguished: dynamics driven by the control actuators (yellow); sensors (orange); navigation filters (green); MPC guidance and control (blue). In this scheme, it is highlighted how the MPC-based guidance and control algorithms not only receive navigation state information but also model parameters estimates, thus reinforcing the model predictive control accuracy. 

The gravity estimation could be enhanced if there are multiple satellites orbiting around the asteroid (e.g. \cite{Stacey2018} considered a swarm of three probes). In this work, since the relative motion is not controlled, a constellation mission concept is considered. Assuming a number of $\eta_{\text{max}}$ satellites, the gravity estimation is updated, by gathering the constellation data, as a weighted mean
	\begin{equation}
		\begin{aligned}
			\hat{C}_{ij}=\sum^{\eta_{\text{max}}}_{\eta=1}w^{[\eta]}_{C_{ij}}\hat{C}^{[\eta]}_{ij},\quad \hat{S}_{ij}=\sum^{\eta_{\text{max}}}_{\eta=1}w^{[\eta]}_{S_{ij}}\hat{S}^{[\eta]}_{ij}\\ w^{[\eta]}_{C_{ij}}=\frac{\left(1/\sigma^{[\eta]}_{C_{ij}}\right)^2}{\sum^{\eta_{\text{max}}}_{\eta=1}\left(1/\sigma^{[\eta]}_{C_{ij}}\right)^2},\quad w^{[\eta]}_{S_{ij}}=\frac{\left(1/\sigma^{[\eta]}_{S_{ij}}\right)^2}{\sum^{\eta_{\text{max}}}_{\eta=1}\left(1/\sigma^{[\eta]}_{S_{ij}}\right)^2},
		\end{aligned}
	\end{equation}
	where more plausability is given to the lesser uncertain estimates. The weighted mean mitigates outliers, thus potentially enhancing the gravity estimation convergence and accuracy. As there is a cross-correlation between the state and spherical harmonics in the filters, the individual covariances are not updated. The UKF requires the covariance extended state matrix to be positive definite. As such, updating the gravity model block of the covariance matrix may not guarantee the resulting extended state covariance matrix to be positive definite in the next call. The complete methodology is shown as pseudocode in Algorithm \ref{alg6:constellation}. The constellation concept arises through step 15, where the joint gravity estimation is computed and shared across the constellation after the output of each satellite individual filter. Note that each satellite carries out its own navigation process and control computation in parallel. 

\begin{algorithm}[t]
	\caption{(Integrated GNC scheme for $\eta_{\text{max}}$ probes constellation)}
	\textbf{Input: $\hat{\mathbf{x}}^{[\eta]}_{\text{orb}}(t_0)$, $\hat{\mathbf{x}}^{[\eta]}_{\text{att}}(t_0)$, $\Delta\hat{\pmb{\omega}}^{[\eta]}_{\text{gyro}}(t_0)$, $\hat{C}^{[\eta]}_{ij}(t_0)$, $\hat{S}^{[\eta]}_{ij}(t_0)$, $\Sigma^{[\eta]}_{\text{orb}}(t_0)$, $\Sigma^{[\eta]}_{\text{att}}(t_0)$, $\hat{Q}^{[\eta]}_{x_{\text{orb}}}(t_0)$, $\hat{Q}^{[\eta]}_{x_{\text{att}}}(t_0)$} \\
	\textbf{Output: orbit and attitude station-keeping with centralized gravity parameters estimation} 
	\begin{algorithmic}[1]
		\While{station-keeping phase = on}
		\State{Obtain reference orbit as Eq.\eqref{eq:orb_reference}: $\bar{\mathbf{x}}^{[\eta]}_{\text{orb}}(t),\bar{\mathbf{a}}_u^{[\eta]}(t)$;}
		\State{Solve the orbit QP control problem \eqref{eq:orb_control_problem_compact_form} obtaining $\mathbf{a}_u^{[\eta]}(t)=\bar{\mathbf{a}}_u^{[\eta]}(t)+\Delta\mathbf{a}^{[\eta]}_u(t)$;} 
		\For{$i=1\hdots N_{\text{UKF,orb}}$}
		\State{Obtain reference attitude as Eq.\eqref{eq:orb_reference}: $\bar{\mathbf{x}}^{[\eta]}_{\text{att}}(t)$;}
		\State{Solve the attitude QP control problem \eqref{eq:att_control_problem_compact_form} obtaining $\mathbf{T}^{[\eta]}_u(t)$;}
		\For{$j=1\hdots N_{\text{UKF,att}}$}
		\State{Update time: $t\gets t+\Delta t_{\text{UKF,att}}$;}
		\State{Collect attitude measurement(t): $\mathbf{z}^{[\eta]}_{\text{att}}(t)$;}
		\State{Call attitude UKF through algorithm \ref{alg1} obtaining: $\hat{\mathbf{x}}^{[\eta]}_{\text{att}}(t)$, $\Delta\hat{\pmb{\omega}}^{[\eta]}_{\text{gyro}}(t)$, $\hat{C}^{[\eta]}_{ij}(t)$, $\hat{S}^{[\eta]}_{ij}(t)$, $\Sigma^{[\eta]}_{\text{att}}(t)$, $\hat{Q}^{[\eta]}_{x_{\text{att}}}(t)$;}
		\EndFor
		\State{Update time: $t\gets t+\Delta t_{\text{UKF,orb}}$;}
		\State{Collect orbit measurement(t): $\mathbf{z}^{[\eta]}_{\text{orb}}(t)$;}
		\State{Call orbit UKF through algorithm \ref{alg1} obtaining: $\hat{\mathbf{x}}^{[\eta]}_{\text{orb}}(t)$, $\hat{C}^{[\eta]}_{ij}(t)$, $\hat{S}^{[\eta]}_{ij}(t)$, $\Sigma^{[\eta]}_{\text{orb}}(t)$, $\hat{Q}^{[\eta]}_{x_{\text{orb}}}(t)$;}
		\State{Average gravity parameters: $\hat{C}^{[\eta]}_{ij}\gets\sum^{\eta_{\text{max}}}_{\eta=1}w^{[\eta]}_{C_{ij}}\hat{C}^{[\eta]}_{ij}$,~~$\hat{S}^{[\eta]}_{ij}\gets\sum^{\eta_{\text{max}}}_{\eta=1}w^{[\eta]}_{S_{ij}}\hat{S}^{[\eta]}_{ij}$;}
		\EndFor
		\EndWhile
	\end{algorithmic}\label{alg6:constellation}
\end{algorithm}

\section{Numerical results}\label{sec:results}

In this section, the numerical results using the proposed GNC strategy are shown. First, the scenario key variables such as the target asteroid, satellite configuration and controller parameters are declared. Then, some useful control performance indexes are defined. Finally, simulations assessing the out-of-plane control nullifying method, the learning-based MPC performance and the gravity estimation through the satellite constellation concept are shown.   

\subsection{Scenario parameters}

The target asteroid is 433 Eros because it is the benchmark for small body missions. This is due to the huge amount of data collected during the NEAR Shoemaker mission. Eros is a shape elongated near-Earth object with a gravitational parameter of $\mu=4.4628\cdot10^5~\text{m}^3/\text{s}^2$ and a rotation period of $T=5.27~\text{h}$ around its major inertia axis. This information is known by navigation purposes. Eros inhomogeneous gravity field is characterized by 15$\times$15 degree and order spherical harmonics from \cite{Konopliv2002}. These coefficients are normalized with $R_e=16~\text{km}$ and are employed for the real dynamics simulation. The Sun position is considered in a simplified way as $\mathbf{r}_{\odot}=[1.46,0,0]^T$ AU, thus assumed constant to its average distance from Eros.

The satellite is equipped with a camera having a resolution of 2048$\times$2048 pixels, a 30$^{\circ}$ field of view, and a focal length of 300 mm. It is assumed the camera landmark recognizition algorithm (not implemented) can track $q_{\text{max}}=3$ landmarks betweeen orbit filter calls. A set of 522 surface landmarks from Eros mission data \cite{Konopliv2002} are considered to be exactly known by the filter. In order to ease LIDAR ranging acquisition (which is not studied), the tracked landmarks are the ones with higher relative elevation with respect to the camera boresight. The camera boresight is aligned with the $-x_B$ axis as
\begin{equation}
	\mathbf{R}^C_B=\begin{bmatrix}
		0 & 0 & -1\\
		0 & 1 & 0\\
		1 & 0 & 0\\
	\end{bmatrix}.
\end{equation} 
Table \ref{table:sensors_datasheet} shows the considered sensors noises datasheet where the camera and LIDAR data is taken similar as in \cite{Vetrisano2016}, the star tracker from \cite{startracker} and the gyroscopes from \cite{Venkateswaran2004}. The star tracker noise is introduced via the rotation angle $\theta_{\text{rot}}$. 
\begin{table}[h]
	\centering
	\begin{tabular}{lcccc}
		\hline \hline
		\multicolumn{1}{l}{Sensor} & Variable & Bias & 1-$\sigma$ noise\\
		\hline
		Camera & $\mathbf{p}_{q}$ & $[0,0]^T~\text{px}$ & $[0.5,0.5]^T~\text{px}$\\
		LIDAR & $\rho_q$ & 0 m & 5 m\\
		Star tracker & $\theta_{\text{rot}}$ & $0~\text{arcsec}$ & $10~\text{arcsec}$\\
		Gyroscopes & $\pmb{\omega}_{\text{gyro}}$ & $[5,5,5]^T~^{\circ}/\text{h}$ & $[0.05,0.05,0.05]^T~^{\circ}/\text{h}$ \\
		\hline \hline
	\end{tabular}
	\caption{Sensors datasheet}
	\label{table:sensors_datasheet}
\end{table}

The control acceleration and torque bounds are taken as $\mathbf{a}_{u_\text{max}}=[1,1,1]^T~\text{cm}/\text{s}^2$ and $\mathbf{T}_{u_{\text{max}}}=[1,1,1]^T~\text{N}\cdot\text{cm}$. Although the GNC algorithms assumes discrete changes of the control signals, see Eq. \eqref{eq:control_problem_compact_form}, the realistic simulation considers acceleration and torque transients in a continuous form as  
\begin{equation}
	\begin{aligned}
		\mathbf{a}_u(t)&=\mathbf{a}_{u,k}+e^{-\tau(t-t_{k-1})}(\mathbf{a}_{u,k-1}-\mathbf{a}_{u,k}),& t\in[t_{k-1},t_k),\\
		\mathbf{T}_u(t)&=\mathbf{T}_{u,k}+e^{-\tau(t-t_{k-1})}(\mathbf{T}_{u,k-1}-\mathbf{T}_{u,k}),& t\in[t_{k-1},t_k),
	\end{aligned}
\end{equation}
where $\tau$ is the time constant assumed as $\tau=0.1~\text{s}^{-1}$. The probe mass distribution is given in Table \ref{table:mass_distribution}. The associated inertia components, in the body frame, are $\{J_{11},~J_{22},~J_{33},~J_{12},~J_{13},~J_{23}\}=\{2000,~16400,~17600,~0,~0,~0\}~\text{kg}\cdot\text{m}^2$. Note that this inertia distribution would allow gravity-gradient stabilization (due to $m_1$ which acts as a boom) in an homogeneous gravity field. 
\begin{table}[h]
	\centering
	\begin{tabular}{lcccc}
		\hline \hline
		\multicolumn{1}{c}{$l$} & $x_l^B~[\text{m}]$ & $y_l^B~[\text{m}]$ & $z_l^B~[\text{m}]$ & $m_l~[\text{kg}]$\\
		\hline
		1 & 8 & 0 & 0 & 200\\ 
		2 & -2 & -2 & 0 & 200\\ 
		3 & -2 & 2 & 0 & 200\\
		4 & -2 & 0 & -1 & 200\\
		5 & -2 & 0 & 1 & 200\\
		\hline \hline
	\end{tabular}
	\caption{Probe mass distribution.}
	\label{table:mass_distribution}
\end{table} 
Finally, the coefficient of reflectivity and SRP exposed area are $C_{R}=1.4$ and $A=10~\textup{m}^2$.

The GNC tuning parameters for the whole section, although stated otherwise, are declared. The estimated spherical harmonics degree and order is chosen as $\{n_{\text{orb}},~n_{\text{att}}\}=\{4,~2\}$. This way, the orbit filter estimates up to 4$\times$4 order and degree gravity parameters while the attitude filter estimates a 2$\times$2 model. The UKF filter parameters are chosen as $\{\alpha,~\theta,~\beta,~\lambda\}=\{0.98,~10^{-3},~2,~(\theta^2-1)n\}$ being $n$ the dimension of the extended state (27 for orbit and 14 for attitude). The fading factor $\alpha$ is chosen through experience (a slow update of process covariance ensured filter stability) while $\{\theta,\beta,\lambda\}$ follows the UKF canonical tuning choice of \cite{Wan2000}. Regarding sampling rates the attitude and orbit UKF are executed each $3.6~\text{s}$ and $36~\text{s}$ respectively. The attitude filter sampling rate is one order of magnitude higher than the orbit as its sensors are able to operate at higher frequencies. The orbit filter has to have enough margin in order to let the camera carry out its feature identification process but it has to also be fast enough in order to capture the highest order spherical harmonics. The guidance and control algorithm parameters (control horizon, discretization intervals, interval duration and tracking error weight) are stated in Table 3. The chosen orbit control horizon accounts for a quarter of the orbital period for a 34 km circular orbit around 433 Eros ($\approx 16~\text{h}$). The number of discretization intervals are chosen to keep a relatively low control program dimensionality in order to reduce computational effort. The tracking error weight is chosen to give a high priority to reduce tracking errors with respect to fuel consumption. This helps to assess the overall station-keeping strategy. All the simulations will last 2 weeks.
	\begin{table}[h]
		\centering
		\begin{tabular}{lcccc}
			\hline \hline
			\multicolumn{1}{r}{} & $\text{Control horizon}~[\text{min}]$ & $N~[-]$ & $\Delta t$ [s] & $\gamma~[-]$\\
			\hline
			Attitude & 6 & 10 & 36 & $10^3$\\ 
			Orbit & 240 & 40 & 360 & $10^3$\\ 
			\hline \hline
		\end{tabular}
		\caption{Guidance and control algorithm parameters.}
		\label{table:guidance_control_parameters}
\end{table}

At all cases, the initial state values are considered to match the reference (successful orbit transfer) and estimated (accurate initial fix from Earth ground segment) ones: $\mathbf{x}_{\text{orb}}(t_0)=\hat{\mathbf{x}}_{\text{orb}}(t_0)=\bar{\mathbf{x}}_{\text{orb}}(t_0)$; $\mathbf{x}_{\text{att}}(t_0)=\hat{\mathbf{x}}_{\text{att}}(t_0)=\bar{\mathbf{x}}_{\text{att}}(t_0)$. However, gravity inhomogeneties and gyroscopes bias are completely unknown as $\hat{C}_{ij}=\hat{S}_{ij}=0$ and $\Delta\hat{\pmb{\omega}}_{\text{gyro}}=\mathbf{0}$. The initial extended state covariances are taken as
	\begin{equation}
		\begin{aligned}
			\pmb{\Sigma}_{\text{orb}}(t_0)&=\begin{bmatrix}
				5^2~\text{m}^2 & \mathbf{0}_{1\times5} & \mathbf{0}_{1\times21}\\
				\mathbf{0}_{5\times1} & (5\cdot10^{-6})^2\mathbf{I} & \mathbf{0}_{5\times21}\\
				\mathbf{0}_{21\times1} & \mathbf{0}_{21\times5} & (5\cdot10^{-3})^2\mathbf{I}\\
			\end{bmatrix},\\
			\pmb{\Sigma}_{\text{att}}(t_0)&=\begin{bmatrix}
				(10^{-6})^2\mathbf{I} & \mathbf{0}_{3\times3} & \mathbf{0}_{3\times5} & \mathbf{0}_{3\times3}\\
				\mathbf{0}_{3\times3} & (10^{-8})^2\mathbf{I}~\text{s}^{-2} & \mathbf{0}_{3\times5} & \mathbf{0}_{3\times3}\\
				\mathbf{0}_{5\times3} & \mathbf{0}_{5\times3}& (5\cdot10^{-3})^2\mathbf{I} & \mathbf{0}_{5\times3}\\
				\mathbf{0}_{3\times3} & \mathbf{0}_{3\times3} & \mathbf{0}_{5\times3} & (2.42\cdot10^{-6})^2\mathbf{I}~\text{s}^{-2} \\
			\end{bmatrix},
		\end{aligned}
	\end{equation}
	which considers an accurate navigation fix while a high uncertainty is assumed for inhomogeneous gravity and gyroscope bias. Let recall that these covariances are referred to the orbit extended state (modified equinoctial elements and gravity parameters), see Eq.\eqref{eq:orb_extended_state}, and the attitude extended state (MRP, angular velocity, gravity parameters and gyroscope bias), see Eq.\eqref{eq:att_extended_state}. The initial process noises are null as they will be progressively estimated, $\mathbf{Q}_{y,\text{orb}}=\mathbf{0}_{27\times27}$ and $\mathbf{Q}_{y,\text{att}}=\mathbf{0}_{14\times14}$.

\subsection{Performance indexes}

Subsequently, performance indexes for the whole scenario timespan will be defined. In this paragraph, $t_0$ and $t_f$ refer to the initial and final scenario times. The orbit control efficiency is measured in terms of fuel consumption as
	\begin{equation}
		m_F=\int^{t_f}_{t_0}{\dot{m}}(t)dt\approx\int^{t_f}_{t_0}\frac{m_0\mathbf{a}_u(t)}{g_0I_{\text{sp}}}dt,
	\end{equation}
	where $g_0=9.8066~\text{m/s}^2$ and the mass is assumed constant. Due to the continuous application of control acceleration, the most suitable propulsion device seems to be electric thrusters, thus $I_{\text{sp}}=2900~\text{s}$. As the target orbits are circular, the orbit control accuracy is measured as the average and maximum tracking error on the orbital radius
	\begin{equation}
		\Delta R =\frac{1}{t_f-t_0}\int^{t_f}_{t_0}|\Delta r(t)|dt,\quad \Delta R_{\text{max}}=\text{max}\{|\Delta r(t)|\},
	\end{equation}
	where $\Delta r=\lVert\mathbf{r}(t)\rVert_2-R$. The attitude control efficiency is measured as
	\begin{equation}
		T_U=\frac{1}{t_f-t_0}\int^{t_f}_{t_0}\lVert\mathbf{T}_u(t)\rVert_2dt,\\
\end{equation}
 while the attitude control accuracy metrics are the average and maximum tracking errors in terms of Euler angles as
	\begin{equation}
		\begin{aligned}
			\Delta\pmb{\Theta}&=\frac{1}{t_f-t_0}\int^{t_f}_{t_0}|\Delta \pmb{\theta}(t)|dt,\quad \Delta\pmb{\Theta}_{\text{max}}=\text{max}\{|\Delta\pmb{\theta}(t)|\},
		\end{aligned}
	\end{equation}  
	where $\pmb{\theta}=[\theta_1,~\theta_2,~\theta_3]^T\equiv\{\text{pitch},\text{roll},\text{yaw}\}$ defines the following rotation sequence from the orbit to the body frame
	\begin{equation}
		O\xrightarrow[x_O]{\theta_3} S' \xrightarrow[y_{S'}]{\theta_2} S'' \xrightarrow[z_{S''}]{\theta_1}B.
\end{equation}

\subsection{Simulations}\label{simulations}

Three issues are assessed in this section. First, a comparison between nullifying or not the out-of-plane control is provided. Then, the learning-based MPC performance is compared to non-learning MPC. Finally, the constellation concept mission is compared to monolithic missions in terms of gravity estimation.

\subsubsection{Impact of nullifying out-of-plane control}\label{sec:nullifying}

First, let assess the impact, on control performance, of nullifying the out-of-plane control, $a_{u_n}(t)=0$, by design. For this purpose, five simulations with and without out-of-plane nullifying are carried out. The initial orbit is circular as $\{a_0,~e_0,~\omega_0,~\Omega_0~,\nu_0\}=\{34~\text{km},~0,~0^{\circ},~0^{\circ},~0^{\circ}\}$ with the initial inclination, $i_0$, being the parameter under study as it masters the asteroid overflight regions. 

The results for both methods are summarized in Table \ref{table:nullifying_method}. The nullifying method reduces fuel consumption needs at all cases. This reduction oscillates from a 3\% ($i_0=30^{\circ}$) to a 45\% ($i_0=60^{\circ}$). Regarding orbit tracking accuracy, no clear trends are established since nullifying the out-of-plane control is more accurate, both in average terms and the peak, for $i_0=60^{\circ},90^{\circ}$ while allowing it is more accurate for the remaining three cases. In conclusion, nullifying the out-of-plane control significantly reduces the required control effort at the possible expense of a slight degradation of the orbit tracking accuracy. Therefore, the nullifying out-of-plane control is subsequently employed.   

\begin{table*}[ht]
	\centering
	\begin{tabular}{lcccccc}
		\hline \hline
		\multicolumn{1}{l}{} & \multicolumn{3}{c}{$a_{u_n}(t)=0$} & \multicolumn{3}{c}{$a_{u_n}(t)\equiv\text{free}$}\\
		\midrule
		\text{Simulation} & $m_F[\text{kg}]$ & $\Delta R[\text{m}]$ & $\Delta R_{\text{max}}[\text{m}]$ & $m_F[\text{kg}]$ & $\Delta R[\text{m}]$ & $\Delta R_{\text{max}}[\text{m}]$\\
		\hline
		$i_0=30^{\circ}$ & 1.7962 & 160.12 & 615.16 & 1.8458 & 146.39 & 636.14\\  
		$i_0=60^{\circ}$ & 1.3505 & 251.39 & 609.87 & 2.4677 & 319.78 & 583.50\\ 
		$i_0=90^{\circ}$ & 1.3926 & 298.72 & 811.80 & 2.4366 & 314.82 & 1026.7\\  
		$i_0=120^{\circ}$ & 1.3392 & 155.36 & 485.75 & 1.4218 & 95.206 & 475.35\\  
		$i_0=150^{\circ}$ & 1.7338 & 123.90 & 465.42 & 1.8441 & 85.028 & 459.81\\    
		\hline \hline
	\end{tabular}
	\caption{Orbit control performance with and without nullifying the out-of-plane control}
	\label{table:nullifying_method}
\end{table*}  

Let do a basic comparison with NEAR Shoemaker orbital corrections between an equivalent period of time (24/01/2001-06/02/2001) in low Eros orbits. NEAR Shoemaker did five impulsive maneuvers of 1.81 m/s \cite{Williams2002} while translating the case $i_0=90^{\circ}$ continuous acceleration to an equivalent form yields 40.4 m/s which is significantly higher. Nonetheless, at that mission timeline, NEAR Shoemaker had accurate knowledge of the 433 Eros gravity field and was inserted in a stable low asteroid orbit. Lacking that information, the proposed approach is conservative by applying active control. In-situ gravity estimation enables the determination of stable orbits that may be subsequently targeted.

\subsubsection{Impact of the gravity parameters in the control performance}

In this section, the benefits of learning-based predictive control are analyzed. Under this purpose, the previous section simulation cases are compared with a basic MPC algorithm not learning the gravity model from the filters (however, the navigation filters estimate inhomogeneous gravity to provide equivalent accuracy on the state). The gravity estimation accuracy of the learning-based MPC is shown in the Tables \ref{table:second_order_gravity}-\ref{table:fourth_order_gravity} (section \ref{sec:results}.\ref{simulations}.\ref{results_constellation}) to ease the comparison with the constellation cases. The same cases as in paragraph \ref{sec:results}.\ref{simulations}.\ref{sec:nullifying} are considered. 

\begin{figure*}[] 
	\begin{center}
		\includegraphics[width=14cm,height=14cm,keepaspectratio]{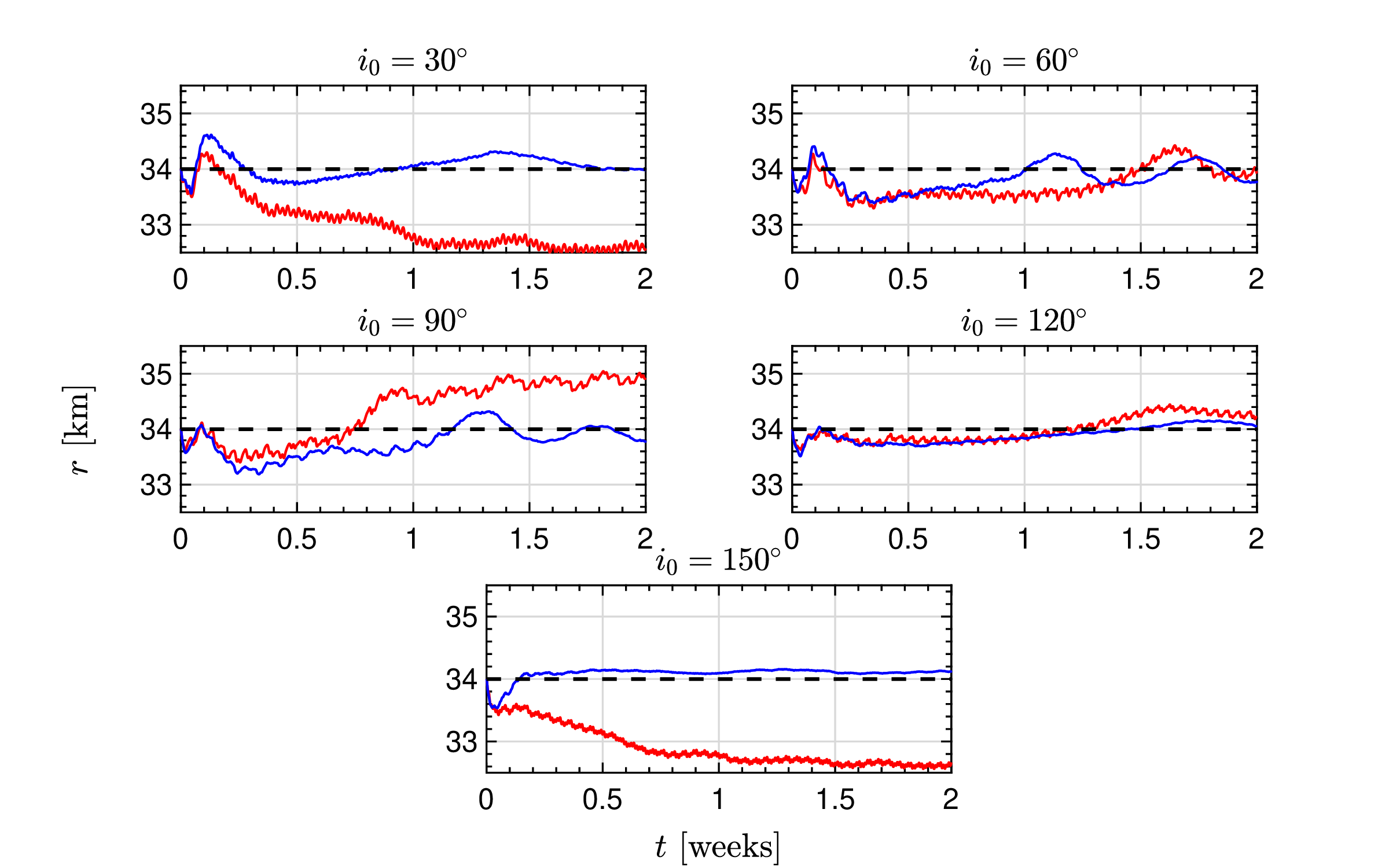}
		\caption{Orbital radius. Blue: learning-based MPC; red: non-learning MPC; black: reference.}	
		\label{fig:orbital_radius}
	\end{center}
\end{figure*}

\begin{figure}[t] 
	\begin{center}
		\subfigure{\includegraphics[width=7cm,height=7cm,keepaspectratio]{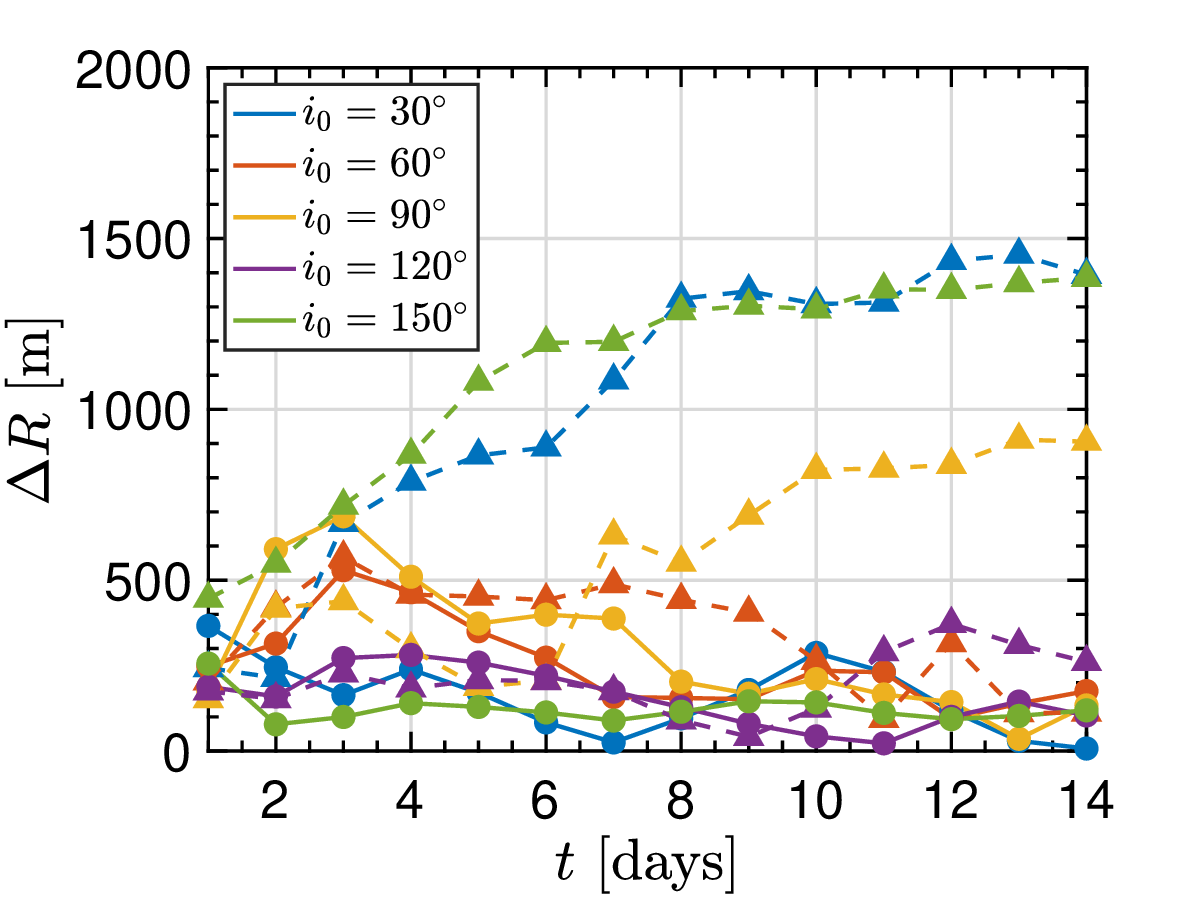} \label{a_comparison}}
		\subfigure{\includegraphics[width=7cm,height=7cm,keepaspectratio]{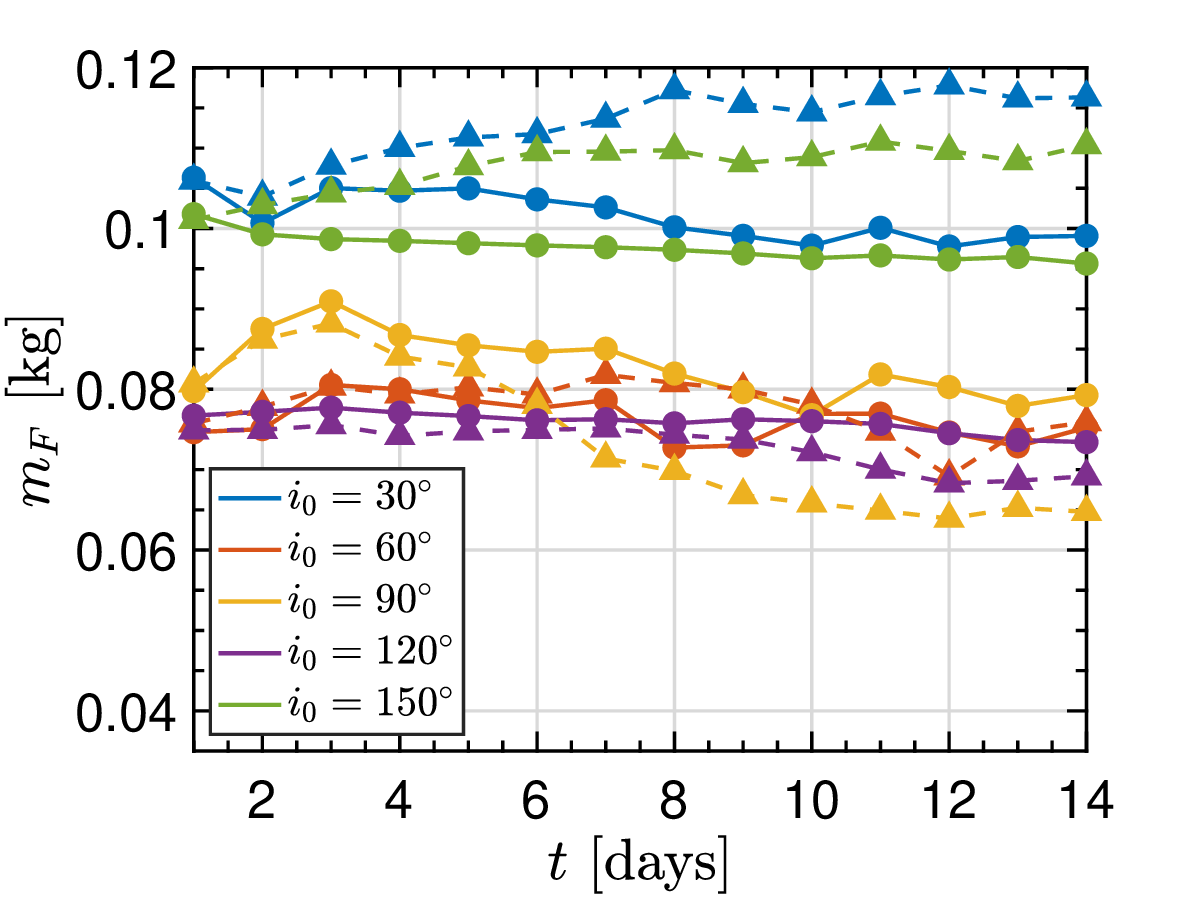} \label{e_comparison}}
		\caption{Average tracking error per day (\textit{top}) and fuel consumption per day (\textit{bottom}). Dots: learning-based MPC; triangles: non-learning MPC.}
		\label{fig:cumulative_control_variables_comparison}
	\end{center}
\end{figure}

\begin{figure}[t] 
	\begin{center}
		\subfigure{\includegraphics[width=7cm,height=7cm,keepaspectratio]{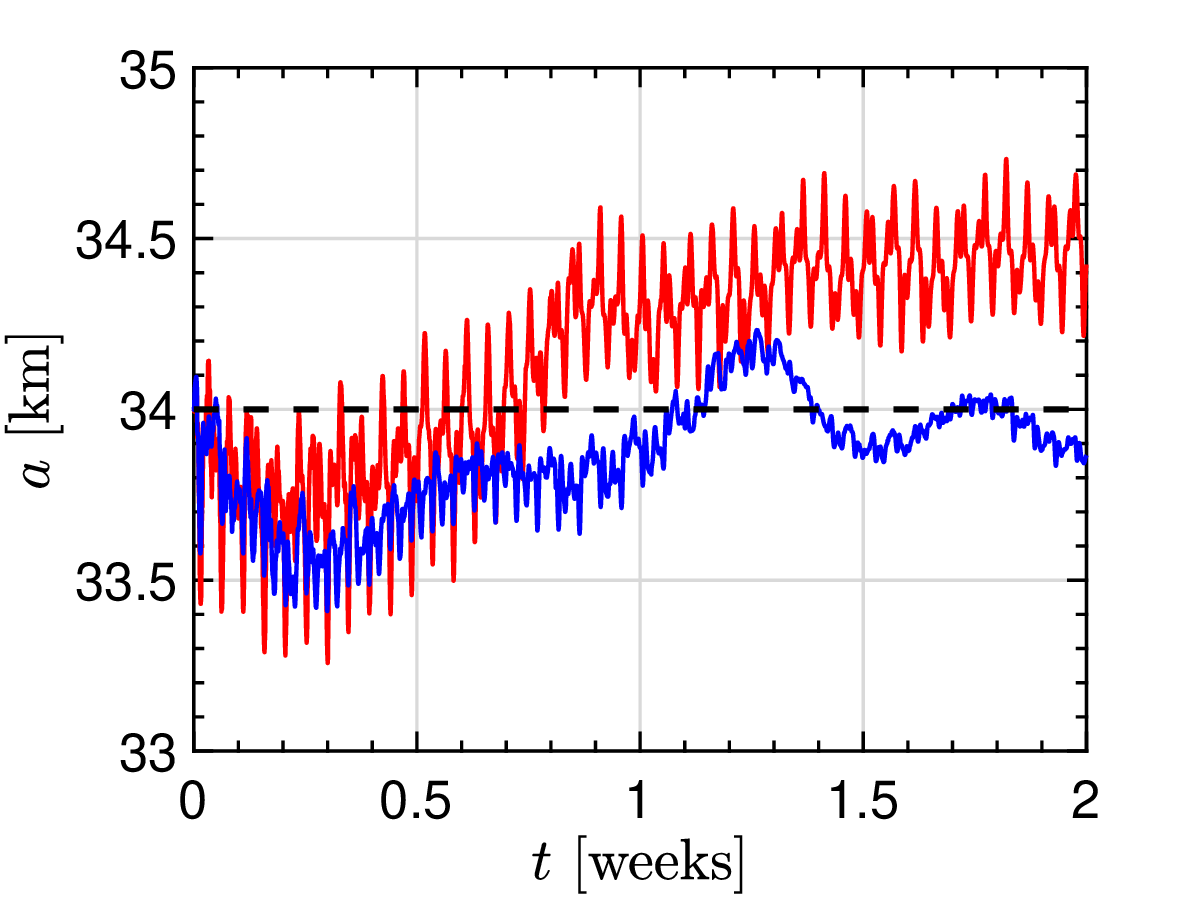} \label{a_comparison}}
		\subfigure{\includegraphics[width=7cm,height=7cm,keepaspectratio]{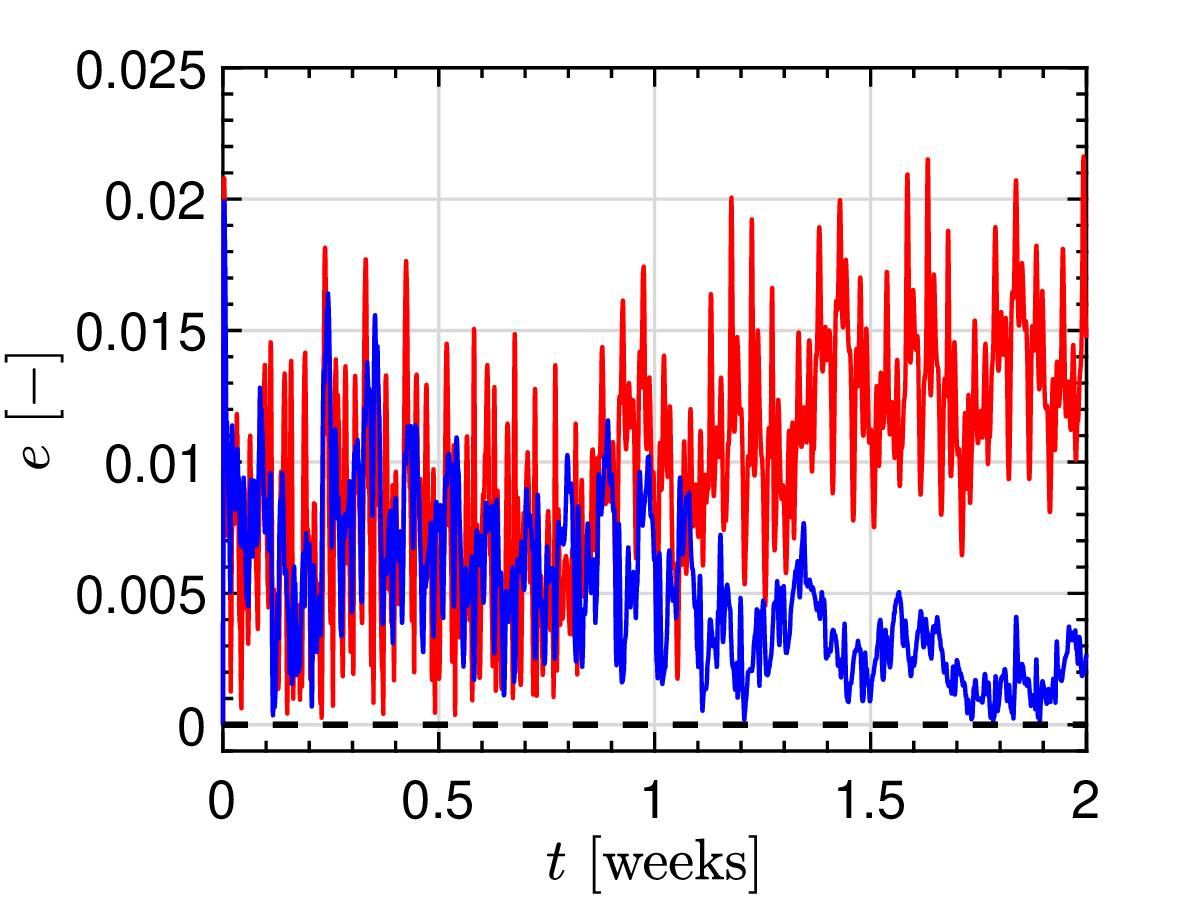} \label{e_comparison}}
		\caption{Semi-major axis (\textit{top}) and eccentricity (\textit{bottom}) for $i_0=90^{\circ}$. Blue: learning-based MPC; red: non-learning MPC.}
		\label{fig:orbit_control_variables_comparison}
	\end{center}
\end{figure}

\begin{figure}[t] 
	\begin{center}
	\includegraphics[width=9cm,height=9cm,keepaspectratio]{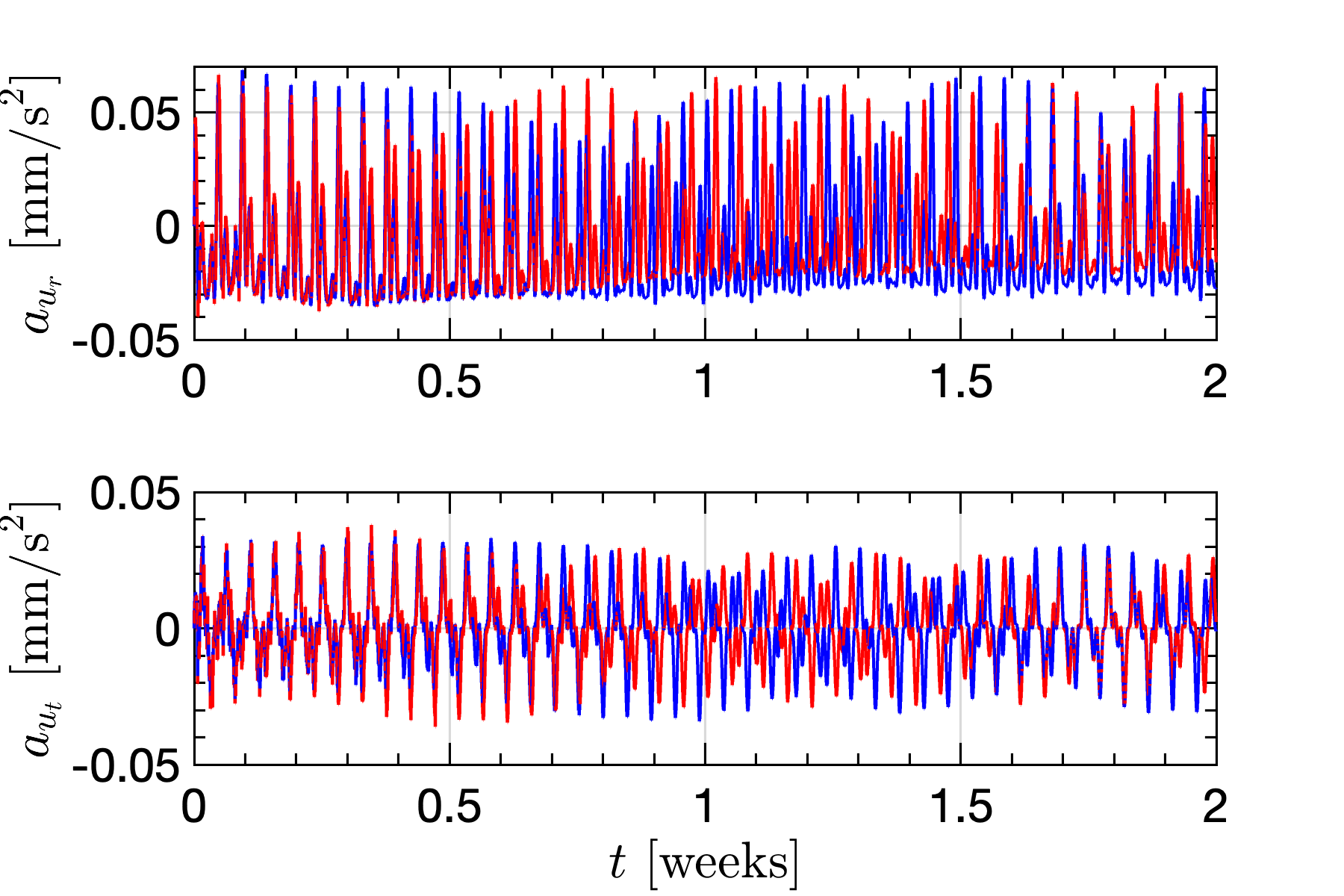}
		\caption{Control acceleration for $i_0=90^{\circ}$. Blue: learning-based MPC; red: non-learning MPC.}	
		\label{fig:u_comparison}
	\end{center}
\end{figure}

\textbf{Orbit results:} the orbital radius evolution of each case is shown in Fig.\ref{fig:orbital_radius}. Visually, it is evident that the learning-based MPC provides a higher degree of tracking accuracy for the cases $i_0=30^{\circ},90^{\circ}, 150^{\circ}$. It is also noted that the maximum tracking error typically arises at the initial simulations instants where the model accuracy is not good enough. Figure \ref{fig:cumulative_control_variables_comparison} shows both the average orbit tracking error and fuel consumption per day. It can be observed that, for the model learning cases, the tracking error per day typically decreases when compared to the initial days, thus demonstrating the goodness of the concept. On the other hand, the evolution of fuel consumption needs per day seems practically stationary (reduction in the long term does not seem significant).  
The previous results are translated to performance metrics, over the whole scenario, in Table \ref{table:orbit_performance_with_gravity}. In all the cases, the learning-based MPC outperforms the non-learning one in terms of tracking accuracy. Specifically, the learning-based approach is more accurate in a factor ranging from 0.23 ($i_0=120^{\circ}$) to 8.9 ($i_0=150^{\circ}$) in average. The maximum tracking error also favours the learning-based approach at the majority of cases reducing the peak by a factor of 3.1 ($i_0=150^{\circ}$) whilst there is a single case ($i_0=120^{\circ}$) where it slighty increases by just 42 m. This is to be expected as the learning-based benefits were seen in the long-term when the model is identified with a higher degree of accuracy. The fuel consumption needs are similar for both approaches as the learning-based MPC consumes more for $i_0=60^{\circ},90^{\circ},120^{\circ}$ and less for $i_0=30^{\circ},150^{\circ}$ when compared to the non-learning MPC. If the fuel of all the scenario is summed up, the learning-based MPC consumes only 219 g more. In conclusion, the superior performance of learning-based MPC in terms of tracking accuracy overcomes the potential slight increase in terms of fuel consumption. 
\begin{table*}[ht]
	\centering
	\begin{tabular}{lcccccc}
		\hline \hline
		\multicolumn{1}{l}{} & \multicolumn{3}{c}{Learning-based MPC} & \multicolumn{3}{c}{MPC with $\hat{C}_{ij}=\hat{S}_{ij}=0$}\\
		\midrule
		Simulation & $m_F[\text{kg}]$ & $\Delta R[\text{m}]$ & $\Delta R_{\text{max}}[\text{m}]$ & $m_F[\text{kg}]$ & $\Delta R[\text{m}]$ & $\Delta R_{\text{max}}[\text{m}]$\\
		\hline
		$i_0=30^{\circ}$ & 1.7962 & 160.12 & 615.16 & 1.9958 & 1023.0 & 1533.4\\  
		$i_0=60^{\circ}$ & 1.3505 & 251.39 & 609.87 & 1.3788 & 341.06 & 704.45\\ 
		$i_0=90^{\circ}$ & 1.3926 & 298.72 & 811.80 & 1.2560 & 562.58 & 1043.1\\  
		$i_0=120^{\circ}$ & 1.3392 & 155.36 & 485.75 & 1.2886 & 201.06 & 443.44\\  
		$i_0=150^{\circ}$ & 1.7338 & 123.90 & 465.42 & 1.9121 & 1099.2 & 1437.5\\       
		\hline \hline
	\end{tabular}
	\caption{Orbit control performance of learning-based MPC and non-learning MPC.}
	\label{table:orbit_performance_with_gravity}
\end{table*}    
It is also of interest to analyze GNC variables for a particular case. Figures \ref{fig:orbit_control_variables_comparison}-\ref{fig:u_comparison} show results of interest for both controllers in the polar orbit case, $i_0=90^{\circ}$. In Fig.\ref{fig:orbit_control_variables_comparison} the semi-major axis and eccentricity for both controllers is shown. In accordance to previous results, the learning-based MPC ends with an orbit of $\approx$33.9 km semi-major axis and a eccentricity of $\approx$0.002  which can be considered quasi-circular. The non-learning MPC orbit ends with $\approx$34.4 km semi-major axis and a eccentricity of $\approx$0.0125. In Fig.\ref{fig:u_comparison}, the tangential and radial control accelerations of both cases are plotted. Roughly, the same pattern is followed until the end of the first simulation week which is when the learning-based MPC control signal seems to be anticipated due to a more accurate prediction over the control horizon. The superior reference tracking accuracy, see Fig.\ref{fig:orbital_radius} and Fig.\ref{fig:orbit_control_variables_comparison}, of the learning-based MPC with respect to its non-learning counterpart can be directly correlated to the accurate estimation of second-order gravity in Table \ref{table:second_order_gravity}.
The orbit filter performance is shown in Tables \ref{table:orbit_navigation_errors}-\ref{table:orbit_navigation_residuals}. Table \ref{table:orbit_navigation_errors} shows the absolute position errors mean and maximum for the whole scenario and its lasts two days. It is observed that the errors for the last two days are below the scenario average, thus confirming the filter is increasing its accuracy. Table \ref{table:orbit_navigation_residuals} provides statistical information (bias and 1-$\sigma$ uncertainty) of the orbit filter residuals. If one compares the bias with the 1-$\sigma$ deviations, it can be deduced that residuals are slightly biased (one/two orders of magnitude lower than uncertainty). This complies with sensors datasheet as per Table \ref{table:sensors_datasheet} except from the fact that the camera residuals uncertainty is higher than the datasheet values. This discrepancy could be explained by the fact that process noise is non-zero, thus augmenting residuals uncertainty.
%\begin{figure}[] 
%	\begin{center}
%		\includegraphics[width=14cm,height=14cm,keepaspectratio]{RTN_poserror}
%		\caption{Radial, tangential and normal position error of learning-based MPC for $i_0=90^{\circ}$.}	
%		\label{fig:orbit_error}
%	\end{center}
%\end{figure}
%\begin{figure}[] 
%	\begin{center}
%		\includegraphics[width=14cm,height=14cm,keepaspectratio]{orbit_residuals}
%		\caption{Pixels and ranging distance residuals of learning-based MPC for $i_0=90^{\circ}$.}	
%		\label{fig:orbit_residuals}
%	\end{center}
%\end{figure}
\begin{table}[h]
	\centering
	\begin{tabular}{lcccccc}
		\hline \hline
		\multicolumn{1}{r}{} & \multicolumn{2}{c}{Days 1-14} & \multicolumn{2}{c}{Days 13-14}\\
		\hline
		\multicolumn{1}{l}{Nav. error} & Mean & max & Mean & max\\
		\hline
		Radial [m] & 1.7455 & 12.243 & 1.3531 & 6.2559\\  
		Tangential [m] & 14.067 & 136.37 & 9.0449 & 33.459\\  
		Normal [m] & 9.7575 & 80.250 & 8.4733 & 26.227\\
		\textbf{Total [m]} & \textbf{19.023} & \textbf{152.53} & \textbf{13.936} & \textbf{34.780}\\
		\hline \hline
	\end{tabular}
	\caption{Absolute navigation errors in position of learning-based MPC for $i_0=90^{\circ}$.}
	\label{table:orbit_navigation_errors}
\end{table}  
\begin{table}[h]
	\centering
	\begin{tabular}{lcccc}
		\hline \hline
		\multicolumn{1}{r}{} & \multicolumn{2}{c}{Days 1-14} & \multicolumn{2}{c}{Days 13-14}\\
		\hline
		\multicolumn{1}{l}{Residual} & Bias & 1-$\sigma$ & Bias & 1-$\sigma$\\
		\hline
		Pixel row [-] & 0.1019 & 2.4239 & -0.0696 & 1.5749\\  
		Pixel column [-] & -0.1121 & 1.7373 & -0.2088 & 1.4419\\  
		Range [m] & 0.0158 & 4.2282 & 0.0598 & 4.4095\\
		\hline \hline
	\end{tabular}
	\caption{Orbit filter residuals statistics of learning-based MPC for $i_0=90^{\circ}$.}
	\label{table:orbit_navigation_residuals}
\end{table}

\begin{figure}[t] 
	\begin{center}
				\includegraphics[width=9cm,height=9cm,keepaspectratio]{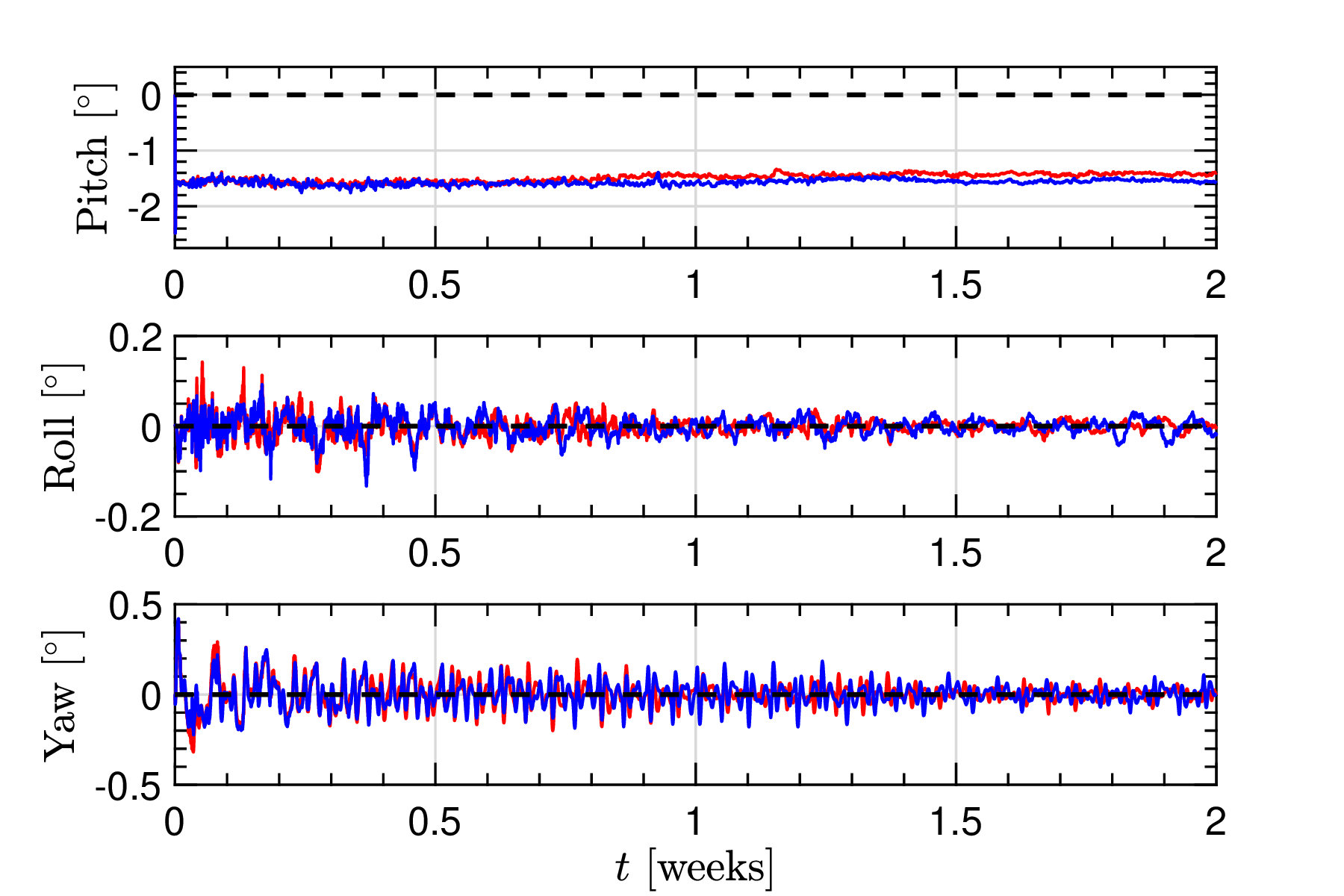}
		\caption{Pitch, roll and yaw for $i_0=90^{\circ}$. Blue: learning-based MPC; red: non-learning MPC.}	
		\label{fig:eulangles_comparison}
	\end{center}
\end{figure}

\begin{figure}[t] 
	\begin{center}
		\includegraphics[width=9cm,height=9cm,keepaspectratio]{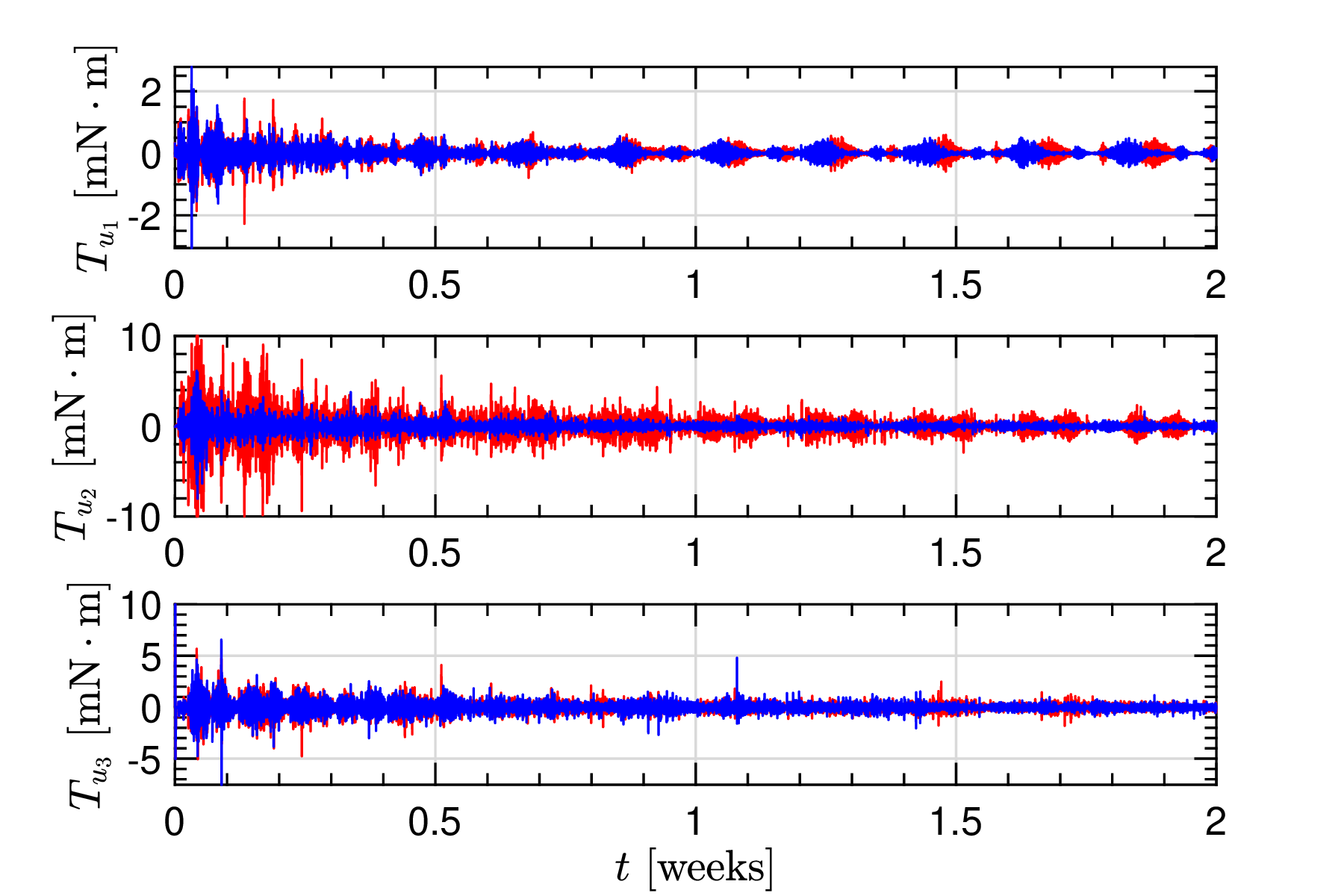}
		\caption{Control torque for $i_0=90^{\circ}$. Blue: learning-based MPC; red: non-learning MPC.}	
		\label{fig:Tu_comparison}
	\end{center}
\end{figure}

\textbf{Attitude results:} the attitude control perfomance for each case is shown in Table \ref{table:attitude_performance_with_gravity}. The attitude control accuracy is practically independent, in terms of mean and maximum tracking errors, from using learning-based MPC or non-learning MPC. The roll and yaw are driven to almost null mean values while the pitch angle presents an offset of $-1.5^{\circ}$ approximately. This may be explained by the fact that the target attitude is not an equilibrium of the system. An offset-free tracking MPC is only guaranteed if the target is an equilibrium. Still, the discrepancy is low enough, thus enabling camera pointing. In terms of control effort, the learning-based MPC is between a 35.2\% ($i_0=90^{\circ}$) and a 42.6\% ($i_0=150^{\circ}$) more efficient when compared to non-learning based MPC. The previous fact highlights the superiority of the learning-based MPC in terms of attitude control efficiency without tracking accuracy losses. 
\begin{table*}[h]
	\footnotesize
	\centering
	\begin{tabular}{lcccccc}
		\hline \hline
		\multicolumn{1}{l}{} & \multicolumn{3}{c}{Learning-based MPC} & \multicolumn{3}{c}{MPC with $\hat{C}_{ij}=\hat{S}_{ij}=0$}\\
		\hline
		\text{Simulation} & $T_U[\text{mN}\cdot\text{m}]$ & $\Delta\pmb{\Theta}[^{\circ}]$ & $\Delta\pmb{\Theta}_{\text{max}}[^{\circ}]$ & $T_U[\text{mN}\cdot\text{m}]$ & $\Delta\pmb{\Theta}[^{\circ}]$ & $\Delta\pmb{\Theta}_{\text{max}}[^{\circ}]$\\
		\hline
		$i_0=30^{\circ}$ & 0.6346 & [1.52, 0.02, 0.05]$^T$ & [2.53, 0.16, 0.39]$^T$ & 1.0519 & [1.67, 0.02, 0.05]$^T$ & [2.53, 0.16, 0.40]$^T$\\  
		$i_0=60^{\circ}$ & 0.5996 & [1.55, 0.02, 0.05]$^T$ & [2.53, 0.15, 0.38]$^T$ & 0.9424 & [1.56, 0.02, 0.05]$^T$ & [2.53, 0.13, 0.36]$^T$\\  
		$i_0=90^{\circ}$ & 0.5005 & [1.57, 0.02, 0.05]$^T$ & [2.54, 0.13, 0.42]$^T$ & 0.7773 & [1.49, 0.02, 0.05]$^T$ & [2.54, 0.12, 0.41]$^T$ \\  
		$i_0=120^{\circ}$ & 0.6027 & [1.55, 0.02, 0.04]$^T$ & [2.54, 0.16, 0.29]$^T$ & 0.9671 & [1.53, 0.02, 0.03]$^T$ & [2.54, 0.11, 0.27]$^T$\\  
		$i_0=150^{\circ}$ & 0.6011 & [1.52, 0.01, 0.04]$^T$ & [2.54, 0.09, 0.38]$^T$ & 1.0465 & [1.68, 0.01, 0.03]$^T$ & [2.54, 0.10, 0.27]$^T$\\  
		\hline \hline
	\end{tabular}
	\caption{Attitude control performance with and without gravity model learning.}
	\label{table:attitude_performance_with_gravity}
\end{table*}
Again, GNC results of interest for the polar orbit case are shown in Fig.  \ref{fig:eulangles_comparison}-\ref{fig:Tu_comparison}. The pitch, roll and yaw evolutions for both controllers are shown in Fig. \ref{fig:eulangles_comparison}. These evolutions are practically the same in accordance with Table \ref{table:attitude_performance_with_gravity}. In Fig. \ref{fig:Tu_comparison}, the applied torque in the body frame is presented where the higher control effort done by the non-learning MPC can be easily seen. 
The attitude filter performance is shown in Tables \ref{table:attitude_navigation_errors}-\ref{table:attitude_navigation_residuals}. Table \ref{table:attitude_navigation_errors} provides the mean and maximum absolute errors for pitch, roll and yaw angles and gyroscope bias. A decreasing trend in the angles navigation errors is clearly observed when the whole scenario is compared with the results for the last two days. The gyroscope bias errors is higher in the last day but its estimation error is still insignificant (0.05\%). Table \ref{table:attitude_navigation_residuals} shows the attitude residuals statistical information where similar conclusions with respect to the orbit filter can be yielded. The 1-$\sigma$ uncertainty is typically one/two orders of magnitude higher than the biases. The biases tends to slightly increase in all the residuals for the last two days. This suggests the presence of persistent model bias which is not being estimated. However, the gyroscope 1-$\sigma$ deviation (0.05 $^{\circ}/\text{h}$) as per Table \ref{table:sensors_datasheet} is approximately obtained.

%\begin{figure}[] 
%	\begin{center}
%		\includegraphics[width=14cm,height=14cm,keepaspectratio]{eulangles_error}
%		\caption{Pitch, yaw and roll navigation errors of learning-based MPC for $i_0=90^{\circ}$.}	
%		\label{fig:attitude_error}
%	\end{center}
%\end{figure}
%\begin{figure}[] 
%	\begin{center}
%		\includegraphics[width=14cm,height=14cm,keepaspectratio]{attitude_residuals}
%		\caption{Attitude filter residuals of learning-based MPC for $i_0=90^{\circ}$.}	
%		\label{fig:attitude_residuals}
%	\end{center}
%\end{figure}
\begin{table}[h]
	\centering
	\begin{tabular}{lcccccc}
		\hline \hline
		\multicolumn{1}{r}{} & \multicolumn{2}{c}{Days 1-14} & \multicolumn{2}{c}{Days 13-14}\\
		\hline
		\multicolumn{1}{l}{Nav. error} & Mean & max & Mean & max\\
		\hline
		Pitch [$^{\circ}$] & 0.0239 & 0.2318 & 0.0153 & 0.0587\\  
		Roll [$^{\circ}$] & 0.0166 & 0.1363 & 0.0143 & 0.0447\\  
		Yaw [$^{\circ}$] & 0.0426 & 0.3857 & 0.0191 & 0.0899\\
		Gyro bias [$^{\circ}/\text{h}$] & 0.0014 & 0.1265 & 0.0025 & 0.0026\\
		\hline \hline
	\end{tabular}
	\caption{Absolute navigation errors in pitch, roll, yaw and gyroscope bias of learning-based MPC for $i_0=90^{\circ}$.}
	\label{table:attitude_navigation_errors}
\end{table}  
\begin{table*}[ht]
	\centering
	\begin{tabular}{lcccccc}
		\hline \hline
		\multicolumn{1}{r}{} & \multicolumn{2}{c}{Days 1-14} & \multicolumn{2}{c}{Days 13-14}\\
		\hline
		\multicolumn{1}{l}{Residual} & Bias & 1-$\sigma$ & Bias & 1-$\sigma$\\
		\hline
		Star-tracker $\sigma_1$ [-] & -2.125$\cdot$10$^{-8}$ & 7.851$\cdot$10$^{-6}$ & -1.052$\cdot$10$^{-7}$ & 9.128$\cdot$10$^{-6}$\\
		Star-tracker $\sigma_2$ [-] & 8.334$\cdot$10$^{-8}$ & 7.830$\cdot$10$^{-6}$ & -9.768$\cdot$10$^{-8}$ & 8.528$\cdot$10$^{-6}$\\
		Star-tracker $\sigma_3$ [-] & -6.633$\cdot$10$^{-8}$ & 7.945$\cdot$10$^{-6}$ & -7.879$\cdot$10$^{-8}$ & 8.469$\cdot$10$^{-6}$\\
		Gyroscope $\omega_1$ [$^{\circ}/\text{h}$] & -4.345$\cdot$10$^{-4}$ & 0.0510 & 5.291$\cdot$10$^{-4}$ & 0.0509\\
		Gyroscope $\omega_2$ [$^{\circ}/\text{h}$] & 4.590$\cdot$10$^{-4}$ & 0.0541 & -1.160$\cdot$10$^{-3}$ & 0.0548\\
		Gyroscope $\omega_3$ [$^{\circ}/\text{h}$] & -2.734$\cdot$10$^{-3}$ & 0.0559 & -4.179$\cdot$10$^{-3}$ & 0.0603\\
		\hline \hline
	\end{tabular}
	\caption{Attitude filter residuals statistics of learning-based MPC for $i_0=90^{\circ}$.}
	\label{table:attitude_navigation_residuals}
\end{table*}  

\textbf{Computational effort:} the computational times (mean, 1-$\sigma$ deviation and maximum) of filters and guidance and control algorithms is shown in Table \ref{table:computational times}. In that table, MPC refers to the guidance and control module. The execution times have been measured in an i7-8700 CPU 3.2 GHz using a MATLAB environment. As expected, the execution of the guidance and control algorithms is the most time-consuming task being two orders of magnitude slower than the filters computation. The orbit modules (UKF and MPC) execution times are higher than its attitude counterparts which is due, in part, to their longer propagations periods. Translating the worst-case computation as a percentage of the sampling rates yields 2.30\% and 0.33\% execution times with respect to the filter calls period for attitude and orbit respectively. The guidance and control execution takes a 6.69\% (attitude) and 1.04\% (orbit) with respect to their sampling rates. These results are promising in terms of justifying the potential mission autonomy, at least for the orbit modules. Nonetheless, the attitude modules computational burden may be reduced if one renounces to estimate gravity within its filter and considers the non-learning attitude MPC (with lower control efficiency).
	\begin{table}[h]
		\centering
		\begin{tabular}{lcccccc}
			\hline \hline
			\multicolumn{1}{r}{} & Mean [s] & 1-$\sigma$ [s] & max [s]\\
			\hline
			Attitude UKF & 0.0482 & 8.242$\cdot$10$^{-4}$ & 0.0827\\  
			Orbit UKF & 0.0698 & 9.421$\cdot$10$^{-4}$ & 0.1171\\  
			Attitude MPC & 1.645 & 0.0134 & 2.408\\
			Orbit MPC & 3.721 & 0.0406 & 4.097\\    
			\hline \hline
		\end{tabular}
		\caption{Computational times of the GNC modules for $i_0=90^{\circ}$.}
		\label{table:computational times}
\end{table}
\subsubsection{Gravity estimation through constellations of satellites}\label{results_constellation}

This section is devoted to demonstrate the enhancement in terms of gravity parameters estimation by considering satellites constellations.Setting as common parameters $\bar{e}^{[\eta]}=\{0,\hdots,0\}$,$\Omega_0^{[\eta]}=\omega_0^{[\eta]}=\nu_0^{[\eta]}=\{0^{\circ},\hdots,0^{\circ}\}$, the following constellation configurations are explored. For 3 satellites, $\bar{a}^{[\eta]}=\{34,~36,~38\}~\text{km}$, whereas $
		i_0^{[\eta]}=\{45,~90,~135\}^{\circ}$. For 6 satellites, $\bar{a}^{[\eta]}=\{31,~33,~35,~37,~39,~41\}~\text{km}$, whereas $i_0^{[\eta]}=\{15,~45,~75,~105,~135,~165\}^{\circ}$. For 9 satellites, $\bar{a}^{[\eta]}=\{28,~30,~32,~34,~36,~38,~40,~42,~44\}~\text{km}$, whereas 
		$i_0^{[\eta]}=\{18,~36,~54,~72,~90,~108,~126,~144,~162\}^{\circ}$, 
where a significant separation between satellites (2 km) is considered to prevent collisions.

The relevant gravity  parameters estimation results for the constellation based configuration, along with previously simulated monolithic missions, are shown in Tables \ref{table:second_order_gravity}-\ref{table:fourth_order_gravity}.A gravity parameter is considered relevant if $|C_{ij}|,|S_{ij}|>2\cdot10^{-3}$. The estimation metrics are the final estimation error and the convergence time $t_{C_{ij}/S_{ij}}$. It is assumed convergence is achieved when the estimation error is under 20\% and the error is maintained below that threshold for the rest of the simulation. If no convergence is achieved it will be marked as "no" in the tables.

As shown in Table \ref{table:second_order_gravity}, the second-order gravity parameters are estimated accurately with estimation errors under 2.5\% except for the $S_{22}$ in the polar orbit case. It can also be observed that constellation-based configurations not neccesarily provide better estimation errors than monolithic missions. However, convergence is achieved faster at all cases (between 2-3.5 times) and the outliers arising in monolithic mission for $S_{22}$ are mitigated. The third and fourth order gravity terms are more difficult to estimate, since they are less observable and will likely absorbe model uncertainty, as shown in Tables \ref{table:third_order_gravity}-\ref{table:fourth_order_gravity}. In that line, monolithic missions (except $i_0=30^{\circ}$) fail to converge the 75\% of relevant third order gravity ($C_{31},C_{33},S_{31}$) overall. A similar trend is observed for fourth order gravity where these missions typically do not achieve convergence in the 60\% of cases ($C_{44}, S_{42}, S_{44}$). Constellation-based missions only fail achieving convergence, in average, for the 25\% of third-order gravity cases and the 40\% of fourth order gravity. This non-convergence only arises for the 3 Sats configuration while 6 Sats and 9 Sats missions achieves convergence for all the relevant spherical harmonics. Let compare the best monolithic mission $i_0=30^{\circ}$, in terms of gravity estimation, with the 9 Sats constellation. The 9 Sats constellation reduces convergence times in a 87.5\%, 45\% and 29\% for second, third and fourth order relevant gravity parameters respectively. The estimation error is more accurate for the 9 Sats constellation in 42\% of parameters, similar (error diference below 2\%) for the 33\% and less accurate for the 25\% of parameters when compared to the more accurate monolithic mission $i_0=30^{\circ}$.       
\begin{table*}[ht]
	\centering
	\begin{tabular}{lcccccc}
		\hline \hline
		\multicolumn{1}{l}{\text{Simulation}} & $C_{20}[\%]$ & $t_{C_{20}}[\text{h}]$ & $C_{22}[\%]$ & $t_{C_{22}}[\text{h}]$ & $S_{22}[\%]$ & $t_{S_{22}}[\text{h}]$\\
		\hline
		$i_0=30^{\circ}$ & 0.2027 & 6.4 & 0.5796 & 6.5 & 2.2642 & 74.8\\  
		$i_0=60^{\circ}$ & 0.6010 & 2.0 & 0.0815 & 1.3 & 1.3459 & 7.8\\  
		$i_0=90^{\circ}$ & 0.1482 & 1.9 & 0.1400 & 1.7 & 7.3956 & 193.1\\  
		$i_0=120^{\circ}$ & 0.1856 & 2.5 & 0.5236 & 0.6 & 0.6305 & 38.5\\  
		$i_0=150^{\circ}$ & 0.3760 & 3.5 & 0.2751 & 0.5 & 2.2060 & 34.1\\    
		\textbf{Average} & \textbf{0.3027} & \textbf{3.3} & \textbf{0.3200} & \textbf{2.1} & \textbf{2.7684} & \textbf{69.7}\\
		\hline
		$\text{3~Sats}$ & 0.3490 & 1.7 & 0.6779 & 0.8 & 2.4628 & 43.2\\  
		$\text{6~Sats}$ & 0.5284 & 1.7 & 0.3093 & 1.0 & 1.1560 & 5.7\\    
		$\text{9~Sats}$ & 0.4747 & 1.7 & 1.6169 & 1.0 & 0.2184 & 6.0\\  
		\textbf{Average} & \textbf{0.4507} & \textbf{1.7} & \textbf{0.8680} & \textbf{0.9} & \textbf{1.2719} & \textbf{18.3}\\
		\hline \hline
	\end{tabular}
	\caption{Relevant second order gravity parameters estimation.}
	\label{table:second_order_gravity}
\end{table*}  
\begin{table*}[h]
	\centering
	\begin{tabular}{lcccccccc}
		\hline \hline
		\multicolumn{1}{l}{\text{Simulation}} & $C_{31}~[\%]$ & $t_{C_{31}}~[\text{h}]$ & $C_{33}~[\%]$ & $t_{C_{33}}~[\text{h}]$ & $S_{31}~[\%]$ & $t_{S_{31}}~[\text{h}]$ & $S_{33}~[\%]$ & $t_{S_{33}}~[\text{h}]$\\
		\hline
		$i_0=30^{\circ}$ & 13.712 & 279.0 & 18.729 & 320.3 & 10.179 & 187.1 & 0.0732 & 80.9\\  
		$i_0=60^{\circ}$ & 4.7739 & 104.2 & 55.491 & no & 14.222 & 172.6 & 0.0951 & 159.7\\  
		$i_0=90^{\circ}$ & 11.943 & 140.2 & 73.011 & no & 21.608 & no & 0.1352 & 185.3\\  
		$i_0=120^{\circ}$ & 30.669 & no & 45.803 & no & 5.9831 & 248.5 & 8.4806 & 134.7\\  
		$i_0=150^{\circ}$ & 47.641 & no & 31.365 & no & 0.0857 & 156.6 & 1.4666 & 37.3\\    
		\textbf{Average} & \textbf{21.748} & \textbf{no} & \textbf{44.880} & \textbf{no} & \textbf{10.4156} & \textbf{no} & \textbf{2.0501} & \textbf{119.58}\\    
		\hline
		$\text{3~Sats}$ & 1.3464 & 122.9 & 40.868 & no & 1.2289 & 120.0 & 10.096 & 180.9\\  
		$\text{6~Sats}$ & 10.7600 & 117.0 & 5.9992 & 254.9 & 9.6157 & 31.2 & 4.6747 & 82.9\\    
		$\text{9~Sats}$ & 7.2493 & 109.7 & 0.3219 & 176.0 & 4.5998 & 26.6 & 3.4082 & 61.7\\  
		\textbf{Average} & \textbf{6.4519} & \textbf{116.5} & \textbf{15.730} & \textbf{no} & \textbf{5.1484} & \textbf{59.3} & \textbf{6.0596} & \textbf{108.5}\\
		\hline \hline
	\end{tabular}
	\caption{Relevant third order gravity parameters estimation.}
	\label{table:third_order_gravity}
\end{table*}
\begin{table*}[h]
	\centering
	\begin{tabular}{lcccccccccc}
		\hline \hline
		\multicolumn{1}{c}{\text{Simulation}} & $C_{40}[\%]$ & $t_{C_{40}}[\text{h}]$ & $C_{42}[\%]$ & $t_{C_{42}}[\text{h}]$ & $C_{44}[\%]$ & $t_{C_{44}}[\text{h}]$ & $S_{42}[\%]$ & $t_{S_{42}}[\text{h}]$ & $S_{44}[\%]$ & $t_{S_{44}}[\text{h}]$\\
		\hline
		$i_0=30^{\circ}$ & 2.4286 & 180.9 & 11.680 & 93.0 & 17.764 & 204.7 & 4.0302 & 187.1 & 1.6548 & 152.7\\  
		$i_0=60^{\circ}$ & 15.562 & 268.9 & 3.3268 & 84.4 & 14.513 & 197.6 & 2.2619 & 39.1 & 31.160 & no\\  
		$i_0=90^{\circ}$ & 8.1015 & 125.0 & 3.3383 & 114.6 & 32.887 & no & 14.194 & 329.5 & 56.613 & no\\  
		$i_0=120^{\circ}$ & 3.8394 & 80.6 & 0.3583 & 97.2 & 24.283 & no & 26.187 & no & 26.900 & no\\   
		$i_0=150^{\circ}$ & 12.150 & 145.9 & 3.9113 & 168.7 & 25.408 & no & 32.276 & no & 7.2245 & 142.7\\    
		\textbf{Average} & \textbf{8.4163} & \textbf{160.3} & \textbf{4.5229} & \textbf{111.6} & \textbf{22.971} & \textbf{no} & \textbf{15.790} & \textbf{no} & \textbf{24.711} & \textbf{no}\\  
		\hline
		$\text{3~Sats}$ & 6.0512 & 172.6 & 1.6459 & 113.3 & 24.165 & no & 3.0307 & 99.7 & 22.265 & no\\  
		$\text{6~Sats}$ & 3.9950 & 105.1 & 4.9765 & 106.6 & 5.8115 & 198.8 & 3.2776 & 150.2 & 7.2308 & 225.4\\  
		$\text{9~Sats}$ & 10.202 & 99.3 & 12.684 & 73.2 & 3.1098 & 171.7 & 11.797 & 57.4 & 1.0366 & 92.5\\   
		\textbf{Average} & \textbf{6.7493} & \textbf{125.7} & \textbf{6.4355} & \textbf{97.7} & \textbf{11.029} & \textbf{no} & \textbf{6.0351} & \textbf{102.4} & \textbf{10.178} & \textbf{no}\\  
		\hline \hline
	\end{tabular}
	\caption{Relevant fourth order gravity parameters estimation.}
	\label{table:fourth_order_gravity}
\end{table*}

\begin{figure}[ht] 
	\begin{center}
		\includegraphics[width=9cm,height=9cm,keepaspectratio]{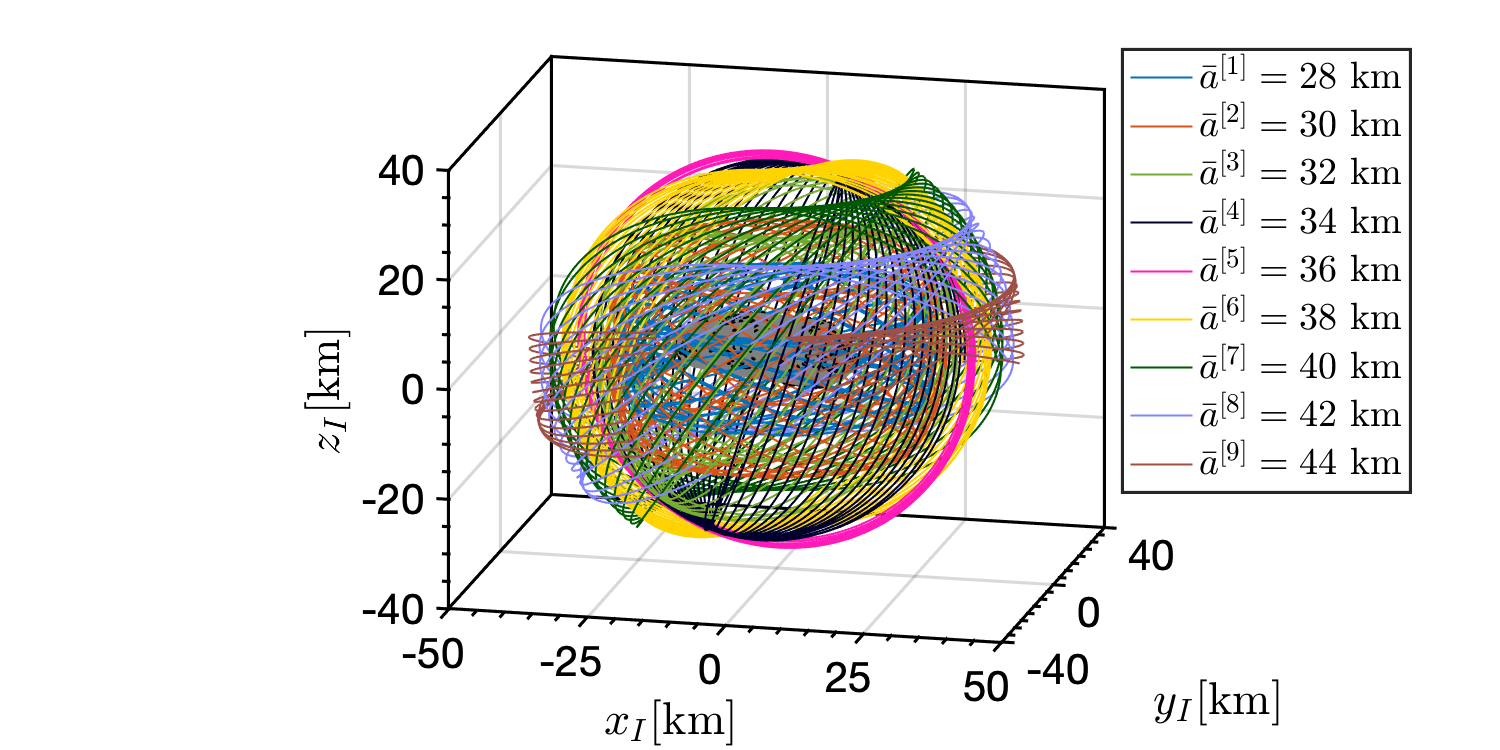}
		\caption{Trajectories in the inertial frame for 9 Sats constellation. Black dots: surface landmarks.}	
		\label{fig:3Dtrajectory_inertial_9sats}
	\end{center}
\end{figure}

\begin{figure*}[ht] 
	\begin{center}
		\includegraphics[width=14cm,height=14cm,keepaspectratio]{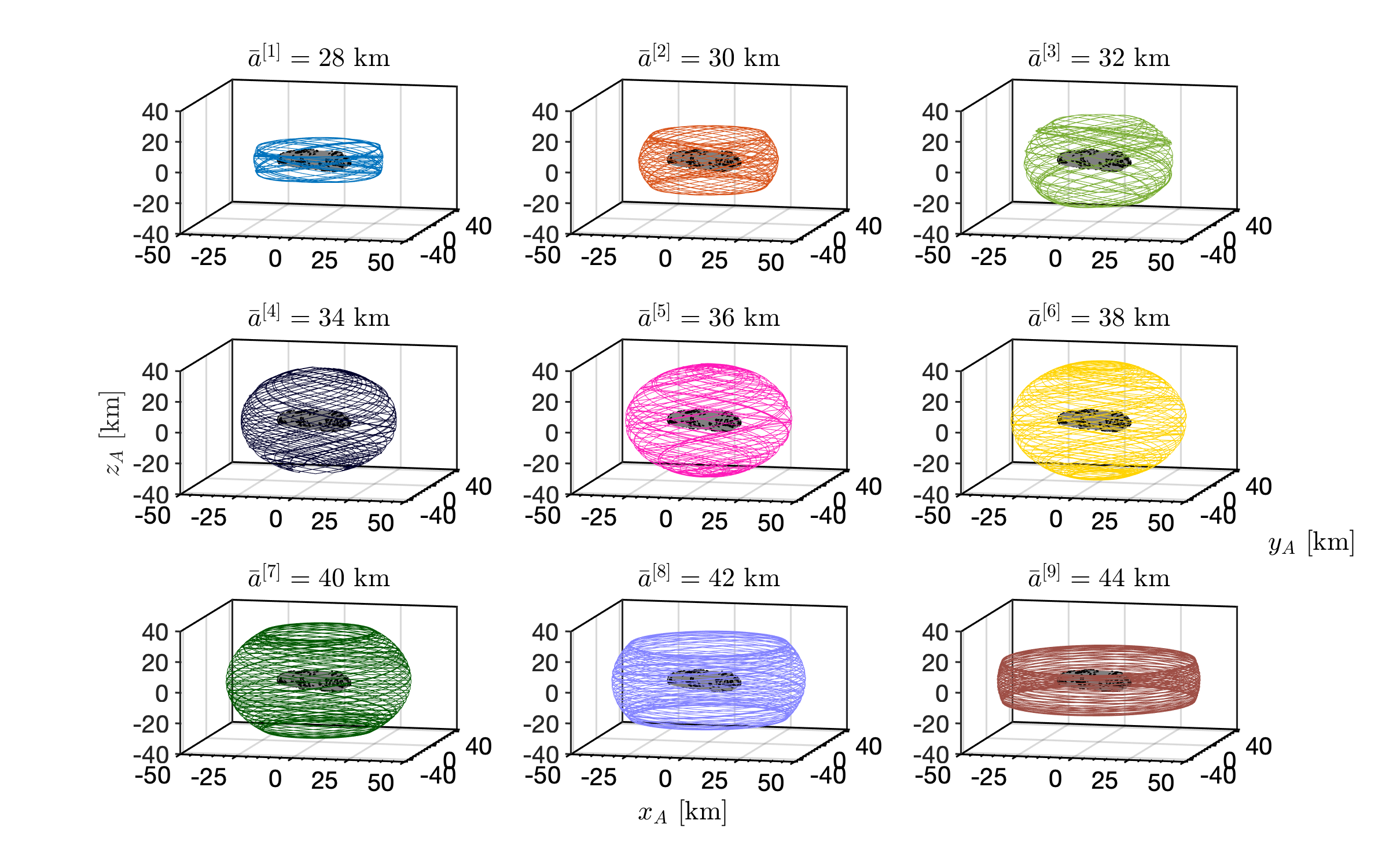}
		\caption{Individual trajectories in the asteroid frame for 9 Sats constellation. Black dots: surface landmarks.}	
		\label{fig:3Dtrajectory_asteroid_9sats}
	\end{center}
\end{figure*}

Some results of interest for the 9 Sats constellation can be seen in Fig.\ref{fig:3Dtrajectory_inertial_9sats}-\ref{fig:eulangles_9sats}. Figure \ref{fig:3Dtrajectory_inertial_9sats} show the probes trajectories in the inertial frame where it can be easily seen that the orbits are closed but affected by nodal precession as the right ascension of the ascending node is not being controlled. In Fig.\ref{fig:3Dtrajectory_asteroid_9sats}, the trajectories are projected into the asteroid frame in order to provide information of what asteroid regions are being overflown. The satellites orbital radius is shown in Fig.\ref{fig:r_9sats} where tracking convergence is achieved at all cases after an initial transient period (poor gravity model). Figure \ref{fig:i_9sats} shows the orbital inclinations which are constant due to the fact that no out-of-plane control is being applied. This helps to predict what asteroid regions will be overflown from the constellation design. The tracking error metric and fuel consumption of the different satellites is shown in Fig.\ref{fig:fuelrange_9sats}. A clear increasing trend of fuel consumption needs is clearly seen as the spacecraft orbits closer to the asteroid. On the other, no clear trends with the tracking error can be seen which is partly due to the fact that the inclination also plays a role in the orbital stability.

The constellation gravity estimation, for the relevant parameters, can be seen in Fig.\ref{fig:2nd3rdordergravity_9sats}-\ref{fig:4ordergravity_9sats}. This confirms previous results such as better accuracies and faster convergence times for the more dominant second order terms than for third and fourth order gravity. The gravity estimation uncertainty is progressively reduced. The constellation 1-$\sigma$ uncertainty is computed as $\sigma^2_{\hat{C}_{ij}}=\sum^{\eta_{\text{max}}}_{\eta=1}w^{[\eta]}_{\hat{C}_{ij}}(\sigma^{[\eta]}_{\hat{C}_{ij}})^2$ and $\sigma^2_{\hat{S}_{ij}}=\sum^{\eta_{\text{max}}}_{\eta=1}w^{[\eta]}_{\hat{S}_{ij}}(\sigma^{[\eta]}_{\hat{S}_{ij}})^2$, though not inserted in the filters as mentioned in Section \ref{sec:constellation}.

Finally, Fig.\ref{fig:eulangles_9sats} show the pitch, roll and yaw angles for all spacecrafts. This figure confirms previous trends as the roll and yaw are effectively driven to a null value while the pitch remains with a considerable offset between 1$^{\circ}$-3$^{\circ}$. In light of Fig. \ref{fig:eulangles_9sats}, it is also concluded that orbits closer to the asteroid surface are more challenging to control. 

\section{Conclusions}

This paper has presented a learning-based predictive algorithm for orbit-attitude asteroid station-keeping. The main ingredients are the unscented Kalman filter (UKF), for navigation and model identification, and model predictive control (MPC), for guidance and control. The numerical results have confirmed the positive impact of learning-based control in terms of orbit control accuracy and attitude control efficiency when compared to a non-learning MPC strategy. The computational times are not prohibitive and the assumed on-board sensors does not rely on Earth ground segment. Consequently, this is a first step towards demonstrating the feasibility and autonomy of the proposed mission concept. Finally, a spacecraft constellation mission has been presented as a proof of concept (since distributed systems practical aspects are not addressed) to improve gravity estimation. When compared to monolithic missions, the constellation concept provides significant gains in terms of estimation convergence without accuracy losses. Actually, the most relevant second-order gravity estimation convergence was reduced in a 50\% of time overall.     

Future work includes exploring emergent machine learning techniques such as neural networks for model identification. This could be combined with a more accurate particle-based approach when compared to the unscented transform. This approach may benefit by differential algebra which have drastically reduced uncertainty propagation computational cost for a high number of samples. The navigation process, for the constellation of satellites, could be evolved to a distributed Kalman filter \cite{Olfati2007} in order to solve the consensus problem, from the uncertainty update perspective, when gathering all the local estimates. From the control perspective, two main future lines are identified. The continuous rejection of orbital perturbations may unnecessarily expend fuel in cancelling the short-period variations of the modified equinoctial elements (MEE). The reformulation of the control in terms of mean modified equinoctial elements could potentially reduce fuel consumption. Finally, the basic MPC formulation can be robustified with a stochastic prediction model, in the spirit of \cite{Gavilan2012}, using the filters uncertainties.

\section*{Acknowledgments}

The authors gratefully acknowledge financial support from Universidad de Sevilla, through its V-PPI US, under grant PP2016-6975 and from the Spanish Ministry of Science and Innovation under grant PGC2018-100680-B-C21.

\bibliography{asteroid_bib}
\bibliographystyle{aiaa}

\appendix
\section{Appendix}\label{appendix}

\begin{figure}[h!] 
	\begin{center}
		\subfigure{\includegraphics[width=8cm,height=8cm,keepaspectratio]{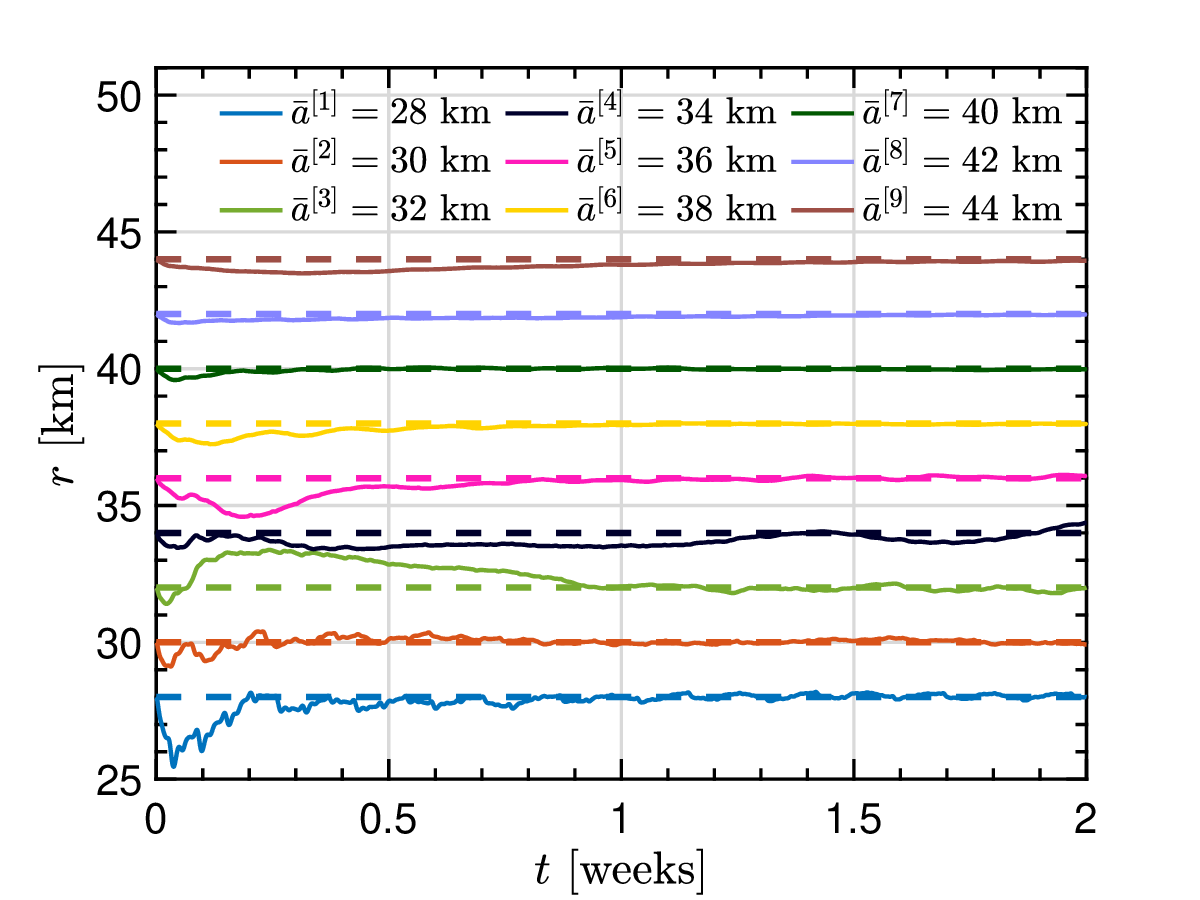}\label{fig:r_9sats}}
		\subfigure{\includegraphics[width=8cm,height=8cm,keepaspectratio]{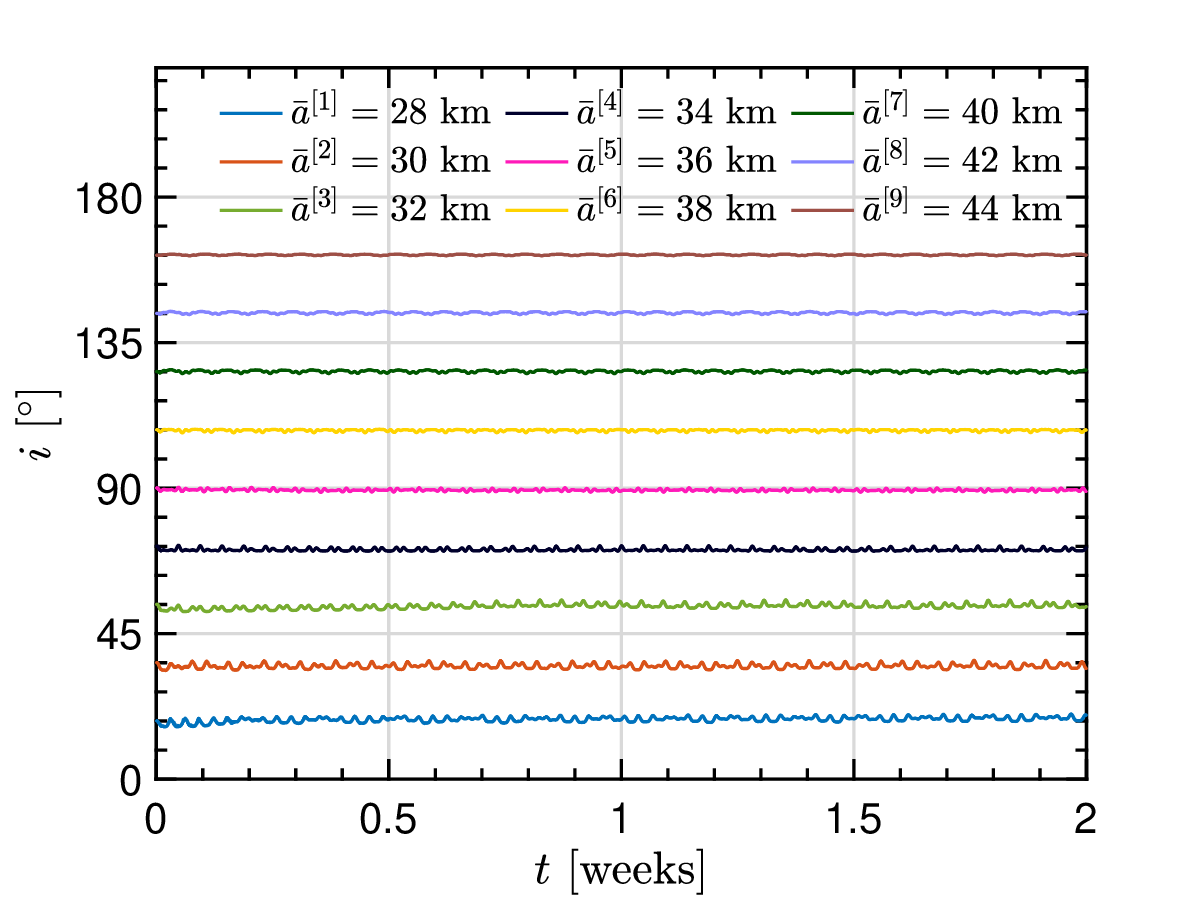} \label{fig:i_9sats}}
		\caption{Radius (\textit{top}) and inclination (\textit{bottom}) for 9 Sats constellation.}
		\label{fig:orbit_control_variables_9sats}
	\end{center}
\end{figure}

\begin{figure}[h!] 
	\begin{center}
		\subfigure{\includegraphics[width=8cm,height=8cm,keepaspectratio]{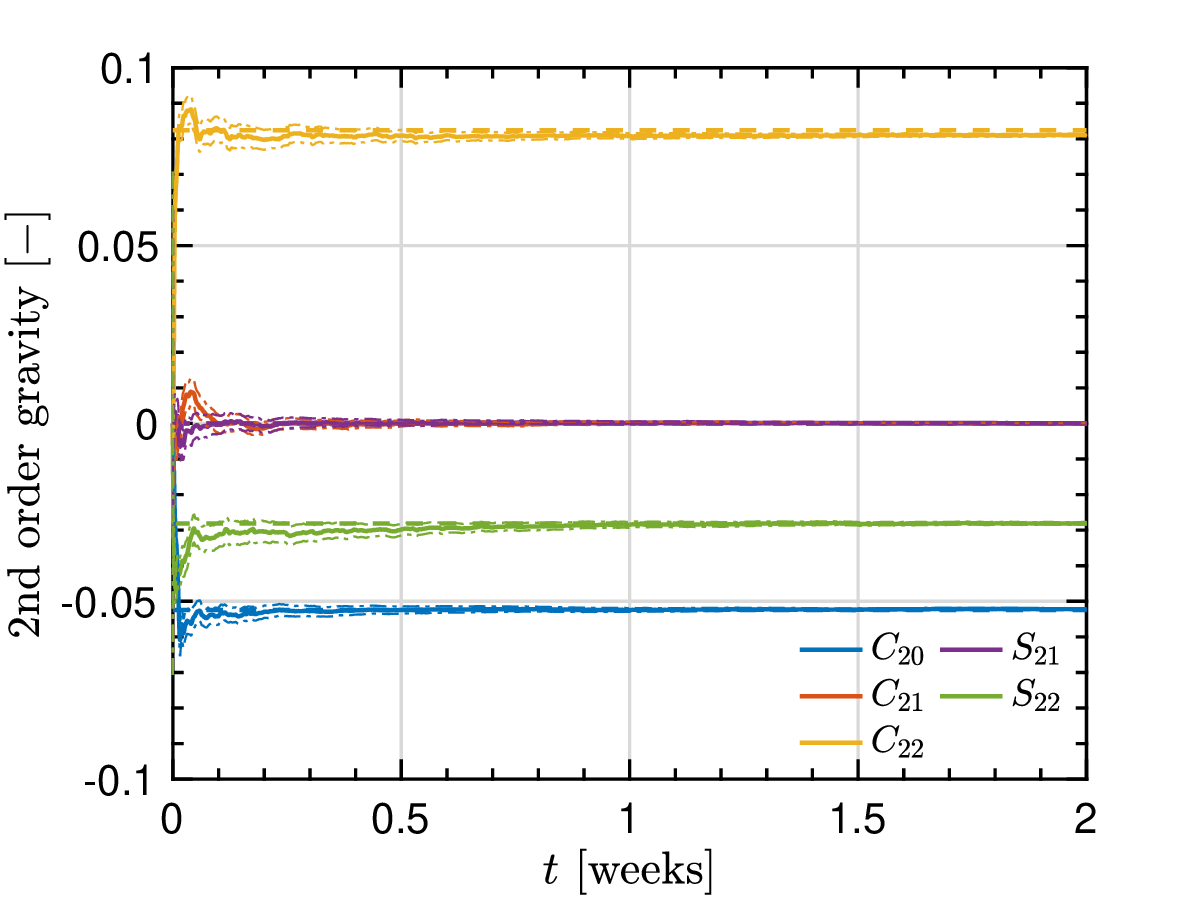}\label{fig:2ndordergravity_9sats}}
		\subfigure{\includegraphics[width=8cm,height=8cm,keepaspectratio]{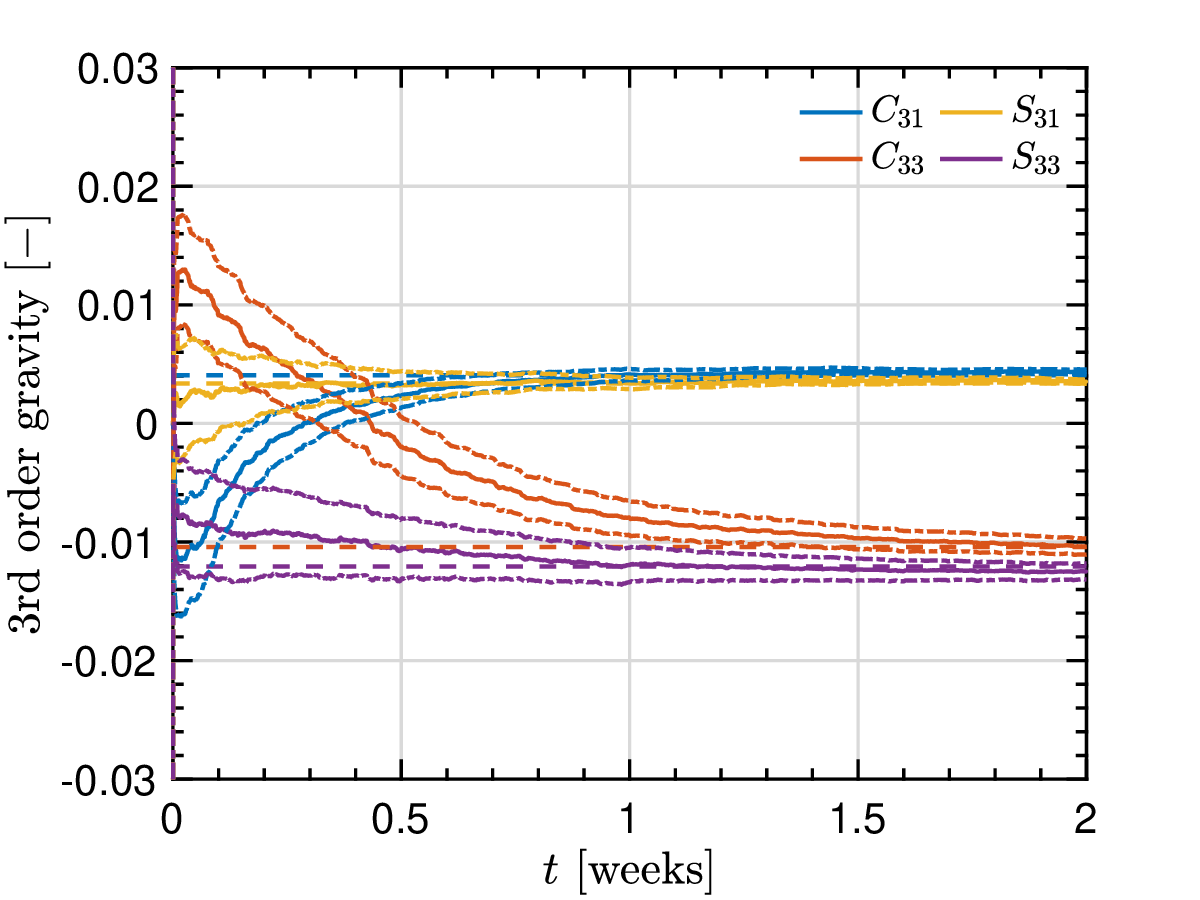}\label{3ordergravity_9sats}}
		\caption{Relevant second-order gravity (\textit{top}) and third-order gravity (\textit{bottom}) parameters for 9 Sats constellation; dashed$\equiv$truth, solid$\equiv$estimation, dot-dashed$\equiv$1-$\sigma$ uncertainty.}
		\label{fig:2nd3rdordergravity_9sats}
	\end{center}
\end{figure}
\begin{figure*}[h!] 
	\begin{center}
			\includegraphics[width=11cm,height=11cm,keepaspectratio]{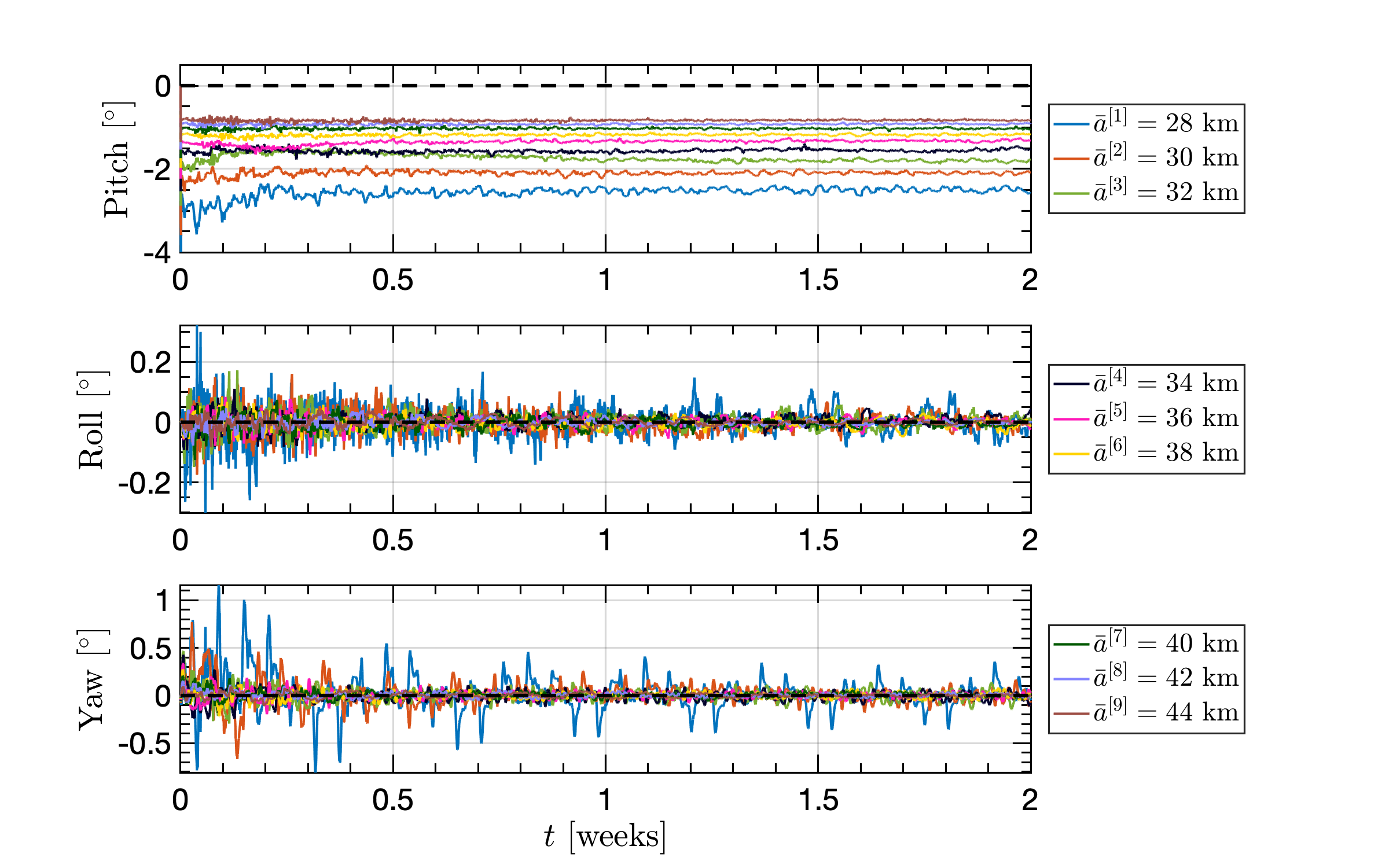}
		\caption{Pitch, roll and yaw for 9 Sats constellation.}	
		\label{fig:eulangles_9sats}
	\end{center}
\end{figure*}
\begin{figure}[h!] 
	\begin{center}
		\subfigure{\includegraphics[width=8cm,height=8cm,keepaspectratio]{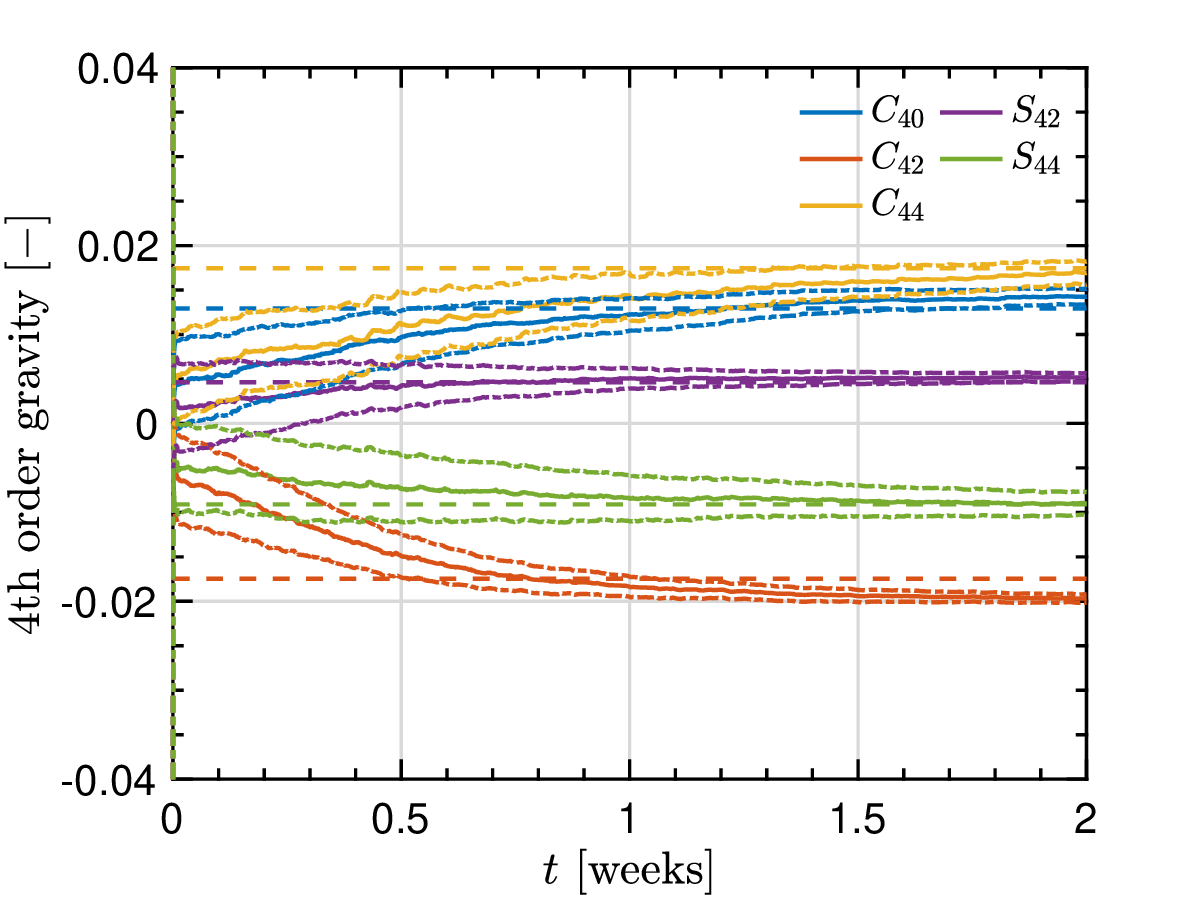}\label{fig:4ordergravity_9sats}}
		\subfigure{\includegraphics[width=8cm,height=8cm,keepaspectratio]{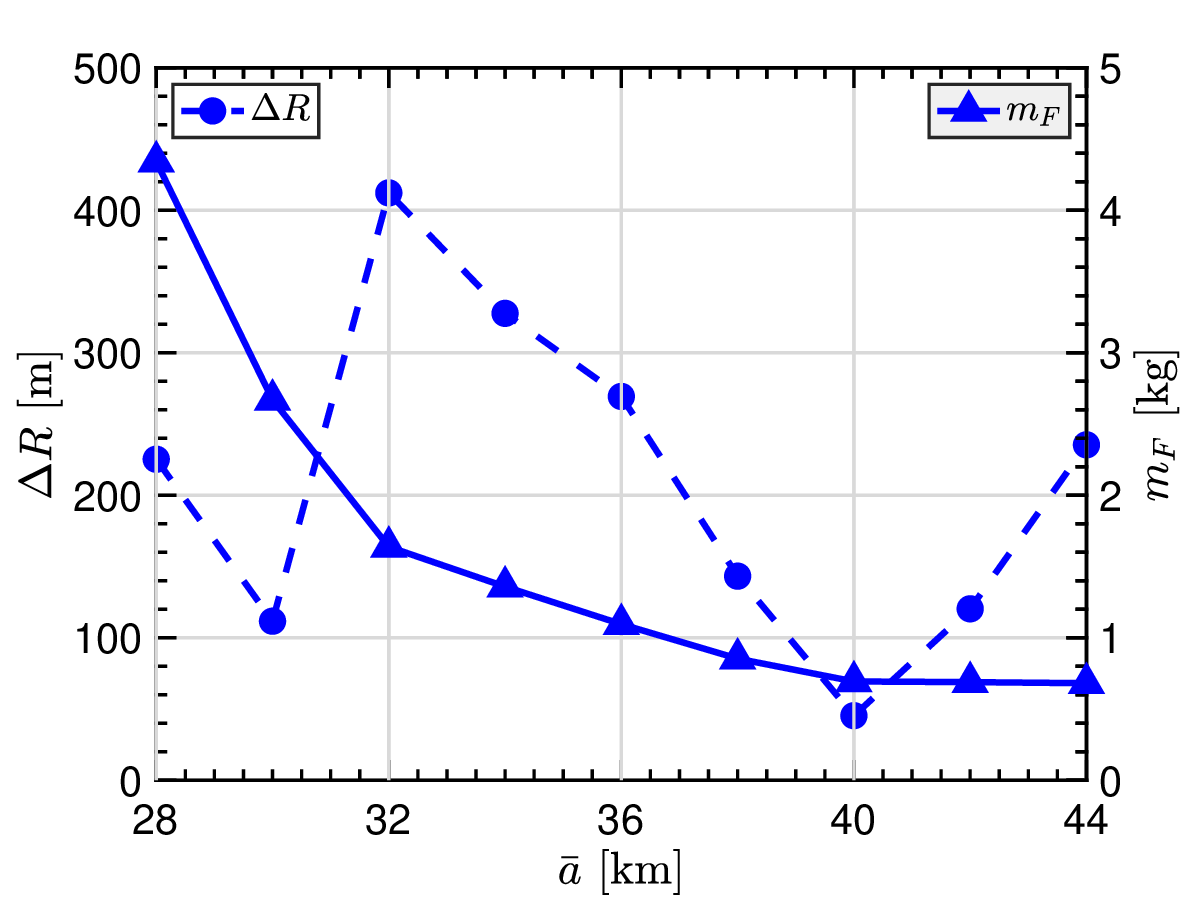}\label{fig:fuelrange_9sats}}
		\caption{Relevant fourth-order gravity parameters (\textit{top}), orbit tracking metric and fuel consumption (\textit{bottom}) for 9 Sats constellation; dashed$\equiv$truth, solid$\equiv$estimation, dot-dashed$\equiv$1-$\sigma$ uncertainty.}
		\label{fig:3th4thordergravity_9sats}
	\end{center}
\end{figure}

In this appendix, the orbit-attitude controllers formulations are briefly presented. The facts that the orbit reference is consistent (thus the orbit reference drift is null $\Delta\dot{\bar{\mathbf{x}}}_{\text{orb}}=\mathbf{0}$) and that the attitude one is not since $\bar{\mathbf{T}}_u=\mathbf{0}$ (thus $\mathbf{T}_u=\Delta\mathbf{T}_u$) have been explicitly taken into account.

\subsection{Orbit control}

\begin{equation}
	\begin{array}{rrclcl}
		\displaystyle \min_{\Delta\mathbf{x}_{\text{orb}}(t),\Delta\mathbf{a}_u(t)} && \multicolumn{3}{l} {J_{\text{orb}}=\frac{1}{t_f-t_0}\int^{t_f}_{t_0}\left(\gamma_{\text{orb}}\Delta\mathbf{x}^T_{\text{orb}}(t)\mathbf{P}_{x_\text{orb}}\Delta\mathbf{x}_{\text{orb}}(t)+\Delta\mathbf{a}_{u}^T(t)\Delta\mathbf{a}_{u}(t)\right)dt,} \\
		\textrm{s.t.} &&&\Delta\dot{\mathbf{x}}_{\text{orb}}(t)=\dot{\mathbf{x}}_{\text{orb}}(\mathbf{x}_{\text{orb}}(t),\mathbf{a}_u(t))-\dot{\bar{\mathbf{x}}}_{\text{orb}}(t),\\
		&&& \mathbf{a}_u(t)=\bar{\mathbf{a}}_u(t) + \Delta\mathbf{a}_u(t),\quad\Delta a_{u_n}(t)=0,\\
		&&&-\mathbf{a}_{u_{\text{max}}} \leq \mathbf{a}_u(t) \leq \mathbf{a}_{u_{\text{max}}}.\\
	\end{array}\label{eq:orb_control_problem_continuous}
\end{equation}
\textbf{Dynamics linearization:}
\begin{equation*}\label{eq:orb_dynamics_linearization}
	\dot{\mathbf{x}}_{\text{orb}}(\mathbf{x}_{\text{orb}}(t),\mathbf{a}_u(t))\approx\dot{\bar{\mathbf{x}}}_{\text{orb}}(t)+\mathbf{A}_{\text{orb}}(\bar{\mathbf{x}}_{\text{orb}}(t),\bar{\mathbf{a}}_{u}(t))\Delta\mathbf{x}_{\text{orb}}(t)+\mathbf{B}_{\text{orb}}(\bar{\mathbf{x}}_{\text{orb}}(t))\Delta\mathbf{a}_u(t),
\end{equation*}
\begin{equation*}
	\mathbf{A}_{\text{orb}}=\frac{\partial \dot{\mathbf{x}}_{\text{orb}}}{\partial \mathbf{x}_{\text{orb}}}\bigg\rvert_{\bar{\mathbf{x}}_{\text{orb}}(t),\bar{\mathbf{a}}_u(t)}+\frac{\partial \dot{\mathbf{x}}_{\text{orb}}}{\partial\mathbf{a}_{\text{grav}}}\frac{\partial\mathbf{a}_{\text{grav}}}{\partial{\mathbf{x}_{\text{orb}}}}\bigg\rvert_{\bar{\mathbf{x}}_{\text{orb}}(t),\bar{\mathbf{a}}_u(t)},\quad\mathbf{B}_{\text{orb}}=\frac{\partial \dot{\mathbf{x}}_{\text{orb}}}{\partial \mathbf{a}_u}\bigg\rvert_{\bar{\mathbf{x}}_{\text{orb}}(t),\bar{\mathbf{a}}_u(t)}. 
\end{equation*}
\begin{equation*}\label{eq:orb_err_state_dynamics_linear}
	\Delta\dot{\mathbf{x}}_{\text{orb}}(t)=\mathbf{A}_{\text{orb}}(\bar{\mathbf{x}}_{\text{orb}}(t),\bar{\mathbf{a}}_{u}(t))\Delta\mathbf{x}_{\text{orb}}(t)+\mathbf{B}_{\text{orb}}(\bar{\mathbf{x}}_{\text{orb}}(t))\Delta\mathbf{a}_u(t),
\end{equation*}
\begin{equation*}\label{eq:orb_err_state_LTV_solution}
	\Delta\mathbf{x}_{\text{orb}}(t)=\pmb{\Phi}_{\text{orb}}(t,t_0)\Delta\mathbf{x}_{\text{orb},0}+\int^{t}_{t_0}\pmb{\Phi}_{\text{orb}}(t,\tau)\mathbf{B}_{\text{orb}}(\tau)\Delta\mathbf{a}_u(\tau)d\tau,
\end{equation*}  
\begin{equation*}
	\dot{\pmb{\Phi}}_{\text{orb}}(t,t_0)=\mathbf{A}_{\text{orb}}(\bar{\mathbf{x}}_{\text{orb}}(t),\bar{\mathbf{a}}_u(t))\pmb{\Phi}_{\text{orb}}(t,t_0),\quad\pmb{\Phi}_{\text{orb}}(t_0,t_0)=\mathbf{I}.\label{eq:STM_orb_dynamics}
\end{equation*}
\textbf{Discretization:}
\begin{eqnarray*}
	\Delta\mathbf{x}_{\text{orb},k}&=&\pmb{\Phi}_{\text{orb}}(t_k,t_0)\Delta\mathbf{x}_{\text{orb},0}\\&&+\sum^{k}_{i=1}\pmb{\Phi}_{\text{orb}}(t_k,t_i)\left(\int^{t_i}_{t_{i-1}}\pmb{\Phi}_{\text{orb}}(t_i,\tau)\mathbf{B}_{\text{orb}}(\tau)d\tau\right)\Delta\mathbf{a}_{u,i}.\\
	J_{\text{orb}}&=&\sum^{N_{\text{orb}}}_{k=1}\left(\gamma_{\text{orb}}\Delta\mathbf{x}^T_{\text{orb},k}\mathbf{P}_{x_{\text{orb}}}\Delta\mathbf{x}_{\text{orb},k}+\Delta\mathbf{a}^T_{u,k}\Delta\mathbf{a}_{u,k}\right).
\end{eqnarray*}
\textbf{Compact formulation:}
\begin{equation*}\label{eq:orb_err_state_compact_propagation}
	\Delta\mathbf{x}_{\mathbf{S}\text{orb}}=\mathbf{D}_{\text{orb}}\Delta\mathbf{x}_{\text{orb},0}+\mathbf{G}_{\text{orb}}\Delta\mathbf{a}_{\mathbf{S}u},
\end{equation*}
\begin{equation*}\label{eq:orb_obj_function_compact}
	J_{\text{orb}}=\gamma_{\text{orb}}\Delta\mathbf{x}^T_{\mathbf{S}\text{orb}}\mathbf{P}_{\mathbf{S}x_{\text{orb}}}\Delta\mathbf{x}_{\mathbf{S}\text{orb}}+\Delta\mathbf{a}^T_{\mathbf{S}u}\Delta\mathbf{a}_{\mathbf{S}u}.
\end{equation*}
\begin{equation}
\begin{array}{rll}
			\displaystyle \min_{\Delta\mathbf{a}_{\mathbf{S}u}} &J_{\text{orb}}=2\gamma_{\text{orb}}\Delta\mathbf{x}^T_{\text{orb},0}\mathbf{D}^T_{\text{orb}}\mathbf{P}_{\mathbf{S}x_{\text{orb}}}\mathbf{G}_{\text{orb}}\Delta\mathbf{a}_{\mathbf{S}u}\\&+\Delta\mathbf{a}^T_{\mathbf{S}u}(\gamma_{\text{orb}}\mathbf{G}_{\text{orb}}^T\mathbf{P}_{\mathbf{S}x_{\text{orb}}}\mathbf{G}_{\text{orb}}+\mathbf{I})\Delta\mathbf{a}_{\mathbf{S}u},\\
			\textrm{s.t.}
			& \mathbf{W}_{\mathbf{S}u_n}\Delta\mathbf{a}_{\mathbf{S}u}=\mathbf{0}_{N_{\text{orb}}\times1},\quad
			-\mathbf{a}_{\mathbf{S}u_{\text{max}}} \leq \bar{\mathbf{a}}_{\mathbf{S}u}+\Delta\mathbf{a}_{\mathbf{S}u} \leq \mathbf{a}_{\mathbf{S}u_{\text{max}}}.\\
	\end{array}\label{eq:orb_control_problem_compact_form}
\end{equation}

\subsection{Attitude control}

\begin{equation}
	\begin{array}{rrclcl}
		\displaystyle \min_{\Delta\mathbf{x}_{\text{att}}(t),\mathbf{T}_u(t)} && \multicolumn{3}{l} {J_{\text{att}}=\frac{1}{t_f-t_0}\int^{t_f}_{t_0}\left(\gamma_{\text{att}}\Delta\mathbf{x}^T_{\text{att}}(t)\mathbf{P}_{x_{\text{att}}}\Delta\mathbf{x}_{\text{att}}(t)+\mathbf{T}_u^T(t)\mathbf{T}_u(t)\right)dt,} \\
		\textrm{s.t.} &&&\Delta\dot{\mathbf{x}}_{\text{att}}(t)=\dot{\mathbf{x}}_{\text{att}}(\mathbf{x}_{\text{att}}(t),\mathbf{T}_u(t))-\dot{\bar{\mathbf{x}}}_{\text{att}}(t)+\Delta\dot{\bar{\mathbf{x}}}_{\text{att}}(t),\\
		&&&-\mathbf{T}_{u_{\text{max}}} \leq \mathbf{T}_u(t) \leq \mathbf{T}_{u_{\text{max}}}.\\
	\end{array}\label{eq:att_control_problem_continuous}
\end{equation}
\textbf{Dynamics linearization:}
\begin{equation*}\label{eq:att_dynamics_linearization}
	\dot{\mathbf{x}}_{\text{att}}(t)\approx\dot{\bar{\mathbf{x}}}_{\text{att}}(t)+\mathbf{A}_{\text{att}}(\bar{\mathbf{x}}_{\text{att}}(t))\Delta\mathbf{x}_{\text{att}}(t)+\mathbf{B}_{\text{att}}(\bar{\mathbf{x}}_{\text{att}}(t))\mathbf{T}_u(t),
\end{equation*}
\begin{equation*}
	\mathbf{A}_{\text{att}}=\frac{\partial \dot{\mathbf{x}}_{\text{att}}}{\partial \mathbf{x}_{\text{att}}}\bigg\rvert_{\bar{\mathbf{x}}_{\text{att}}(t)}+\frac{\partial \dot{\mathbf{x}}_{\text{att}}}{\partial\mathbf{T}_{\text{grav}}}\frac{\partial\mathbf{T}_{\text{grav}}}{\partial{\mathbf{x}_{\text{att}}}}\bigg\rvert_{\bar{\mathbf{x}}_{\text{att}}(t)},\quad\mathbf{B}_{\text{att}}=\frac{\partial \dot{\mathbf{x}}_{\text{att}}}{\partial \mathbf{T}_u}\bigg\rvert_{\bar{\mathbf{x}}_{\text{att}}(t)},
\end{equation*}
\begin{equation*}\label{eq:att_err_state_dynamics_linear}
	\Delta\dot{\mathbf{x}}_{\text{att}}(t)=\mathbf{A}_{\text{att}}(\bar{\mathbf{x}}_{\text{att}}(t))\Delta\mathbf{x}_{\text{att}}(t)+\mathbf{B}_{\text{att}}(\bar{\mathbf{x}}_{\text{att}}(t))\mathbf{T}_u(t)+\Delta\dot{\bar{\mathbf{x}}}_{\text{att}}(t),
\end{equation*}
\begin{equation*}\label{eq:att_err_state_LTV_solution}
	\Delta\mathbf{x}_{\text{att}}(t)=\pmb{\Phi}_{\text{att}}(t,t_0)\Delta\mathbf{x}_{\text{att},0}+\int^{t}_{t_0}\pmb{\Phi}_{\text{att}}(t,\tau)\mathbf{B}_{\text{att}}(\tau)\mathbf{T}_u(\tau)d\tau+\Delta\bar{\mathbf{x}}_{\text{att}}(t),
\end{equation*}  
\begin{equation*}
	\dot{\pmb{\Phi}}_{\text{att}}(t,t_0)=\mathbf{A}_{\text{att}}(\bar{\mathbf{x}}_{\text{att}}(t))\pmb{\Phi}_{\text{att}}(t,t_0),\quad\pmb{\Phi}_{\text{att}}(t_0,t_0)=\mathbf{I}.\label{eq:STM_att_dynamics}
\end{equation*}
\textbf{Discretization:}
\begin{eqnarray*}
	\Delta\mathbf{x}_{\text{att},k}&=&\pmb{\Phi}_{\text{att}}(t_k,t_0)\Delta\mathbf{x}_{\text{att},0}
	%\\&&
	+\sum^{k}_{i=1}\pmb{\Phi}_{\text{att}}(t_k,t_i)\left(\int^{t_i}_{t_{i-1}}\pmb{\Phi}_{\text{att}}(t_i,\tau)\mathbf{B}_{\text{att}}(\tau)d\tau\right)\mathbf{T}_{u,i}+\Delta\bar{\mathbf{x}}_{\text{att},k},\\
	J_{\text{att}}&=&\sum^{N_{\text{att}}}_{k=1}\left(\gamma_{\text{att}}\Delta\mathbf{x}^T_{\text{att},k}\mathbf{P}_{x_{\text{att}}}\Delta\mathbf{x}_{\text{att},k}+\mathbf{T}_{u,k}^T\mathbf{T}_{u,k}\right).
\end{eqnarray*}
\textbf{Compact formulation:}
\begin{equation*}\label{eq:att_err_state_compact_propagation}
	\Delta\mathbf{x}_{\mathbf{S}\text{att}}=\mathbf{D}_{\text{att}}\Delta\mathbf{x}_{\text{att},0}+\mathbf{G}_{\text{att}}\mathbf{T}_{\mathbf{S}u}+\Delta\bar{\mathbf{x}}_{\mathbf{S}\text{att}},
\end{equation*}
\begin{equation*}\label{eq:att_obj_function_compact}
	J_{\text{att}}=\gamma_{\text{att}}\Delta\mathbf{x}^T_{\mathbf{S}\text{att}}\mathbf{P}_{\mathbf{S}x_{\text{att}}}\Delta\mathbf{x}_{\mathbf{S}\text{att}}+\mathbf{T}^T_{\mathbf{S}u}\mathbf{T}_{\mathbf{S}u},
\end{equation*}
\begin{equation}
\begin{array}{rll}
			\displaystyle \min_{\mathbf{T}_{\mathbf{S}u}} &J_{\text{att}}=2\gamma_{\text{att}}(\mathbf{D}_{\text{att}}\Delta\mathbf{x}_{\text{att},0}+\Delta\bar{\mathbf{x}}_{\mathbf{S}\text{att}})^T\mathbf{P}_{\mathbf{S}x_{\text{att}}}\mathbf{G}_{\text{att}}\mathbf{T}_{\mathbf{S}u}
			%\\&
			+\mathbf{T}^T_{\mathbf{S}u}(\gamma_{\text{att}}\mathbf{G}_{\text{att}}^T\mathbf{P}_{\mathbf{S}x_{\text{att}}}\mathbf{G}_{\text{att}}+\mathbf{I})\mathbf{T}_{\mathbf{S}u}, \\
			\textrm{s.t.}
			&-\mathbf{T}_{\mathbf{S}u_{\text{max}}} \leq \mathbf{T}_{\mathbf{S}u} \leq \mathbf{T}_{\mathbf{S}u_{\text{max}}}.\\
	\end{array}\label{eq:att_control_problem_compact_form}
\end{equation}

%\bibliographystyle{aiaa}
%\bibliography{asteroid_bib}

\end{document}